\documentclass[a4paper,12pt,openany]{article}
\usepackage[utf8x]{inputenc}
\usepackage[english]{babel}
\usepackage{a4wide}
\usepackage{amssymb}
\usepackage{amsmath}
\usepackage{accents}
\usepackage{multirow}
\usepackage{bbding}
\usepackage{graphicx}
\usepackage{pstricks}
\usepackage{comment}
\usepackage{subfig}
\usepackage{hyperref}
\usepackage{ytableau}
\usepackage{colortbl}
\usepackage{tabularx}
\usepackage{subfloat}
\usepackage{placeins}
\usepackage{cite}
\usepackage{cleveref}
\usepackage{rotating}

\usepackage{lscape}

\hypersetup{
    bookmarks=false,         % show bookmarks bar?
    unicode=false,          % non-Latin characters in Acrobat’s bookmarks
    pdftoolbar=true,        % show Acrobat’s toolbar?
    pdfmenubar=true,        % show Acrobat’s menu?
    pdffitwindow=false,     % window fit to page when opened
    pdfstartview={FitH},    % fits the width of the page to the window
    pdftitle={},    % title
    pdfauthor={Alessio Marrani and Luca Romano},     % author
    pdfsubject={},   % subject of the document
    pdfnewwindow=true,      % links in new window
    colorlinks=false,       % false: boxed links; true: colored links
    linkcolor=red,          % color of internal links
    citecolor=red,        % color of links to bibliography
    filecolor=red,      % color of file links
    urlcolor=red           % color of external links
    linkbordercolor={red},
    citebordercolor={red},
    urlbordercolor={red}
}

\def\chapterautorefname~#1\null{Chap.~(#1)\null}
\def\sectionautorefname~#1\null{Sec.~(#1)\null}
\def\subsectionautorefname~#1\null{sub--Sec.~(#1)\null}
\def\figureautorefname~#1\null{Fig.~(#1)\null}
\def\tableautorefname~#1\null{Tab.~(#1)\null}
\def\equationautorefname~#1\null{eq.~(#1)\null}

\def\equationautorefname~#1\null{eq.~(#1)\null}

\begin{document}
\def\equationautorefname~#1\null{eq.~(#1)\null}
\def\tableautorefname~#1\null{tab.~(#1)\null}
\def\figureautorefname~#1\null{fig.~(#1)\null}

\begin{flushright}
\small
IFT-UAM/CSIC-19-85\\
%\texttt{arXiv:yymm.nnnnn [hep-th]}\\
%June 14 \textsuperscript{th}, 2019\\
\normalsize
\end{flushright}

\vspace{0.5cm}

\begin{center}

{\Large {\bf Orbits in Non-Supersymmetric Magic Theories}}

\vspace{1.5cm}

\renewcommand{\thefootnote}{\alph{footnote}}
{\sl\large Alessio Marrani$^{~1,2}$}\footnote{Email: {\tt alessio.marrani@pd.infn.it}}
{\sl\large and Luca Romano$^{~3}$}\footnote{Email: {\tt lucaromano2607@gmail.com}}

\setcounter{footnote}{0}
\renewcommand{\thefootnote}{\arabic{footnote}}

\vspace{0.8cm}

${}^{1}${\small \it Museo Storico della Fisica e Centro Studi e Ricerche “Enrico Fermi”,\\
Via Panisperna 89A, I-00184, Roma, Italy}

\vspace{0.2cm}
${}^{2}${\small \it Dipartimento di Fisica e Astronomia “Galileo Galilei”, Università di Padova, and INFN, sezione di Padova, Via Marzolo 8, I-35131 Padova, Italy}

\vspace{0.2cm}

${}^{3}${\small \it Instituto de F\'{\i}sica Te\'orica UAM/CSIC\\
C/ Nicol\'as Cabrera, 13--15,  C.U.~Cantoblanco, E-28049 Madrid, Spain}\\

\vspace{.8cm}

%%%%%%%%%%%%%%%%%%%%%%%%%%%%%%%%%%%%%%%%%%%%%%%%%%%%%%%%%%%%%%%%%%%%%%

{\bf Abstract}
\end{center}
\begin{quotation}
We determine and classify the electric-magnetic duality orbits of fluxes
supporting asymptotically flat, extremal black branes in $D=4,5,6$
space-time dimensions in the so-called non-supersymmetric magic
Maxwell-Einstein theories, which are consistent truncations of maximal
supergravity and which can be related to Jordan algebras (and related
Freudenthal triple systems) over the split complex numbers $\mathbb{C}_{s}$
and the split quaternions $\mathbb{H}_{s}$. By studying the stabilizing
subalgebras of suitable representatives, realized as bound states of
specific weight vectors of the corresponding representation of the
electric-magnetic duality symmetry group, we obtain that, as for the case of
maximal supergravity, in magic non-supersymmetric Maxwell-Einstein theories
there is no splitting of orbits, namely there is only one orbit for each
non-maximal rank element of the relevant Jordan algebra (in $D=5$ and $6$)
or of the relevant Freudenthal triple system (in $D=4$).
\end{quotation}

\newpage

\pagestyle{plain}

\tableofcontents

\newpage

\section{\label{intro}Introduction}

The so-called \textit{magic} Maxwell-Einstein supergravity theories
(MESGT's) were discovered in \cite%
{Gunaydin:1983rk,Gunaydin:1983bi,Gunaydin:1984ak}; they are endowed with
eighth-maximal ($\mathcal{N}=2$) local supersymmetry, and they can be
described in terms of Euclidean\footnote{%
The Lorentzian version of the exceptional, simple, Lorentzian cubic Jordan
algebras can also be defined; for its definition and symmetries, see \cite%
{Squaring-Magic}, and for its relation to MESGT's, see \cite{GZ}.} simple
Jordan algebras $J_{3}^{\mathbb{A}}$ of rank $3$, generated by $3\times 3$
Hermitian matrices over the four normed division algebras $\mathbb{A}=%
\mathbb{R}$ (real numbers), $\mathbb{C}$ (complex numbers), $\mathbb{H}$
(quaternions), and $\mathbb{O}$ (octonions). The \textquotedblleft magic" of
these theories can be traced back to the fact that their electric-magnetic ($%
U$-)duality\footnote{%
Here duality is referred to as the analogue in a non-supersymmetric context
of the \textquotedblleft continuous\textquotedblright\ symmetries of \cite%
{Cremmer:1978ds, Cremmer:1979up}, whose discrete versions in the
supersymmetric case are the $U$-dualities of non-perturbative string theory
introduced by Hull and Townsend \cite{Hull:1994ys}.} Lie groups in $D=3,4,5$
Lorentzian space-time dimensions can respectively be arranged into the
fourth, the third, and the second row of the \textit{single split}
(non-symmetric) real form of the \textit{magic square} of Freudenthal,
Rozenfeld and Tits (\cite{FRT}; see also \cite{Squaring-Magic}); it is worth
noting that the fourth row is made up only of exceptional Lie groups : $%
F_{4(4)}$, $E_{6(2)}$, $E_{7(-5)}$ and $E_{8(-24)}$, respectively.

The remarkable relation of magic MESGT's to simple Jordan algebras has
allowed for a quite elegant algebraic classification of the non-transitive
action of their global duality symmetries on the representation spaces of
fluxes supporting asymptotically flat black brane solutions; this has been
exploited firstly in \cite{Ferrara:1997uz}, and then in a number of papers
(see e.g. \cite{Borsten:2011ai} for a comprehensive treatment, and for list
of Refs.). In recent years, also other approaches yielded to deep insights
in the orbit stratifications; for instance, in \cite{Marrani-Riccioni-Romano}
the duality orbits were investigated by studying the stabilizing subalgebras
of suitable representatives, realized as bound states of specific weight
vectors of the corresponding duality representations.

For what concerns the massless spectrum of magic MESGT's, in \cite%
{Riccioni:2008jz} its bosonic sector (also including the $(D-1)$-forms and
the $D$-forms) was obtained from the very-extended Kac-Moody algebras $%
\mathfrak{g}^{+++}$ (where $\mathfrak{g}$ denotes the suitable non-compact,
real form of the Lie algebra of the $D=3$ $U$-duality group). More
specifically, the deletion of a suitable node in the Tits-Satake diagram of $%
\mathfrak{g}^{+++}$ determines the global symmetry of the theory in all the
dimensions in which the corresponding MESGT can be consistently defined; the
fact that the nodes of the Tits-Satake diagrams corresponding to compact
Cartan generators cannot be deleted explains why the magic MESGT's can be
defined only up to $D=6$.\bigskip

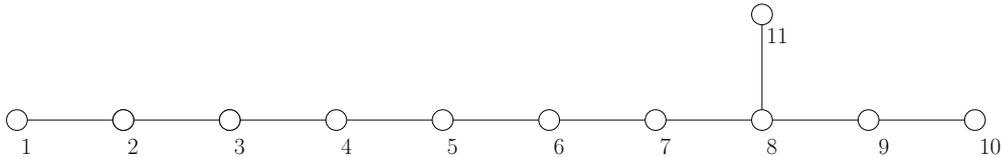
\begin{figure}[h]
\centering
% Generated with LaTeXDraw 2.0.8
% Thu Jun 16 10:12:40 CEST 2016
% \usepackage[usenames,dvipsnames]{pstricks}
% \usepackage{epsfig}
% \usepackage{pst-grad} % For gradients
% \usepackage{pst-plot} % For axes
\scalebox{0.35} % Change this value to rescale the drawing.
{\
\begin{pspicture}(0,-1.14375)(39.1875,4.10375)
% \definecolor{color392b}{rgb}{1.0,0.7843137254901961,0.0}
% \definecolor{color400b}{rgb}{0.996078431372549,0.996078431372549,0.996078431372549}
% \definecolor{color438b}{rgb}{0.8509803921568627,0.0,0.0}
\psline[linewidth=0.02cm](1.926875,-0.29625)(38.126877,-0.29625)
\psline[linewidth=0.02cm](30.126875,3.70375)(30.126875,-0.29625)
\pscircle[linewidth=0.02,dimen=outer,fillstyle=solid](2.126875,-0.29625){0.4}
\pscircle[linewidth=0.02,dimen=outer,fillstyle=solid](6.126875,-0.29625){0.4}
\pscircle[linewidth=0.02,dimen=outer,fillstyle=solid](10.126875,-0.29625){0.4}
\pscircle[linewidth=0.02,dimen=outer,fillstyle=solid](14.126875,-0.29625){0.4}
\pscircle[linewidth=0.02,dimen=outer,fillstyle=solid](18.126875,-0.29625){0.4}
\pscircle[linewidth=0.02,dimen=outer,fillstyle=solid](30.126875,3.70375){0.4}
\pscircle[linewidth=0.02,dimen=outer,fillstyle=solid](26.126875,-0.29625){0.4}
\pscircle[linewidth=0.02,dimen=outer,fillstyle=solid](22.126875,-0.29625){0.4}
\rput(2.47875,-1.28625){\huge 1}
\rput(26.490938,-1.28625){\huge 7}
\rput(22.503124,-1.28625){\huge 6}
\rput(18.491875,-1.28625){\huge 5}
\rput(14.512813,-1.28625){\huge 4}
\rput(10.4948435,-1.28625){\huge 3}
\rput(6.4889064,-1.28625){\huge 2}
\rput(38.69875,-1.28625){\huge 10}
\pscircle[linewidth=0.02,dimen=outer,fillstyle=solid](30.126875,-0.29625){0.4}
\pscircle[linewidth=0.02,dimen=outer,fillstyle=solid](38.126877,-0.29625){0.4}
\pscircle[linewidth=0.02,dimen=outer,fillstyle=solid](34.126877,-0.29625){0.4}
\rput(30.501875,-1.28625){\huge 8}
\rput(34.707813,-1.28625){\huge 9}
\rput(30.702969,2.91375){\huge 11}
\pscircle[linewidth=0.02,dimen=outer,fillstyle=solid](10.126875,-0.29625){0.4}
\pscircle[linewidth=0.02,dimen=outer,fillstyle=solid](6.126875,-0.29625){0.4}
% \rput(4.31375,-3.63625){\Huge $Gl(3,\mathds{R})$}
% \rput(25.35375,-3.63625){\Huge $E_{8}$}
% \psline[linewidth=0.08](1.526875,-1.89625)(1.526875,-2.49625)(6.726875,-2.49625)(6.726875,-1.89625)(6.726875,-1.89625)
% \psline[linewidth=0.08](13.526875,-1.89625)(13.526875,-2.49625)(38.726875,-2.49625)(38.726875,-1.89625)
% \rput(9.966406,0.76375){\Huge l}
\end{pspicture}
}
\caption{{\protect\footnotesize The }$\mathfrak{e}${\protect\footnotesize $%
_{8(8)}^{+++}$ Dynkin diagram.}}
\label{E88+++dynkindiagram}
\end{figure}

One can also consider the split versions of complex numbers, quaternions and
octonions, respectively denoted by $\mathbb{C}_{s}$, $\mathbb{H}_{s}$, and $%
\mathbb{O}_{s}$; such normed composition algebras are no more division,
because they contain non-trivial zero divisors. Correspondingly, a doubly
split (symmetric) magic square can be constructed using $\mathbb{R}$, $%
\mathbb{C}_{s}$, $\mathbb{H}_{s}$, $\mathbb{O}_{s}$ \cite{Barton:2000ki}
(also \textit{cfr.} \cite{Squaring-Magic}), and it is given in Table \ref%
{doublysplitmagicsquare}.
\begin{table}[t]
\renewcommand{\arraystretch}{1.1}
\par
\begin{center}
\begin{tabular}{|c||c|c|c|c|}
\hline
& $\mathbb{R}$ & $\mathbb{C}_s$ & ${\mathbb{H}}_{s}$ & ${\mathbb{O}}_{s}$ \\
\hline\hline
$\mathbb{R}$ & $SO(3)$ & $SL(3,\mathbb{R})$ & $Sp(6,\mathbb{R})$ & $F_{4(4)}$
\\ \hline
$\mathbb{C}_s$ & $SL(3,\mathbb{R})$ & $SL(3,\mathbb{R}) \times SL(3,\mathbb{R%
})$ & $SL(6,\mathbb{R})$ & $E_{6(6)}$ \\ \hline
${\mathbb{H}}_{s}$ & $Sp(6,\mathbb{R})$ & $SL(6,\mathbb{R})$ & $SO(6,6)$ & $%
E_{7(7)}$ \\ \hline
${\mathbb{O}}_{s}$ & $F_{4(4)}$ & $E_{6(6)}$ & $E_{7(7)}$ & $E_{8(8)}$ \\
\hline
\end{tabular}%
\end{center}
\caption{{\protect\footnotesize The doubly split magic square \protect\cite%
{Barton:2000ki}.}}
\label{doublysplitmagicsquare}
\end{table}
The theories based on the corresponding Euclidean simple cubic Jordan
algebras $J_{3}^{\mathbb{A}_{s}}$ are an interesting class of
Maxwell-Einstein theories; in fact, they all share the feature that upon
dimensional reduction to $D=3$, the resulting $U$-duality group (which can
be realized as quasiconformal \cite{GNK} symmetry of the corresponding $%
J_{3}^{\mathbb{A}_{s}}$) is real and maximally non-compact (i.e., split),
and of exceptional type $E_{n(n)}$, namely $E_{6(6)}$ for $\mathbb{C}_{s}$, $%
E_{7(7)}$ for $\mathbb{H}_{s}$ and $E_{8(8)}$ for $\mathbb{O}_{s}$. The
theory based on split octonions $\mathbb{O}_{s}$ (and thus on the
corresponding split form $J_{3}^{\mathbb{O}_{s}}$ of the Albert algebra) has
$E_{8(8)}$ $U$-duality symmetry in $D=3$, and it is \textit{maximal}
supergravity. Maximally supersymmetric MESGT's in any dimension are related
to the very-extended Kac-Moody algebra $\mathfrak{e}_{8(8)}^{+++}$ (also
dubbed $\mathfrak{e}_{11}$) \cite{West:2001as}; since this is a split form,
the Tits-Satake diagram coincides with the Dynkin diagram, with all
non-compact white nodes (corresponding to non-compact Cartan generators), as
given in fig. \ref{E88+++dynkindiagram}.
The supergravity theory in $D$ Lorentzian space-time dimensions is related
to the decomposition of $\mathfrak{e}_{8(8)}^{+++}$ in which the
\textquotedblleft gravity line\textquotedblright\ is identified with the $%
\mathfrak{a}_{D-1}\approx \mathfrak{sl}(D,\mathbb{R})$ subalgebra containing
node $1$; on the other hand, the portion of the diagram not connected to the
$\mathfrak{a}_{D-1}$ subalgebra gives rise to the internal ($U$-duality)
symmetry. Thus, from fig. \ref{E88+++dynkindiagram} one can realize that
maximal supergravity can consistently be defined for any $3\leqslant
D\leqslant 11$, with two different theories existing in $D=10$, namely the
IIA non-chiral theory (with $\mathfrak{a}_{9}$ given by nodes from $1$ to $9$%
) and the IIB chiral theory (with $\mathfrak{a}_{9}$ given by nodes from $1$
to $8$, plus node 11) \cite{Schnakenburg:2001he}. Moreover, the resulting
decomposition of $\mathfrak{e}_{8(8)}^{+++}$ yields all the $p$-forms of the
maximally supersymmetric theory in the corresponding dimension $D$, sitting
in the pertaining representation(s) of the $U$-duality group, as given in
Table \ref{E8+++spectrum} \cite{Riccioni:2007au,Bergshoeff:2007qi}.
\begin{table}[h]
\renewcommand{\arraystretch}{1.1}
\par
\begin{center}
\scalebox{.75}{
\begin{tabular}{|c|c||c|c|c|c|c|c|c|c|c|c|}
\hline
Dim & Symmetry  & $p=1$  & $p=2$ & $p=3$ & $p=4$ & $p=5$ & $p=6$ & $p=7$ & $p=8$ & $p=9$ & $p=10$ \\
\hline
\hline
11 & $-$ &  &   &  ${\bf 1}$  &  &  & ${\bf 1}$ &  & & & \\
\hline
10A & $\mathbb{R}^+$ & ${\bf 1}$ & ${\bf 1}$ & ${\bf 1}$ & & ${\bf 1}$ & ${\bf 1}$  & ${\bf 1}$ & ${\bf 1}$ & ${\bf 1}$ & $2\times {\bf 1}$\\
\hline
\multirow{2}{*}{ 10B }& \multirow{2}{*}{$SL(2,\mathbb{R})$} &  &\multirow{2}{*}{ ${\bf 2}$} &  & \multirow{2}{*}{${\bf 1}$}  & & \multirow{2}{*}{${\bf 2}$}  & & \multirow{2}{*}{${\bf 3}$}  &  & $ {\bf 4}$\\
 &  &  &  &  & & &   & &   &  & $ {\bf 2}$\\
\hline \multirow{2}{*}{$9$} & \multirow{2}{*}{$GL(2,\mathbb{R})$} & ${\bf 2}$ & \multirow{2}{*}{${\bf 2}$} & \multirow{2}{*}{${\bf 1}$} &  \multirow{2}{*}{${\bf 1}$}  & \multirow{2}{*}{${\bf 2}$} & ${\bf 2}$ &${\bf 3}$ & ${\bf 3}$ & ${\bf 4}$
 \\
 & & ${\bf 1}$& & &&  & ${\bf 1}$ & ${\bf 1}$& ${\bf 2}$& $2 \times {\bf 2}$ \\
 \cline{1-11}
\multirow{3}{*}{$8$} & \multirow{3}{*}{$SL(3,\mathbb{R})\times SL(2,\mathbb{R})$ } & \multirow{3}{*}{${\bf (\overline{3},2)}$} & \multirow{3}{*}{${\bf ({3},1)}$}  & \multirow{3}{*}{${\bf ({1},2)}$}  & \multirow{3}{*}{${\bf (\overline{3},1)}$}  & \multirow{3}{*}{${\bf ({3},2)}$} & ${\bf (8,1)}$& ${\bf (6,2)}$ & ${\bf (15,1)}$\\
& & & & & & &  & & ${\bf (3,3)}$\\
&&&&&&& ${\bf ({1},3)}$& ${\bf (\overline{3},2)}$& $2\times {\bf (3,1)}$\\
\cline{1-10}
\multirow{3}{*}{$7$} & \multirow{3}{*}{$SL(5,\mathbb{R})$} &\multirow{3}{*}{${\bf \overline{10}}$} & \multirow{3}{*}{${\bf 5}$} & \multirow{3}{*}{${\bf \overline{5}}$} & \multirow{3}{*}{${\bf 10}$} & \multirow{3}{*}{${\bf 24}$} & ${\bf \overline{40}}$&  $ {\bf 70}$  \\
& & &  & &  &  & & ${\bf {45}}$  \\
& & & & &  &  & ${\bf \overline{15}}$& ${\bf {5}}$  \\
\cline{1-9}
\multirow{3}{*}{$6$} & \multirow{3}{*}{$SO(5,5)$} &\multirow{3}{*}{${\bf 16}$} & \multirow{3}{*}{${\bf 10}$} & \multirow{3}{*}{${\bf \overline{16}}$} & \multirow{3}{*}{${\bf 45}$} &\multirow{3}{*}{${\bf 144}$} & $ {\bf 320}$ \\
& & &  & &  & & ${\bf \overline{126}}$ \\
& & & & &  &  & ${\bf {10}}$ \\
\cline{1-8}
\multirow{2}{*}{$5$} & \multirow{2}{*}{$E_{6(6)}$} &\multirow{2}{*}{${\bf 27}$} & \multirow{2}{*}{${\bf \overline{27}}$} & \multirow{2}{*}{${\bf 78}$} & \multirow{2}{*}{${\bf 351}$} &${\bf  \overline{1728}}$  \\
 & & &  & &  &${\bf \overline{27}}$  \\
\cline{1-7}
\multirow{2}{*}{$4$} & \multirow{2}{*}{$E_{7(7)}$} &\multirow{2}{*}{${\bf 56}$} & \multirow{2}{*}{${\bf 133}$} & \multirow{2}{*}{${\bf 912}$} & ${\bf 8645}$ \\
 & & &  & & ${\bf {133}}$ \\
\cline{1-6}
\multirow{3}{*}{$3$} & \multirow{3}{*}{$E_{8(8)}$} &\multirow{3}{*}{${\bf 248}$} & ${\bf 3875}$ & ${\bf 147250}$  \\
 & & & & ${\bf 3875}$ \\
& & & ${\bf 1}$& ${\bf 248}$ \\
\cline{1-5}
\end{tabular}
}
\end{center}
\caption{{\protect\footnotesize All the $p$-forms of the }$\mathfrak{e}%
_{8(8)}^{+++}$ {\protect\footnotesize theory in any dimension.}}
\label{E8+++spectrum}
\end{table}
\bigskip

\begin{figure}[h]
\centering
% Generated with LaTeXDraw 2.0.8
% Thu Jun 16 10:12:40 CEST 2016
% \usepackage[usenames,dvipsnames]{pstricks}
% \usepackage{epsfig}
% \usepackage{pst-grad} % For gradients
% \usepackage{pst-plot} % For axes
\scalebox{0.35} % Change this value to rescale the drawing.
{\
\begin{pspicture}(0,-1.14375)(35.1875,8.10375)
% \definecolor{color392b}{rgb}{1.0,0.7843137254901961,0.0}
% \definecolor{color400b}{rgb}{0.996078431372549,0.996078431372549,0.996078431372549}
% \definecolor{color438b}{rgb}{0.8509803921568627,0.0,0.0}
\psline[linewidth=0.02cm](5.926875,-0.29625)(30.126877,-0.29625)
\psline[linewidth=0.02cm](22.126875,7.70375)(22.126875,-0.29625)
% \pscircle[linewidth=0.02,dimen=outer,fillstyle=solid](2.126875,-0.29625){0.4}
\pscircle[linewidth=0.02,dimen=outer,fillstyle=solid](6.126875,-0.29625){0.4}
\pscircle[linewidth=0.02,dimen=outer,fillstyle=solid](10.126875,-0.29625){0.4}
\pscircle[linewidth=0.02,dimen=outer,fillstyle=solid](14.126875,-0.29625){0.4}
\pscircle[linewidth=0.02,dimen=outer,fillstyle=solid](18.126875,-0.29625){0.4}
\pscircle[linewidth=0.02,dimen=outer,fillstyle=solid](22.126875,3.70375){0.4}
\pscircle[linewidth=0.02,dimen=outer,fillstyle=solid](22.126875,7.70375){0.4}
\pscircle[linewidth=0.02,dimen=outer,fillstyle=solid](26.126875,-0.29625){0.4}
\pscircle[linewidth=0.02,dimen=outer,fillstyle=solid](22.126875,-0.29625){0.4}
% \rput(2.47875,-1.28625){\huge 1}
\rput(26.490938,-1.28625){\huge 6}
\rput(22.503124,-1.28625){\huge 5}
\rput(18.491875,-1.28625){\huge 4}
\rput(14.512813,-1.28625){\huge 3}
\rput(10.4948435,-1.28625){\huge 2}
\rput(6.4889064,-1.28625){\huge 1}
% \rput(38.69875,-1.28625){\huge 10}
\pscircle[linewidth=0.02,dimen=outer,fillstyle=solid](30.126875,-0.29625){0.4}
% \pscircle[linewidth=0.02,dimen=outer,fillstyle=solid](38.126877,-0.29625){0.4}
% \pscircle[linewidth=0.02,dimen=outer,fillstyle=solid](34.126877,-0.29625){0.4}
\rput(30.501875,-1.28625){\huge 7}
% \rput(34.707813,-1.28625){\huge 9}
\rput(22.702969,2.91375){\huge 8}
\rput(22.702969,6.91375){\huge 9}
\pscircle[linewidth=0.02,dimen=outer,fillstyle=solid](10.126875,-0.29625){0.4}
\pscircle[linewidth=0.02,dimen=outer,fillstyle=solid](6.126875,-0.29625){0.4}
% \rput(4.31375,-3.63625){\Huge $Gl(3,\mathds{R})$}
% \rput(25.35375,-3.63625){\Huge $E_{8}$}
% \psline[linewidth=0.08](1.526875,-1.89625)(1.526875,-2.49625)(6.726875,-2.49625)(6.726875,-1.89625)(6.726875,-1.89625)
% \psline[linewidth=0.08](13.526875,-1.89625)(13.526875,-2.49625)(38.726875,-2.49625)(38.726875,-1.89625)
% \rput(9.966406,0.76375){\Huge l}
\end{pspicture}
}
\caption{{\protect\footnotesize The }$\mathfrak{e}${\protect\footnotesize $%
_{6(6)}^{+++}$ Dynkin diagram.}}
\label{E66+++dynkindiagram}
\end{figure}
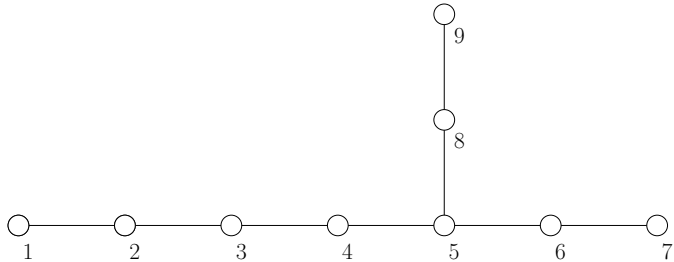

The Maxwell-Einstein theories based on split quaternions ${\mathbb{H}}_{s}$
and on split complex (also named hypercomplex) numbers $\mathbb{C}_{s}$ (and
thus on $J_{3}^{\mathbb{H}_{s}}$ resp. $J_{3}^{\mathbb{C}_{s}}$) are
consistent truncations \cite{Kleinschmidt:2003mf} of maximal supergravity,
but they are non-supersymmetric: namely, their field content cannot be
regarded as the bosonic sector of a theory endowed with any amount of local
supersymmetry; while maximal supergravity has been the object of intense
study along the years (for an analysis of its $U$-orbit structure, see \cite%
{FG-1,LPS-1,F-Maldacena-1,ICL-1,FG-2}), its ${\mathbb{H}}_{s}$- and ${%
\mathbb{C}}_{s}$- based truncations are much less known, and they have been
analyzed in some detail only recently, in \cite%
{Marrani-Pradisi-Riccioni-Romano}. Remarkably, the electric-magnetic ($U$%
-)duality groups of such theories in $D=3,4,5$ Lorentzian space-time
dimensions correspond to the (complex, quaternionic and octonionic columns
of the) fourth, the third and the second rows of the aforementioned \textit{%
doubly split }(symmetric) \textit{magic square}, given in Table \ref%
{doublysplitmagicsquare}. Thus, such theories will be named \textit{magic
non-supersymmetric} Maxwell-Einstein theories. Analogously to their
aforementioned maximal $\mathbb{O}_{s}$-based counterpart, magic
non-supersymmetric theories correspond to the very-extended Kac-Moody
algebras $\mathfrak{e}_{7(7)}^{+++}$ resp. $\mathfrak{e}_{6(6)}^{+++}$ \cite%
{Englert:2003zs,Kleinschmidt:2003mf}; such split forms were firstly
investigated in \cite{Englert:2003zs}, and a progressive deletion of nodes
of the corresponding Tits-Satake diagrams (respectively given by figs. \ref%
{E77+++dynkindiagram} and \ref{E66+++dynkindiagram}), it turns out that they
can be consistently defined up to $D=10$ resp. $D=8$; in such maximal
dimensions, the bosonic spectrum of these theories has been determined in
\cite{Kleinschmidt:2003mf}.

An analysis of the $\mathbb{H}_{s}$- and $\mathbb{C}_{s}$- based magic
non-supersymmetric theories has been carried out in \cite%
{Marrani-Pradisi-Riccioni-Romano}, in which it was shown that they arise as
\textquotedblleft Ehlers\textquotedblright\ $sl(2,\mathbb{R})$- and $sl(3,%
\mathbb{R})$- truncations \cite{Ferrara:2012zc} of maximal
supergravity.\medskip

The magic non-supersymmetric Maxwell-Einstein theory based on $\mathbb{H}%
_{s} $ (and thus on $J_{3}^{\mathbb{H}_{s}}$), is related to $\mathfrak{e}%
_{7(7)}^{+++}$, whose Dynkin diagram is drawn in fig. \ref%
{E77+++dynkindiagram}.
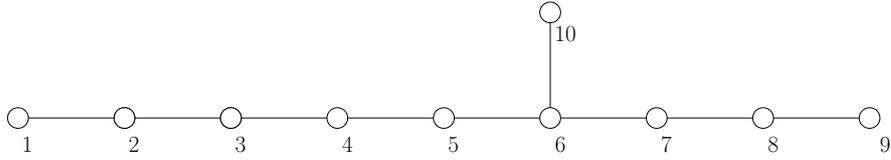
\begin{figure}[h]
\centering
% Generated with LaTeXDraw 2.0.8
% Thu Jun 16 10:12:40 CEST 2016
% \usepackage[usenames,dvipsnames]{pstricks}
% \usepackage{epsfig}
% \usepackage{pst-grad} % For gradients
% \usepackage{pst-plot} % For axes
\scalebox{0.35} % Change this value to rescale the drawing.
{\
\begin{pspicture}(0,-1.14375)(35.1875,4.10375)
% \definecolor{color392b}{rgb}{1.0,0.7843137254901961,0.0}
% \definecolor{color400b}{rgb}{0.996078431372549,0.996078431372549,0.996078431372549}
% \definecolor{color438b}{rgb}{0.8509803921568627,0.0,0.0}
\psline[linewidth=0.02cm](1.926875,-0.29625)(34.126877,-0.29625)
\psline[linewidth=0.02cm](22.126875,3.70375)(22.126875,-0.29625)
\pscircle[linewidth=0.02,dimen=outer,fillstyle=solid](2.126875,-0.29625){0.4}
\pscircle[linewidth=0.02,dimen=outer,fillstyle=solid](6.126875,-0.29625){0.4}
\pscircle[linewidth=0.02,dimen=outer,fillstyle=solid](10.126875,-0.29625){0.4}
\pscircle[linewidth=0.02,dimen=outer,fillstyle=solid](14.126875,-0.29625){0.4}
\pscircle[linewidth=0.02,dimen=outer,fillstyle=solid](18.126875,-0.29625){0.4}
\pscircle[linewidth=0.02,dimen=outer,fillstyle=solid](22.126875,3.70375){0.4}
\pscircle[linewidth=0.02,dimen=outer,fillstyle=solid](26.126875,-0.29625){0.4}
\pscircle[linewidth=0.02,dimen=outer,fillstyle=solid](22.126875,-0.29625){0.4}
\rput(2.47875,-1.28625){\huge 1}
\rput(26.490938,-1.28625){\huge 7}
\rput(22.503124,-1.28625){\huge 6}
\rput(18.491875,-1.28625){\huge 5}
\rput(14.512813,-1.28625){\huge 4}
\rput(10.4948435,-1.28625){\huge 3}
\rput(6.4889064,-1.28625){\huge 2}
% \rput(38.69875,-1.28625){\huge 10}
\pscircle[linewidth=0.02,dimen=outer,fillstyle=solid](30.126875,-0.29625){0.4}
% \pscircle[linewidth=0.02,dimen=outer,fillstyle=solid](38.126877,-0.29625){0.4}
\pscircle[linewidth=0.02,dimen=outer,fillstyle=solid](34.126877,-0.29625){0.4}
\rput(30.501875,-1.28625){\huge 8}
\rput(34.707813,-1.28625){\huge 9}
\rput(22.702969,2.91375){\huge 10}
\pscircle[linewidth=0.02,dimen=outer,fillstyle=solid](10.126875,-0.29625){0.4}
\pscircle[linewidth=0.02,dimen=outer,fillstyle=solid](6.126875,-0.29625){0.4}
% \rput(4.31375,-3.63625){\Huge $Gl(3,\mathds{R})$}
% \rput(25.35375,-3.63625){\Huge $E_{8}$}
% \psline[linewidth=0.08](1.526875,-1.89625)(1.526875,-2.49625)(6.726875,-2.49625)(6.726875,-1.89625)(6.726875,-1.89625)
% \psline[linewidth=0.08](13.526875,-1.89625)(13.526875,-2.49625)(38.726875,-2.49625)(38.726875,-1.89625)
% \rput(9.966406,0.76375){\Huge l}
\end{pspicture}
}
\caption{{\protect\footnotesize The }$\mathfrak{e}${\protect\footnotesize $%
_{7(7)}^{+++}$ Dynkin diagram.}}
\label{E77+++dynkindiagram}
\end{figure}
As mentioned above, the $D$-dimensional theory corresponds to an $sl(D,%
\mathbb{R})$ symmetry in the diagram, which involves the nodes with labels $%
1 $ to $D-1$, whereas those nodes which are not connected to any of the $%
sl(D,\mathbb{R})$ nodes determine the global, electric-magnetic duality Lie
algebra.

\begin{table}[t]
\renewcommand{\arraystretch}{1.1}
\par
\begin{center}
\scalebox{.9}{
\begin{tabular}{|c|c||c|c|c|c|c|c|c|c|}
\hline
Dim & Symmetry  & $p=1$  & $p=2$ & $p=3$ & $p=4$ & $p=5$ & $p=6$ & $p=7$ & $p=8$  \\
\hline
\hline
10 & $-$ &  &   &    & ${\bf 1}$ &  & &  & \\
\hline 9 & $\mathbb{R}^+$ & ${\bf 1}$ & &${\bf 1}$& ${\bf 1}$ &  &${\bf 1}$ & ${\bf 1}$ &
 \\
 \hline
\multirow{2}{*}{8A} & \multirow{2}{*}{$GL(2,\mathbb{R})$} & \multirow{2}{*}{${\bf 2}$} & \multirow{2}{*}{${\bf 1}$} & \multirow{2}{*}{${\bf 2}$} & \multirow{2}{*}{${\bf 1}$} & \multirow{2}{*}{${\bf 2}$}& ${\bf 3}$& \multirow{2}{*}{$2\times{\bf 2}$}&   ${\bf 3}$\\
& & & & & & & ${\bf 1}$ & & $2\times{\bf 1}$\\
\hline
\multirow{2}{*}{8B }& \multirow{2}{*}{$SL(3,\mathbb{R})$} & & \multirow{2}{*}{${\bf 3}$} &  & \multirow{2}{*}{${\bf \overline{3}}$ }& & \multirow{2}{*}{${\bf 8}$}& & ${\bf 15}$\\
& & & & & & &  & & ${\bf 3}$\\
\hline
& & ${\bf {3}}$ & & & ${\bf \overline{3}}$ &  ${\bf 8}$& ${\bf {8}}$& ${\bf {15}}$ \\
$7$& $GL(3,\mathbb{R})$ & & ${\bf 3}$ &${\bf \overline{3}}$&  &  & ${\bf \overline{6}}$& ${\bf \overline{6}}$ \\
 &  &${\bf 1}$ &  &  & ${\bf 1}$ & ${\bf 1}$& ${\bf 3}$&  $2 \times {\bf 3}$  \\
\cline{1-9}
\multirow{4}{*}{$6$} & \multirow{4}{*}{$SL(4,\mathbb{R})\times SL(2,\mathbb{R})$} &\multirow{4}{*}{${\bf (4,2)}$} & \multirow{4}{*}{${\bf (6,1)}$} & \multirow{4}{*}{${\bf (\overline{4},2)}$} &  & &  ${\bf (64,1)}$\\
& & &  & & ${\bf (15,1)}$ &${\bf (\overline{20},2)}$  & ${\bf (\overline{10},3)}$ \\
& & & & &${\bf (1,3)}$  & ${\bf (4,2)}$ &  ${\bf (6,3)}$ \\
& & & & &  &  &  $2\times {\bf (6,1)}$ \\
\cline{1-8}
& & & & &
${\bf \overline{105}}$  & ${\bf \overline{384}}$  \\
$5$ & $SL(6,\mathbb{R})$ &${\bf 15}$ & ${\bf \overline{15}}$ & ${\bf 35}$ &  & ${\bf 105}$ \\
 & & &  & & ${\bf 21}$ & ${\bf  \overline{15}}$ \\
\cline{1-7}
& & & & & ${\bf 2079}$  \\
$4$ & $SO(6,6)$ &${\bf 32}$ & ${\bf 66}$ & ${\bf 352}$ & ${\bf {462}}$ \\
 & & &  & &  ${\bf 66}$ \\
\cline{1-6}
& & & ${\bf 1539}$& ${\bf 40755}$ \\
$3$ & $E_{7(7)}$ &${\bf 133}$ & &  ${\bf 1539}$ \\
 & & &  ${\bf 1}$ & ${\bf 1}$\\
\cline{1-5}
\end{tabular}
}
\end{center}
\caption{{\protect\footnotesize All the $p$-forms of the }$\mathfrak{e}$%
{\protect\footnotesize $_{7(7)}^{+++}$ theory in any dimension.}}
\label{E7+++spectrum}
\end{table}
Table \ref{E7+++spectrum} lists all $p$-forms (fitting into representations
of the electric-magnetic duality) of the split quaternionic magic
non-supersymmetric theories in any dimension \cite%
{Marrani-Pradisi-Riccioni-Romano}. In order to get the full bosonic
spectrum, one has to add (Einstein) gravity as well the scalar fields,
coordinatizing a target space which is a symmetric manifold $G/H$, where $G$
is the electric-magnetic duality Lie group and $H$ its maximal compact
subgroup. Note that the aforementioned fact that the $\mathfrak{e}%
_{7(7)}^{+++}$ theory is an $\mathfrak{sl}(2,\mathbb{R})$-truncation of the
maximal $\mathfrak{e}_{8(8)}^{+++}$ theory also explains the occurrence of
two different theories (8A and 8B) in $D=8$ \cite%
{Marrani-Pradisi-Riccioni-Romano}; moreover, it should be noted that in all
dimensions $3\leqslant D\leqslant 10$, the representations of Table \ref%
{E7+++spectrum} result from $\mathfrak{sl}(2,\mathbb{R})$-invariant
truncations of the representations of Table \ref{E8+++spectrum}, with the
only exceptions being given by $D$-forms in $D$ dimensions and by $2$-forms
in $D=3$, in which cases the $\mathfrak{sl}(2,\mathbb{R})$-invariant
truncation overestimates the number of forms \cite%
{Marrani-Pradisi-Riccioni-Romano}. As explained in \cite%
{Marrani-Pradisi-Riccioni-Romano}, the perturbative symmetry of the $\mathbb{%
H}_{s}$-based magic theory in $D$ dimensions is $\mathfrak{so}%
(8-D,8-D)\oplus \mathfrak{sl}(2,\mathbb{R})$, where $\mathfrak{sl}(2,\mathbb{%
R})=\mathfrak{tri}(\mathbb{H}_{s})\ominus \mathfrak{so}(\mathbb{H}_{s})$,
with $\mathfrak{tri}(\mathbb{H}_{s})$ and $\mathfrak{so}(\mathbb{H}_{s})$
respectively denoting the \textit{triality} and the \textit{norm-preserving}
symmetries of $\mathbb{H}_{s}$ \cite{triality}.\medskip

A similar analysis can be done for the magic non-supersymmetric
Maxwell-Einstein theory based on $\mathbb{C}_{s}$ (and thus on $J_{3}^{%
\mathbb{C}_{s}}$), which is related to $\mathfrak{e}_{6(6)}^{+++}$, whose
Dynkin diagram is drawn in fig. \ref{E66+++dynkindiagram}. In any dimension,
the $p$-form massless spectrum of the theory can thus be obtained from the
very extended Kac-Moody algebra $\mathfrak{e}_{6(6)}^{+++}$ \cite%
{Englert:2003zs,Kleinschmidt:2003mf}, whose Dynkin diagram is given in fig. %
\ref{E66+++dynkindiagram}.
Table \ref{E6+++spectrum} lists all $p$-forms (fitting into representations
of the electric-magnetic duality) of the split complex magic
non-supersymmetric theories in any dimension \cite%
{Marrani-Pradisi-Riccioni-Romano}. As for the split quaternionic theories
treated above, the full bosonic massless spectrum is then obtained by adding
(Einstein) gravity as well the scalar fields, whose target space is the
symmetric coset $G/H$. Note that in all dimensions $3\leqslant D\leqslant 8$%
, the representations of Table \ref{E6+++spectrum} result from $\mathfrak{sl}%
(3,\mathbb{R})$-invariant truncations of the representations of Table \ref%
{E8+++spectrum}, again with the only exceptions being given by $D$-forms in $%
D$ dimensions and by $2$-forms in $D=3$, in which cases the $\mathfrak{sl}(3,%
\mathbb{R})$-invariant truncation overestimates the number of forms \cite%
{Marrani-Pradisi-Riccioni-Romano}. Again, as explained in \cite%
{Marrani-Pradisi-Riccioni-Romano}, the perturbative symmetry of the $\mathbb{%
C}_{s}$-based magic theory in $D$ dimensions is $\mathfrak{so}%
(7-D,7-D)\oplus \mathbb{R}$, where $\mathbb{R}\sim \mathfrak{so}(1,1)=%
\mathfrak{tri}(\mathbb{C}_{s})\ominus \mathfrak{so}(\mathbb{C}_{s})$, with $%
\mathfrak{tri}(\mathbb{C}_{s})$ and $\mathfrak{so}(\mathbb{C}_{s})$
respectively denoting the \textit{triality} and the \textit{norm-preserving}
symmetries of $\mathbb{C}_{s}$ \cite{triality}.

\begin{table}[t]
\renewcommand{\arraystretch}{1.1}
\par
\begin{center}
\scalebox{.9}{
\begin{tabular}{|c|c||c|c|c|c|c|c|c|}
\hline
Dim & Symmetry  & $p=1$  & $p=2$ & $p=3$ & $p=4$ & $p=5$ & $p=6$ & $p=7$  \\
\hline
\hline
$8$ & $SL(2,\mathbb{R})$ &  &  & ${\bf 2}$ &  & & ${\bf 3}$&  \\
\hline
\multirow{2}{*}{$7$ }& \multirow{2}{*}{$GL(2,\mathbb{R})$} & \multirow{2}{*}{${\bf 1}$} & \multirow{2}{*}{${\bf 2}$} & \multirow{2}{*}{${\bf {2}}$} & \multirow{2}{*}{${\bf 1}$} &${\bf 3}$ & ${\bf 3}$& ${\bf {4}}$ \\
& & &  & &  &${\bf 1}$  & ${\bf {2}}$&   $2 \times {\bf 2}$  \\
\hline
& & & & &  & ${\bf (3,2)}$   & ${\bf (4,2)}$   \\
& & ${\bf (2,1)}$ & & ${\bf ({2},1)}$ & ${\bf (3,1)}$ & ${\bf (2,3)}$ & ${\bf (2,4)}$  \\
$6$ & $(SL(2,\mathbb{R}))^2\times \mathbb{R}^+$ & & ${\bf (2,2)}$ & & ${\bf (1,1)}$ & &  $3\times{\bf (2,2)}$   \\
& & ${\bf (1,2)}$&  & ${\bf (1,2)}$& ${\bf (1,3)}$ & ${\bf (1,2)}$ & ${\bf (3,1)}$   \\
& & & & &  & ${\bf (2,1)}$ &${\bf (1,3)}$  \\
\cline{1-8}
& & & & & & ${\bf (\overline{15},\overline{3})}$    \\
& & & &${\bf (8,1)}$ & ${\bf (\overline{6},3)}$  & ${\bf (\overline{3},\overline{15})}$   \\
$5$ & $(SL(3,\mathbb{R}))^2$ &${\bf (3,3)}$ & ${\bf (\overline{3},\overline{3})}$ &  & ${\bf (3,3)}$ &$2\times {\bf (\overline{3},\overline{3})}$    \\
 & & &  & ${\bf (1,8)}$ & ${\bf (3,\overline{6})}$ &${\bf (6,\overline{3})}$   \\
& & & & & & ${\bf (\overline{3},6)}$   \\
\cline{1-7}
& & & & ${\bf 70}$ & ${\bf \overline{280}}$  \\
$4$ & $SL(6,\mathbb{R})$ &${\bf 20}$ & ${\bf 35}$ & &  ${\bf {280}}$  \\
 & & &  & ${\bf \overline{70}}$ & ${\bf 189}$   \\

\cline{1-6}
\multirow{4}{*}{$3$} & \multirow{4}{*}{$E_{6(6)}$} &\multirow{4}{*}{${\bf 78}$} &  &  ${\bf \overline{5824}}$ \\
 & & & ${\bf 650}$ & ${\bf 5824}$ \\
& & &${\bf 1}$ & ${\bf 650}$ \\
& & & & ${\bf 78}$ \\
\cline{1-5}
\end{tabular}
}
\end{center}
\caption{{\protect\footnotesize All the $p$-forms of the $e_{6(6)}^{+++}$
theory in any dimension.}}
\label{E6+++spectrum}
\end{table}
\bigskip

The present paper, expanding on \cite{Marrani-Pradisi-Riccioni-Romano}, is
devoted to the detailed analysis of the $U$-duality orbits of asymptotically
flat, extremal black branes in the aforementioned magic non-supersymmetric
Maxwell-Einstein theories, based on $\mathbb{H}_{s}$ and on $\mathbb{C}_{s}$%
, in $D=4,5,6$ Lorentzian space-time dimensions. Such theories, despite
being present in the classification of symmetric non-linear sigma models
coupled to Maxwell-Einstein gravity (\textit{cfr.} Table 2 of \cite{BGM-1}),
did not receive much attention in literature, as mentioned above. Symmetries
of Freudenthal triple systems and cubic Jordan algebras defined over split
algebras were studied in \cite{GNK} and \cite{G-Pavlyk-1} (see also table 1
of \cite{F-Dual}, and Refs. therein); theories over split algebras have been
quite recently considered, in a different context, in \cite{Bossard-N=8},
while $\mathbb{C}_{s}$- and $\mathbb{H}_{s}$- valued scalar fields have also
been recently considered in cosmology \cite{Cosmo}. However, to the best of
our knowledge, magic non-supersymmetric theories were investigated only
recently, in \cite{Marrani-Pradisi-Riccioni-Romano}. In such a paper, an
analysis of the $U$-duality orbit stratification was performed only for the $%
\mathbb{C}_{s}$-based theory in $D=4$ (for the electric-magnetic charges of
dyonic extremal black holes). The present investigation details the study of
the $\mathbb{C}_{s}$-based magic non-supersymmetric theory in $D=5$ and $6$,
as well as of the $\mathbb{H}_{s}$-based magic non-supersymmetric theory in $%
D=4$, $5$ and $6$. This will result in a detailed algebraic classification
of the asymptotically flat, extremal black brane solutions of such
theories.\bigskip

The results of our analysis rigorously prove and illustrate in detail a
conjecture made in \cite{Marrani-Pradisi-Riccioni-Romano} : as for the case
of maximal supergravity (whose the considered theories are
non-supersymmetric truncations related to suitable Jordan subalgebras of $%
J_{3}(\mathbb{O}_{s})$ and $J_{2}(\mathbb{O}_{s})$), in magic
non-supersymmetric Maxwell-Einstein theories there is no splitting of
orbits, namely there is only one orbit for each non-maximal rank element of
the relevant algebraic structure (\textit{i.e.}, of $J_{2}$ - or of ts
non-singlet algebraic complement to $J_{3}$ - in $D=6$, of $J_{3}$ in $D=5$,
and of the Freudenthal triple system over $J_{3}$ in $D=4$). This to be
contrasted to the magic quarter-maximal Maxwell-Einstein supergravity
theories \cite{Marrani-Riccioni-Romano}, in which the orbit splitting for non-maximal-rank elements
generally takes place, depending on the real form and on the relevant
representations of the duality group\footnote{%
Note that the orbit of rank-1 elements is generally the highest weight
orbit, which is unique, regardless the theory under consideration.}.\bigskip

The plane of the paper is as follows.

In Sec. \ref{D=6}, we start and analyze the $U$-orbit structure of the $%
\mathbb{C}_{s}$- (Sec. \ref{D=6-Cs}) and $\mathbb{H}_{s}$- (Sec. \ref{D=6-Hs}%
) based magic non-supersymmetric theories in $D=6$, which, as pointed out
above, is the highest dimension in which they can be consistently defined
(as such they can be regarded as non-supersymmetric truncations of the
non-chiral $(2,2)$ maximal supergravity in $D=6$); the analysis is based on
quadratic (i.e., rank-$2$) Euclidean Jordan algebras over $\mathbb{C}_{s}$
and $\mathbb{H}_{s}$ and on the non-transitive action of their reduced
structure symmetry. Then, in Sec. \ref{D=5}, we determine the $U$-orbit
structure of the $\mathbb{C}_{s}$- (Sec. \ref{D=5-Cs}) and $\mathbb{H}_{s}$-
(Sec. \ref{D=5-Hs}) based magic non-supersymmetric theories in $D=5$ (they
can be regarded as non-supersymmetric truncations of maximal $\mathcal{N}=8$
supergravity); in this case, the analysis is based on cubic (i.e., rank-$3$)
Euclidean Jordan algebras over $\mathbb{C}_{s}$ and $\mathbb{H}_{s}$ and on
the non-transitive action of their reduced structure symmetry. Finally, in
Sec. \ref{D=4}, we determine the $U$-orbit structure of the $\mathbb{C}_{s}$%
- (Sec\footnote{%
As noted, this is the unique case considered in \cite%
{Marrani-Pradisi-Riccioni-Romano}.}. \ref{section:20ofsl(6)}) and $\mathbb{H}%
_{s}$- (Sec. \ref{section:32ofso(6,6)}) based magic non-supersymmetric
theories in $D=4$, which is the lowest dimension in which asymptotically
flat black branes exist; also in this case, such theories can be regarded as
truncations of the maximal $\mathcal{N}=8$ supergravity. In $D=4$, our
analysis is based on (reduced) Freudenthal triple systems \cite{FTS} based
on cubic Euclidean Jordan algebras over $\mathbb{C}_{s}$ and $\mathbb{H}_{s}$
and on the non-transitive action of their derivations (which act as
conformal symmetry of the corresponding cubic Jordan algebra, as well).
Three Appendices, containing technical details, conclude the paper.

\FloatBarrier
\section{\label{D=6}$\mathbf{D=6}$}
\FloatBarrier

In $D=6$ dimensions, the asymptotically flat branes are black holes (electric $0$-branes), their duals, the magnetic black $2$-branes (they are respectively associated to vectors and their duals), and the dyonic black strings ($1$-branes, associated to tensors).
\subsection{\label{D=6-Cs}$\mathbb{C}_{s}$}

The $U$-duality\footnote{Due to their non-supersymmetric nature, both $\mathbb{H}_{s}$- and $\mathbb{C}_{s}$- based non-supersymmetric magic theories are generally anomalous in $D=6$.} group is
\begin{eqnarray}
SO(2,2)\times SO(1,1) &\simeq &SL(2,\mathbb{R})\times SL(2,\mathbb{R})\times SO(1,1) \\
&=&Str_{0}\left( J_{2}^{\mathbb{C}_{s}}\right) \times \frac{Tri\left(\mathbb{C}_{s}\right) }{SO(\mathbb{C}_{s})}\simeq SL(2,\mathbb{C}_{s})\times SO(1,1),
\end{eqnarray}
where $SL(2,\mathbb{R})\times SL(2,\mathbb{R})\simeq SL(2,\mathbb{C}_{s})$ \cite{Rios}, and $Tri\left( \mathbb{C}_{s}\right) $ and $SO(\mathbb{C}_{s})$ respectively denote the triality and norm-preserving symmetries of $\mathbb{C}_{s}$ (see \textit{e.g.} \cite{triality}). Therefore, the scalar manifold reads
\begin{equation}
\frac{Str_{0}\left( J_{2}^{\mathbb{C}_{s}}\right) }{mcs\left( Str_{0}\left(J_{2}^{\mathbb{C}_{s}}\right) \right) }=\frac{SO(2,2)}{SO(2)\times SO(2)}\times SO(1,1).  \label{sm-Cs-D=6}
\end{equation}%
$Str_{0}\left( J_{2}^{\mathbb{C}_{s}}\right) \simeq SL(2,\mathbb{C}_{s})$ is the reduced structure group of the quadratic Jordan algebra $J_{2}^{\mathbb{C}_{s}}$.\\

The electric $0$-brane and magnetic $2$-brane irreps. are the $\left(\mathbf{2,1}\right) _{1}+\left( \mathbf{1},\mathbf{2}\right) _{-1}$, and its conjugate $\left( \mathbf{2,1}\right) _{-1}+\left( \mathbf{1},\mathbf{2}\right) _{1}$, of $SO(2,2)\times SO(1,1)$ respectively, whereas the dyonic $1 $-brane irrep. is the $\left( \mathbf{2},\mathbf{2}\right) _{0}$. While the $\left( \mathbf{2},\mathbf{2}\right) _{0}$ admits a unique independent quadratic invariant polynomial $I_{2}$, the $\left( \mathbf{2,1}\right)_{1}+\left( \mathbf{1},\mathbf{2}\right) _{-1}$ and $\left( \mathbf{2,1}\right) _{-1}+\left(\mathbf{1},\mathbf{2}\right) _{1}$ do \textit{not} admit any quadratic invariant polynomial. Consequently, while (dyonic) strings can be \textit{\textquotedblleft large"} (\textit{i.e.}, with non-vanishing entropy), electric black holes and magnetic black $2$-branes are necessarily \textit{\textquotedblleft small"} (\textit{i.e.}, with a vanishing entropy), \textit{at least} at Einsteinian (two-derivative) level. In this framework, black strings are thus the unique asymptotically flat objects which may exhibit an attractor behaviour \cite{AM-Refs} in their near-horizon region (\textit{mutatis mutandis}, the same considerations holds for the $\mathbb{H}_{s}$-based theory; see further below).

\subsubsection{1-Branes}
While the non-linear action of $SO(2,2)\times SO(1,1)$ on the scalar manifold (\autoref{sm-Cs-D=6}) is transitive, the linear action of $SO(2,2)\times SO(1,1)$ on the $\left( \mathbf{2},\mathbf{2}\right) _{0}$ trivially determines the stratification into the following orbits, classified in terms of invariant constraints on $I_{2}$, or equivalently in terms of the \textit{rank} of the corresponding Jordan algebra $J_{2}^{\mathbb{C}_{s}}$ (\textit{cfr. e.g.} \cite{Gunaydin-rev, Rios}, and Refs. therein).\\

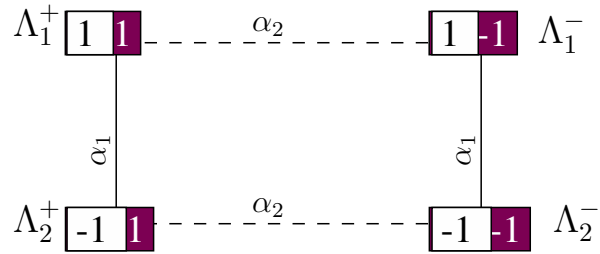
\begin{figure}[h]
\centering
% Generated with LaTeXDraw 2.0.8
% Thu Sep 22 17:34:55 CEST 2016
% \usepackage[usenames,dvipsnames]{pstricks}
% \usepackage{epsfig}
% \usepackage{pst-grad} % For gradients
% \usepackage{pst-plot} % For axes
\scalebox{1} % Change this value to rescale the drawing.
{\
\begin{pspicture}(0,-1.6145834)(11.766875,1.625625)
\definecolor{color1038}{rgb}{0.996078431372549,0.996078431372549,0.996078431372549}
\definecolor{color1057}{rgb}{0.00392156862745098,0.00392156862745098,0.00392156862745098}
\definecolor{color418b}{rgb}{0.48627450980392156,0.0,0.3254901960784314}
\usefont{T1}{ppl}{m}{n}
\rput(2.326875,1.38625){\large $\Lambda^{+}_{1}$}
\psline[linewidth=0.02cm,fillcolor=black,linestyle=dashed,dash=0.16cm 0.16cm,dotsize=0.07055555cm 2.0]{*-*}(3.3634374,-1.24875)(8.163438,-1.24875)
\usefont{T1}{ppl}{m}{n}
\rput(5.3679686,-1.03875){$\alpha_2$}
\usefont{T1}{ppl}{m}{n}
\rput(2.326875,-1.21375){\large $\Lambda^{+}_{2}$}
\usefont{T1}{ppl}{m}{n}
\rput(9.226875,1.28625){\large $\Lambda^{-}_{1}$}
\psline[linewidth=0.02cm,fillcolor=black,dotsize=0.07055555cm 2.0]{*-*}(3.3634374,1.15125)(3.3634374,-1.24875)
\usefont{T1}{ppl}{m}{n}
\rput{90.0}(2.9434376,-3.4125){\rput(3.1479688,-0.21875){$\alpha_1$}}
\usefont{T1}{ppl}{m}{n}
\rput(9.426875,-1.21375){\large $\Lambda^{-}_{2}$}
\usefont{T1}{ptm}{m}{n}
\rput(3.2776563,-1.31875){\large \color{color1038}\psframebox[linewidth=0.02,fillstyle=solid,fillcolor=color418b]{ -1 1}}
\usefont{T1}{ptm}{m}{n}
\rput(3.1076562,-1.31875){\large \color{color1057}\psframebox[linewidth=0.02,fillstyle=solid,fillcolor=color1038]{-1 }}
\psline[linewidth=0.02cm,fillcolor=black,dotsize=0.07055555cm 2.0]{*-*}(8.163438,1.15125)(8.163438,-1.24875)
\usefont{T1}{ppl}{m}{n}
\rput{90.0}(7.7434373,-8.2125){\rput(7.947969,-0.21875){$\alpha_1$}}
\psline[linewidth=0.02cm,fillcolor=black,linestyle=dashed,dash=0.16cm 0.16cm,dotsize=0.07055555cm 2.0]{*-*}(3.3634374,1.15125)(8.163438,1.15125)
\usefont{T1}{ppl}{m}{n}
\rput(5.3679686,1.36125){$\alpha_2$}
\usefont{T1}{ptm}{m}{n}
\rput(3.1907814,1.28125){\large \color{color1038}\psframebox[linewidth=0.02,fillstyle=solid,fillcolor=color418b]{ 1 1}}
\usefont{T1}{ptm}{m}{n}
\rput(3.0207813,1.28125){\large \color{color1057}\psframebox[linewidth=0.02,fillstyle=solid,fillcolor=color1038]{1 }}
\usefont{T1}{ptm}{m}{n}
\rput(8.0607815,1.28125){\large \color{color1038}\psframebox[linewidth=0.02,fillstyle=solid,fillcolor=color418b]{ 1 -1}}
\usefont{T1}{ptm}{m}{n}
\rput(7.820781,1.28125){\large \color{color1057}\psframebox[linewidth=0.02,fillstyle=solid,fillcolor=color1038]{1 }}
\usefont{T1}{ptm}{m}{n}
\rput(8.147656,-1.31875){\large \color{color1038}\psframebox[linewidth=0.02,fillstyle=solid,fillcolor=color418b]{ -1 -1}}
\usefont{T1}{ptm}{m}{n}
\rput(7.907656,-1.31875){\large \color{color1057}\psframebox[linewidth=0.02,fillstyle=solid,fillcolor=color1038]{-1 }}
\end{pspicture}
}
\caption{Dynkin tree of the $\mathbf{(2,2)}_{0}$ of $\mathfrak{sl}(2,\mathbb{R})\oplus \mathfrak{sl}(2,\mathbb{R})\oplus \mathfrak{so}(1,1)$. We denote with different color the Dynkin labels corresponding to the two $\mathfrak{sl}(2,\mathbb{R})$'s.}
\label{jazz}
\end{figure}

By exploiting the method developed in \cite{Marrani-Riccioni-Romano}, we are now going to explicitly determine the U-duality orbits, by studying the stabilizers of bound states of the weights of the $\left( \mathbf{2},\mathbf{2}\right) _{0}$ of the $U$-duality Lie algebra $\mathfrak{sl}(2,\mathbb{R})\oplus \mathfrak{sl}(2,\mathbb{R})\oplus \mathfrak{so}(1,1)$.\\
The action of the Cartan involution, $\theta$, on the simple roots $\alpha _{1}$ and $\alpha _{2}$ of the two $\mathfrak{sl}(2,\mathbb{R})$  is given by
\begin{equation}
\theta \alpha _{i}=-\alpha _{i}\qquad i=1,2.  \label{CI}
\end{equation}\\
Our analysis of the orbits starts by reporting the Dynkin tree of the\footnote{Subscripts denote weights with respect to $SO(1,1)$ throughout.} $\left( \mathbf{2},\mathbf{2}\right) _{0}$ of $\mathfrak{sl}(2,\mathbb{R})\oplus \mathfrak{sl}(2,\mathbb{R})\oplus \mathfrak{so}(1,1)$ in \autoref{jazz}. We note that all the weights in this representation have the same length and are noncompact.
\begin{table}[h!]
\renewcommand{\arraystretch}{1.6}
\begin{center}
\begin{tabular}{|c|c|}
\hline
\multicolumn{2}{|c|}{\textcolor{black}{\bf Stabilizer}}\\
\hline
\textcolor{black}{$\Lambda_{1}^+$}&\textcolor{black}{$\Lambda_{2}^-$}\\
\hline\hline
$\alpha_{1}\quad \alpha_{2}$&\\
$H_{\alpha_{1}}-H_{\alpha_{2}}$&$H_{\alpha_{1}}-H_{\alpha_{2}}$\\
&$-\alpha_{1}\quad -\alpha_{2}$\\
\hline
\end{tabular}

\caption{Stabilizers for the weights in the ${\bf (2,2)}_{0}$ of $\mathfrak{sl}(2,\mathbb{R})\oplus \mathfrak{sl}(2,\mathbb{R})\oplus \mathfrak{so}(1,1)$.}\label{stab22barsl2}
\end{center}
\end{table}

To find the different orbits for single brane and bound states we should consider single weight or linear combinations of them and study their behaviors under the action of the algebra.

\paragraph{1-weight}
To determine the 1-weight stabilizer, one simply can look at the Dynkin tree in \autoref{jazz}, and the results, listed in \autoref{stab33barsl3c} immediately follow. The corresponding 1-weight orbit reads%
\begin{align}
I_{2}=0:&&\mathcal{O}_{I_{2}=0,q=2}:=\frac{SO(2,2)\times SO(1,1)}{\left(SO(1,1)_{\ast }\times SO(1,1)\right) \ltimes \mathbb{R}^{2}}=\frac{SO(2,2)}{SO(1,1)\ltimes \mathbb{R}^{2}},  \label{p-sppp-1-II}
\end{align}%
where $\mathbb{R}^{2}\simeq \mathbf{2}$ of $SO(1,1)_{\ast }\subset SO(2,2)$. We also reported the corresponding $(SO(2,2)\times SO(1,1))$ - invariant constraint on the quadratic invariant $I_{2}$ of the $\left(\mathbf{2},\mathbf{2}\right) _{0}$; note that the orbit (\autoref{p-sppp-1-II}) is the orbit of rank-1 elements (lightlike vectors) of the quadratic Jordan algebra $J_{2}^{\mathbb{C}_{s}}$.

\paragraph{2-weights}
In order to define a bound state we have to identify two weights not connected by the action of a single generator in the algebra. Looking at \autoref{jazz} we realize that we could take $\Lambda_{1}^{+}$ and $\Lambda_{2}^{-}$. These weights could be combined into two different bound states, whose stabilizers are listed in \autoref{L1pmL2stabssl2sl2}.
\begin{table}[h!]
\renewcommand{\arraystretch}{1.6}
\par
\begin{center}
\begin{tabular}{|c|c|c|}
\hline
\textcolor{black}{\bf Common} & \multicolumn{2}{|c|}{\textcolor{black}{\bf Conjunction}} \\ \hline
\textcolor{black}{$\Lambda_{1}^{+},\Lambda_{2}^{-}$} & \textcolor{black}{$\Lambda_{1}^{+}+\Lambda_{2}^{-}$} & \textcolor{black}{$\Lambda_{1}^{+}-\Lambda_{2}^{-}$} \\ \hline\hline
\begin{tabular}{c}
$H_{\alpha_{1}}-H_{\alpha_{2}}$ \\
\end{tabular}
&
\begin{tabular}{c}
$E_{\alpha_{1}}-E_{-\alpha_{2}}$ \\
$E_{\alpha_{2}}-E_{-\alpha_{1}}$ \\
\end{tabular}
&
\begin{tabular}{c}
$E_{\alpha_{1}}+E_{-\alpha_{2}}$ \\
$E_{\alpha_{2}}+E_{-\alpha_{1}}$ \\
\end{tabular}
\\ \hline
\end{tabular}
\end{center}
\caption[$\Lambda_1^{+}\pm\Lambda_2^{-}$ stabilizers]{$\Lambda_1^{+}\pm\Lambda_2^{-}$ stabilizers.}
\label{L1pmL2stabssl2sl2}
\end{table}\\
The corresponding 2-weights bound states orbit reads\footnote{The subscript \textquotedblleft $d$\textquotedblright\ stands for \textit{diagonal} throughout.}, for both choices
\textquotedblleft $\pm $" :
\begin{multline}
I_{2}\neq 0:\mathcal{O}_{I_{2}\neq 0,q=2}:=\frac{SO(2,2)\times SO(1,1)}{SO(2,1)\times SO(1,1)}\\
\simeq \frac{SL(2,\mathbb{R})\times SL(2,\mathbb{R})\times SO(1,1)}{SL(2,\mathbb{R})_{d}\times SO(1,1)}=\frac{SL(2,\mathbb{R})\times SL(2,\mathbb{R})}{SL(2,\mathbb{R})_{d}}.  \label{p-sppp-2-II}
\end{multline}
Note that this is a symmetric manifold. We also reported the corresponding $(SO(2,2)\times SO(1,1)) $-invariant constraint on the quadratic invariant $I_{2}$ of the $(\mathbf{2},\mathbf{2}) _{0}$; note that the orbit (\autoref{p-sppp-2-II}) is the orbit of rank-2 elements (non-lightlike vectors) of the quadratic Jordan algebra $J_{2}^{\mathbb{C}_{s}}$.

\subsubsection{0- and 2- Branes}

On the other hand, the linear action of $SO(2,2)\times SO(1,1)$ on the $\left( \mathbf{2,1}\right) _{1}+\left( \mathbf{1},\mathbf{2}\right) _{-1}$ and $\left( \mathbf{2,1}\right) _{-1}+\left( \mathbf{1},\mathbf{2}\right)_{1}$ determines the stratification into orbits, classified in terms of invariant constraints on the $\left( \mathbf{2},\mathbf{2}\right) _{0}$%
-covariant vector\footnote{We consider throughout only the case of electric black holes ($q_{\alpha}\in \left( \mathbf{2,1}\right) _{1}+\left( \mathbf{1},\mathbf{2}\right)_{-1}$, in this case); the treatment and results for magnetic black $2$-branes ($p^{\alpha }\in \left( \mathbf{2,1}\right) _{-1}+\left( \mathbf{1},\mathbf{2}\right) _{1}$, in this case) are identical.}
\begin{equation}
V^{I}:=\left( \gamma ^{I}\right) ^{\alpha \beta }q_{\alpha }q_{\beta },\label{VV}
\end{equation}
($I=1,...,4$, $\alpha ,\beta =1,...,4$), where $\gamma ^{I}$ here denotes the gamma matrices of $SO(2,2)(\times SO(1,1))$. As a consequence of the result of \cite{Sudbery} and of the algebraic embedding $\mathbb{C}_{s}\subset \mathbb{O}_{s}$, it should be remarked that both the $\left(\mathbf{2,1}\right) _{1}+\left( \mathbf{1},\mathbf{2}\right) _{-1}$ and $\left( \mathbf{2,1}\right) _{-1}+\left( \mathbf{1},\mathbf{2}\right) _{1}$ of $Str_{0}\left( J_{2}^{\mathbb{C}_{s}}\right) \times \frac{Tri\left(\mathbb{C}_{s}\right) }{SO(\mathbb{C}_{s})}=SO(2,2)\times SO(1,1)$ can be represented as a pair of split complex numbers $\mathbb{C}_{s}$.\\

By applying the same methods used in the analysis of the black 1-brane orbits above, the following stratification is determined \footnote{These orbits recently appeared in \cite{FLM-1} (apart from $SO(1,1)\simeq \frac{Tri\left( \mathbb{C}_{s}\right) }{SO(\mathbb{C}_{s})}$ - see further below - in the numerator\ therein).}:

\begin{description}

\item[1-weight orbit] ($V^{I}=0,$ $\forall I$) :
\begin{equation}
\mathcal{O}_{V^{I}=0,q=2}:=\frac{SO(2,2)\times SO(1,1)}{\left( SL(2,\mathbb{R})\times SO(1,1)\right) \ltimes \mathbb{R}},  \label{ps-3}
\end{equation}
where $\mathbb{R}\simeq \mathbf{1}$ is a singlet of $SL(2,\mathbb{R})$. This is the orbit of \textit{pure}, non-generic\footnote{The orbits (\autoref{ps-3}) and (\autoref{ps-2-2}) are the same as the ones obtained in \cite{Pure-Spinors}.} spinors.

\item[2-weights orbit] ($V^{I}\neq 0$ for some $I$) :
\begin{equation}
\mathcal{O}_{V^{I}\neq 0,q=2}:=\frac{SO(2,2)\times SO(1,1)}{SO(1,1)\times \mathbb{R}^{2}}.
\end{equation}%
This is the generic orbit of \textit{pure}, generic spinors.
\end{description}

\subsection{\label{D=6-Hs}$\mathbb{H}_{s}$}

The $U$-duality group is
\begin{eqnarray}
SO(3,3)\times SO(2,1) &\simeq &SL(4,\mathbb{R})\times SL(2,\mathbb{R}) \\
&=&Str_{0}\left( J_{2}^{\mathbb{H}_{s}}\right) \times \frac{Tri\left(\mathbb{H}_{s}\right) }{SO(\mathbb{H}_{s})}\simeq SL(2,\mathbb{H}_{s})\times SL(2,\mathbb{R}),
\end{eqnarray}%
where $SL(4,\mathbb{R})\simeq SL(2,\mathbb{H}_{s})$ \cite{Rios}, and $Tri\left( \mathbb{H}_{s}\right) $ and $SO(\mathbb{H}_{s})$ respectively denote the triality and norm-preserving symmetries of $\mathbb{H}_{s}$ (see \textit{e.g.} \cite{triality}, and Refs. therein). Therefore, the scalar manifold reads
\begin{equation}
\frac{Str_{0}\left( J_{2}^{\mathbb{H}_{s}}\right) }{mcs\left( Str_{0}\left(J_{2}^{\mathbb{H}_{s}}\right) \right) }=\frac{SO(3,3)}{SO(3)\times SO(3)}\times \frac{SO(2,1)}{SO(2)}.  \label{sm-Hs-D=6}
\end{equation}
$Str_{0}\left( J_{2}^{\mathbb{H}_{s}}\right) \simeq SL(2,\mathbb{H}_{s})$ is the reduced structure group of the quadratic Jordan algebra $J_{2}^{\mathbb{H}_{s}}$. The electric $0$-brane and magnetic $2$-brane irreps. are the chiral bi-spinor $\left( \mathbf{4,2}\right) $, and its conjugate $\left( \mathbf{4}^{\prime }\mathbf{,2}\right) $ of $SO(3,3)\times SO(2,1)$ respectively, whereas the dyonic $1$-brane sit in the irrep.\footnote{By virtue of the isomorphism $SO(3,3)\simeq SL(4,\mathbb{R})$, the chiral spinor $\mathbf{4}$, conjugate spinor $\mathbf{4}^{\prime }$, and fundamental $\mathbf{6}$ irreps. of $SO(3,3)$ respectively correspond to the fundamental, rank-$3$ antisymmetric and rank-$2$ antisymmetric irreps. of $%
SL(4,\mathbb{R})$.} $\left( \mathbf{6},\mathbf{1}\right) $. While the $\left( \mathbf{6},\mathbf{1}\right) $ admits a unique independent quadratic invariant polynomial $I_{2}$, the $\left( \mathbf{4,2}\right) $ and $\left(\mathbf{4}^{\prime}\mathbf{,2}\right) $ do \textit{not} admit any quadratic invariant polynomial.\\

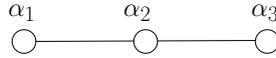
\begin{figure}[h!]
\centering
% \usepackage[usenames,dvipsnames]{pstricks}
% \usepackage{epsfig}
% \usepackage{pst-grad} % For gradients
% \usepackage{pst-plot} % For axes
% \usepackage[space]{grffile} % For spaces in paths
% \usepackage{etoolbox} % For spaces in paths
% \makeatletter % For spaces in paths
% \patchcmd\Gread@eps{\@inputcheck#1 }{\@inputcheck"#1"\relax}{}{}
% \makeatother
% % User Packages:
%
%

\psscalebox{0.4} % Change this value to rescale the drawing.
{
\begin{pspicture}(0,-0.71708983)(10.78,0.71708983)
\psline[linecolor=black, linewidth=0.02](1.4428126,-0.31640235)(9.6796875,-0.28240234)
\pscircle[linecolor=black, linewidth=0.02, fillstyle=solid, dimen=outer](5.4528127,-0.31708986){0.4}
\pscircle[linecolor=black, linewidth=0.02, fillstyle=solid, dimen=outer](1.4528126,-0.31708974){0.4}
\pscircle[linecolor=black, linewidth=0.02, fillstyle=solid, dimen=outer](9.452812,-0.31708986){0.4}
\rput(1.3817188,0.61759764){\huge $\alpha_1$}
\rput(5.1817183,0.61759764){\huge $\alpha_2$}
\rput(9.381719,0.61759764){\huge $\alpha_3$}
\end{pspicture}
}
\caption{Tits-Satake diagram of $\mathfrak{sl}(4,\mathbb{R})$.}\label{fig:sl4titssatakediagram}
\end{figure}

Our conventions for the simple roots of $\mathfrak{sl}(4,\mathbb{R})$ appear in \autoref{fig:sl4titssatakediagram}.

\subsubsection{1-Branes}

While the non-linear action of $SO(3,3)\times SO(2,1)$ on the scalar manifold (\autoref{sm-Hs-D=6}) is transitive, the linear action of $SO(3,3)\times SO(2,1)$ on the $\left( \mathbf{6},\mathbf{1}\right)$ trivially determines the stratification into the following orbits, classified in terms of invariant constraints on $I_{2}$, or equivalently in terms of the \textit{rank} of the corresponding Jordan algebra $J_{2}^{\mathbb{H}_{s}}$.\bigskip
\begin{figure}[h!]
\centering
% Generated with LaTeXDraw 2.0.8
% Sat May 06 17:21:28 CEST 2017
% \usepackage[usenames,dvipsnames]{pstricks}
% \usepackage{epsfig}
% \usepackage{pst-grad} % For gradients
% \usepackage{pst-plot} % For axes
\scalebox{0.8} % Change this value to rescale the drawing.
{
\begin{pspicture}(0,-4.3973956)(5.6001043,4.3973956)
\usefont{T1}{ppl}{m}{n}
\rput{-44.22789}(1.1969377,1.0408303){\rput(1.8492025,-0.93662614){$\alpha_{1}$}}
\psline[linewidth=0.02cm,fillcolor=black,dotsize=0.07055555cm 2.0]{*-*}(0.66489565,-0.0026041667)(2.7648957,-2.0026042)
\psline[linewidth=0.02cm,fillcolor=black,dotsize=0.07055555cm 2.0]{*-*}(2.7648957,-2.0026042)(2.7648957,-4.002604)
\usefont{T1}{ppl}{m}{n}
\rput{-90.0}(6.0618343,-0.24298048){\rput(2.879427,-3.1366262){$\alpha_{2}$}}
\psline[linewidth=0.02cm,fillcolor=black,dotsize=0.07055555cm 2.0]{*-*}(2.7648957,1.9973959)(0.66489565,-0.0026041667)
\usefont{T1}{ppl}{m}{n}
\rput{44.2}(1.2754259,-0.8458102){\rput(1.6492026,1.1633738){$\alpha_{3}$}}
\usefont{T1}{ppl}{m}{n}
\rput{-44.22789}(0.39711022,3.0724673){\rput(3.9492025,1.0633738){$\alpha_{1}$}}
\psline[linewidth=0.02cm,fillcolor=black,dotsize=0.07055555cm 2.0]{*-*}(2.7648957,1.9973959)(4.864896,-0.0026041667)
\psline[linewidth=0.02cm,fillcolor=black,dotsize=0.07055555cm 2.0]{*-*}(2.7648957,3.9973958)(2.7648957,1.9973959)
\usefont{T1}{ppl}{m}{n}
\rput{-90.0}(0.061834335,5.7570195){\rput(2.879427,2.8633738){$\alpha_{2}$}}
\psline[linewidth=0.02cm,fillcolor=black,dotsize=0.07055555cm 2.0]{*-*}(4.864896,-0.0026041667)(2.7648957,-2.0026042)
\usefont{T1}{ppl}{m}{n}
\rput{44.2}(0.47558343,-2.8760355){\rput(3.7492025,-0.8366262){$\alpha_{3}$}}
\usefont{T1}{ppl}{m}{n}
\rput(2.7539582,4.0203123){\Large \psframebox[linewidth=0.02,fillstyle=solid]{0 1 0}}
\usefont{T1}{ppl}{m}{n}
\rput(2.7272396,2.0203125){\Large \psframebox[linewidth=0.02,fillstyle=solid]{1 -1 1}}
\usefont{T1}{ppl}{m}{n}
\rput(4.741927,0.0203125){\Large \psframebox[linewidth=0.02,fillstyle=solid]{-1 0 1}}
\usefont{T1}{ppl}{m}{n}
\rput(2.7319272,-1.8796875){\Large \psframebox[linewidth=0.02,fillstyle=solid]{-1 1 -1}}
\usefont{T1}{ppl}{m}{n}
\rput(0.7272396,0.0203125){\Large \psframebox[linewidth=0.02,fillstyle=solid]{1 0 -1}}
\usefont{T1}{ppl}{m}{n}
\rput(2.6439583,-4.0796876){\Large \psframebox[linewidth=0.02,fillstyle=solid]{0 -1 0}}
\usefont{T1}{ppl}{m}{n}
\rput(4.209427,4.0073957){\begin{Large}$\Lambda_1$\end{Large}}
\usefont{T1}{ppl}{m}{n}
\rput(4.209427,-4.092604){\begin{Large}$\Lambda_2$\end{Large}}
\end{pspicture}
}

\caption{Dynkin tree of the {\bf (6,1)} of $SL(4,\mathbb{R})\times SL(2,\mathbb{R})$. $\alpha_{1}, \alpha_{2}$ and $\alpha_{3}$ are the three simple roots of $\mathfrak{sl}(4,\mathbb{R})$. }\label{fig:61sl4sl2dynkimtree}
\end{figure}
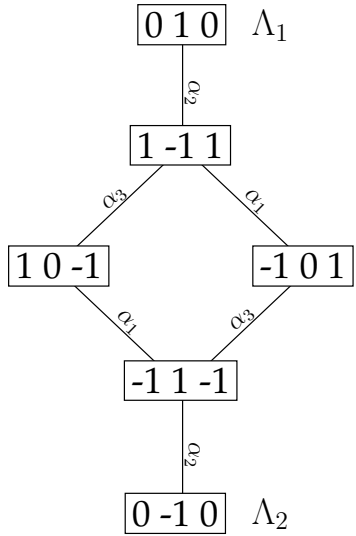

\begin{table}[h!]
\renewcommand{\arraystretch}{1.6}
\begin{center}
%\resizebox{\textwidth}{!}{
\begin{tabular}{|c|c|}
\hline
\multicolumn{2}{|c|}{\textbf{Stabilizers}}\\
\hline
$\Lambda_{1}$&$\Lambda_{2}$\\ \hline\hline
$\alpha_1+\alpha_2+\alpha_3$&\\
$\alpha_1+\alpha_2\quad\alpha_2+\alpha_3$&\\
$\alpha_1\quad\alpha_2\quad\alpha_3$&$\alpha_1\quad\alpha_3$\\
$H_{\alpha_1}\quad H_{\alpha_3}$&$H_{\alpha_1}\quad H_{\alpha_3}$\\
$-\alpha_1\quad-\alpha_3$&$-\alpha_1\quad-\alpha_2\quad-\alpha_3$\\
&$-\alpha_1-\alpha_2\quad-\alpha_2-\alpha_3$\\
&$-\alpha_1-\alpha_2-\alpha_3$\\
\hline
\end{tabular}
%}
\caption{Stabilizers for the weights in the ${\bf (6,1)}$ of $SL(4,\mathbb{R})\times SL(2,\mathbb{R})$}\label{tab:61sl4sl2stabilizers}
\end{center}
\end{table}

\begin{table}[h!]
\renewcommand{\arraystretch}{1.6}
\begin{center}
%\resizebox{\textwidth}{!}{
\begin{tabular}{|c|c|}
\hline
{\bf Common Stabilizers}&{\bf Conjunction Stabilizers}\\
\hline\hline
&$E_{\alpha_1+\alpha_2+\alpha_3}-E_{-\alpha_2}$\\
$\alpha_1\quad\alpha_3$&$E_{\alpha_1+\alpha_2}-E_{-\alpha_2-\alpha_3}$\\
$H_{\alpha_1}\quad H_{\alpha_3}$&$E_{\alpha_2+\alpha_3}-E_{-\alpha_1-\alpha_2}$\\
$-\alpha_1\quad-\alpha_3$&$E_{\alpha_1}-E_{-\alpha_1-\alpha_2-\alpha_3}$\\
\hline
\end{tabular}
%}
\caption{Stabilizers for the weights in the ${\bf (6,1)}$ of $SL(4,\mathbb{R})\times SL(2,\mathbb{R})$. We list only the elements in $SL(4,\mathbb{R})$.}\label{tab:61sl4sl2commonconjucntion}
\end{center}
\end{table}
\begin{description}

\item[Rank-1 (1-weight orbit)] from \autoref{tab:61sl4sl2stabilizers} we recognize
\begin{flalign}
&I_{2} =0:\mathcal{O}_{I_{2}=0,q=4}:=\frac{SO(3,3)\times SO(2,1)}{\left(SO(2,2)\times SO(2,1)\right) \ltimes \mathbb{R}^{4}}  \notag \\
&\simeq\frac{SL(4,\mathbb{R})\times SL\left( 2,\mathbb{R}\right) }{\left[SL(2,\mathbb{R})\times SL(2,\mathbb{R})\times SL\left( 2,\mathbb{R}\right)\right] \ltimes \mathbb{R}^{\left( 2,2,1\right) }}=\frac{SL(4,\mathbb{R})}{\left[ SL(2,\mathbb{R})\times SL(2,\mathbb{R})\right] \ltimes \mathbb{R}^{\left( 2,2\right) }},  \label{p-spppp-1-II}
\end{flalign}
where $\mathbb{R}^{4}\simeq \mathbf{4}$ of $SO(2,2)$, or equivalently $\mathbb{R}^{\left( 2,2,1\right) }\simeq \left( \mathbf{2,2}\right) $ of $SL(2,\mathbb{R})\times SL(2,\mathbb{R})\times SL\left( 2,\mathbb{R}\right) $. We also reported the corresponding $\left( SL(4,\mathbb{R})\times SL\left(2,\mathbb{R}\right) \right) $-invariant constraint on the quadratic invariant $I_{2}$ of the $\left( \mathbf{6},\mathbf{1}\right) $; note that the orbit (\autoref{p-spppp-1-II}) is the orbit of rank-1 elements (lightlike vectors) of the quadratic Jordan algebra $J_{2}^{\mathbb{H}_{s}}$.

\item[Rank-2 (2-weights orbit)] the rank-2 orbit could be defined by combining the weight $\Lambda_1$ and $\Lambda_2$, with the notation of \autoref{fig:61sl4sl2dynkimtree}. Their common and conjunction stabilizers are listed in \autoref{tab:61sl4sl2commonconjucntion} and we obtain an orbit
\begin{equation}
I_{2}\neq 0:\mathcal{O}_{I_{2}\neq 0,q=4}:=\frac{SO(3,3)\times SO(2,1)}{SO(3,2)\times SO(2,1)}\simeq \frac{SL(4,\mathbb{R})\times SL\left( 2,\mathbb{R}\right) }{Sp(4,\mathbb{R})\times SL\left( 2,\mathbb{R}\right) }=\frac{SL(4,\mathbb{R})}{Sp(4,\mathbb{R})}.  \label{p-spppp-2-II}
\end{equation}
Note that this is a symmetric manifold. We also reported the corresponding\\ $\left( SL(4,\mathbb{R})\times SL\left( 2,\mathbb{R}\right) \right) $-invariant constraint on the quadratic invariant $I_{2}$ of the $\left(\mathbf{6},\mathbf{1}\right) $; note that the orbit (\autoref{p-spppp-2-II}) is the orbit of rank-2 elements (non-lightlike vectors) of the quadratic Jordan algebra $J_{2}^{\mathbb{H}_{s}}$. We summarize our results in \autoref{tab:61sl4sl2summary}.
\end{description}

\begin{table}[h!]
\renewcommand{\arraystretch}{1.8}
\begin{center}
\resizebox{\textwidth}{!}{
\begin{tabular}{|c|c|c|c|c|c|c|c|}
\hline
\multicolumn{2}{|c|}{}&\multicolumn{2}{|c|}{{$\mathbf{\theta}$}}&&&\\ \cline{3-4}
\multicolumn{2}{|c|}{\multirow{-2}{*}{{\textbf{States}}}}&{\bf +}&{\bf -}&\multirow{-2}{*}{{\textbf{Semisimple Stabilizer}}}&\multirow{-2}{*}{{\textbf{Stabilizer}}}&\multirow{-2}{*}{{\bf rank}}\\ \hline\hline

{ 1-w}&$\Lambda_{1}$&5&8&$SO(3,2)\times SO(2,1)$&$SO(3,2)\times SO(2,1)$&1\\ \hline

{2-w}&$\Lambda_{1}\pm\Lambda_{2}$&5&8& $Sp(4,\mathbb{R})\times Sp(2,\mathbb{R}) $&$[Sp(4,\mathbb{R})\times Sp( 2,\mathbb{R})]\ltimes \mathbb{R}^{(4,2)})$&2\\
\hline

\hline
\end{tabular}
}
\caption{Summary of the orbits in the {\bf (6,1)} of $SL(4,\mathbb{R})\times SL(2,\mathbb{R}) $}\label{tab:61sl4sl2summary}
\end{center}
\end{table}

\subsubsection{0- and 2- Branes}

On the other hand, the linear action of $SO(3,3)\times SO(2,1)$ on the $\left( \mathbf{4,2}\right) $ and $\left( \mathbf{4}^{\prime }\mathbf{,2}\right) $ determines the stratification into orbits, classified in terms of invariant constraints on the $\left( \mathbf{6},\mathbf{1}\right) $-covariant vector $V^{I}$ (\autoref{VV}) (where in this case $I=1,...,6$,$\alpha ,\beta =1,...,8$, with the $\gamma ^{I}$'s now denoting the gamma matrices of $SO(3,3)\times SO(2,1)$. As a consequence of the result of \cite{Sudbery} and of the algebraic embedding $\mathbb{H}_{s}\subset \mathbb{O}_{s}$, it should be remarked that both the $\left( \mathbf{4,2}\right) $ and $\left( \mathbf{4}^{\prime }\mathbf{,2}\right) $ of $Str_{0}\left( J_{2}^{\mathbb{H}_{s}}\right) \times \frac{Tri\left( \mathbb{H}_{s}\right) }{SO(\mathbb{H}_{s})}=SO(3,3)\times SO(2,1)$ can be represented as a pair of split quaternions $\mathbb{H}_{s}$.\bigskip

By exploiting the same methods used above for the $\mathbb{C}_{s}$-based theory, the following stratification is determined :

\begin{description}

\item[1-weight orbit] ($V^{I}=0,$ $\forall I$) :
\begin{equation}
\mathcal{O}_{V^{I}=0,q=4}:=\frac{SO(3,3)\times SL(2,\mathbb{R})}{\left( SL(3,\mathbb{R})\times SO(1,1)\right) \ltimes \left( \mathbb{R}^{3}\times \mathbb{R}\right) }\cong \frac{SO(3,3)}{SL(3,\mathbb{R})\ltimes \mathbb{R}^{3}}\times \frac{SL(2,\mathbb{R})}{SO(1,1)\ltimes \mathbb{R}},\label{ps-2-2}
\end{equation}
where $\mathbb{R}^{3}\simeq \mathbf{3}^{\prime }$ denotes the rank-$2$ antisymmetric irrep. of $SL(3,\mathbb{R})$, and $SO(1,1)\ltimes \mathbb{R}$ is the maximal triangular subgroup of $SL(2,\mathbb{R})$ itself. This is the orbit of \textit{pure }, non-generic spinors.
\item[2-weights orbit] ($V^{I}\neq 0$ for some $I$) :
\begin{equation}
\mathcal{O}_{V^{I}\neq 0,q=4}:=\frac{SO(3,3)\times SL(2,\mathbb{R})}{\left(SO(2,1)\times SL(2,\mathbb{R})\right) \ltimes \mathbb{R}^{4}},
\end{equation}
where $\mathbb{R}^{4}\simeq \mathbf{4}$ denotes the spinor of $SO(2,1)$. This is the generic orbit of \textit{pure}, generic spinors.
\end{description}

\subsection{\label{Summary}Remark on the Orbit Dimensions}

From the above analysis (as well as from the result of \cite{FG-1,LPS-1,F-Maldacena-1,ICL-1,FG-2} concerning the $U$-orbit stratification in maximal supergravity) and \cite{Sudbery}, one can conclude that for ($q:=\dim_{\mathbb{R}}\mathbb{A}_{s}=8,4,2$ for $\mathbb{O}_{s}$, $\mathbb{H}_{s}$ and $\mathbb{C}_{s}$, respectively; also recall that $SO(5,5)\simeq SL(2,\mathbb{O}_{s})$ \cite{Rios})
\begin{equation}
SL(2,\mathbb{A}_{s})\times \frac{Tri\left( \mathbb{A}_{s}\right) }{SO(\mathbb{A}_{s})}\simeq Spin\left( \frac{q}{2}+1,\frac{q}{2}+1\right) \times \frac{Tri\left( \mathbb{A}_{s}\right) }{SO(\mathbb{A}_{s})},
\end{equation}
it holds that
\begin{eqnarray}
\text{dim}\left( \mathcal{O}_{I_{2}\neq 0,q}\right)  &=&q+2~\text{[\textit{non-lightlike}~vectors}=\text{\textit{non-lightlike}~elements~of~}J_{2}^{\mathbb{A}_{s}}\text{]}; \\
\text{dim}\left( \mathcal{O}_{I_{2}=0,q}\right)  &=&q+2~\text{[\textit{lightlike}~vectors}=\text{\textit{lightlike}~elements~of~}J_{2}^{\mathbb{A}_{s}}\text{]}; \\
\text{dim}\left( \mathcal{O}_{V^{I}\neq 0,q}\right)  &=&2q~\text{[\textit{generic}spinors}=\text{\textit{generic}~elements~of~}\mathbb{A}_{s}^{2}\text{]}; \\
\text{dim}\left( \mathcal{O}_{V^{I}=0,q}\right)  &=&\frac{3}{2}q-1~\text{[\textit{non-generic} spinors}=\text{\textit{non-generic}~elements~of~}\mathbb{A}_{s}^{2}\text{]}.
\end{eqnarray}

\FloatBarrier

\section{\label{D=5}$\mathbf{D=5}$}
\FloatBarrier
In $D=5$ dimensions, the asymptotically flat branes are black holes (electric $0$-branes), and their duals, the black strings (magnetic $1$-branes).

\subsection{\label{D=5-Cs}$\mathbb{C}_{s}$}

The $U$-duality group is $SL(3,\mathbb{R})\times SL(3,\mathbb{R})$ (split form of $SU(3)\times SU(3)$), and the scalar manifold reads
\begin{equation}
\frac{Str_{0}\left( J_{3}^{\mathbb{C}_{s}}\right) }{mcs\left( Str_{0}\left(J_{3}^{\mathbb{C}_{s}}\right) \right) }=\frac{SL(3,\mathbb{R})\times SL(3,\mathbb{R})}{SO(3)\times SO(3)},  \label{sm-Cs-D=5}
\end{equation}%
where $Str_{0}\left( J_{3}^{\mathbb{H}_{s}}\right) $ is the reduced structure group of the cubic Jordan algebra $J_{3}^{\mathbb{C}_{s}}$ (\textit{cfr. e.g.} \cite{Gunaydin-rev}, and Refs. therein).

The electric $0$-brane (black hole) and magnetic $1$-brane (black string) irreps. are the $\left( \mathbf{3},\mathbf{3}^{\prime }\right) $, and its conjugate $\left( \mathbf{3}^{\prime },\mathbf{3}\right) $, respectively. They are both characterized by a unique independent cubic invariant polynomial, which we will denote by $I_{3,el}$ and $I_{3,mag}$, respectively \cite{Sato-Kimura, Kac-80, Garibaldi-2, Garibaldi-2}.

While the non-linear action of $SL(3,\mathbb{R})\times SL(3,\mathbb{R})$ on the scalar manifold (\autoref{sm-Cs-D=5}) is transitive, the linear action of $SL(3,\mathbb{R})\times SL(3,\mathbb{R})$ on the $\left( \mathbf{3},\mathbf{3}^{\prime }\right) $ and $\left( \mathbf{3}^{\prime },\mathbf{3}\right) $ , classified in terms of invariant constraints\footnote{%
In the following treatment, we will consider only magnetic black strings; the treatment and the results are identical for electric black holes.} on $I_{3,el}$ and $I_{3,mag}$, respectively, or equivalently in terms of the \textit{rank} of the corresponding Jordan algebra $J_{3}^{\mathbb{C}_{s}}$ \cite{rank-J,rank-FTS}.\bigskip

\begin{figure}[h]
\centering
% Generated with LaTeXDraw 2.0.8
% Sun Oct 25 16:09:17 CET 2015
% \usepackage[usenames,dvipsnames]{pstricks}
% \usepackage{epsfig}
% \usepackage{pst-grad} % For gradients
% \usepackage{pst-plot} % For axes
\scalebox{0.4} % Change this value to rescale the drawing.
{\
\begin{pspicture}(0,0.17125)(14.317813,2.19125)
\psline[linewidth=0.02cm](1.563125,0.92475)(5.763125,0.92475)
\pscircle[linewidth=0.02,dimen=outer,fillstyle=solid](5.573125,0.9240625){0.4}
\pscircle[linewidth=0.02,dimen=outer,fillstyle=solid](1.573125,0.9240626){0.4}
\psline[linewidth=0.02cm](8.763125,0.92475)(12.963125,0.92475)
\pscircle[linewidth=0.02,dimen=outer,fillstyle=solid](12.773125,0.9240625){0.4}
\pscircle[linewidth=0.02,dimen=outer,fillstyle=solid](8.773125,0.9240626){0.4}
\usefont{T1}{ptm}{m}{n}
\rput(1.5020312,1.85875){\huge $\alpha_1$}
\usefont{T1}{ptm}{m}{n}
\rput(5.302031,1.85875){\huge $\alpha_2$}
\usefont{T1}{ptm}{m}{n}
\rput(8.902031,1.85875){\huge $\alpha_3$}
\usefont{T1}{ptm}{m}{n}
\rput(12.702031,1.85875){\huge $\alpha_4$}
\end{pspicture}
}
\caption{Tits-Satake diagram of $\mathfrak{sl}(3,\mathbb{R})\oplus \mathfrak{sl}(3,\mathbb{R})$.}
\label{sl3ctitsatake1}
\end{figure}
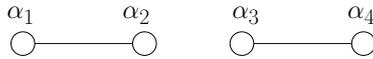

We are now going to determine such orbits, by studying the stabilizers of bound states of the weights of the $\left( \mathbf{3}^{\prime },\mathbf{3}\right) $ of the $U$-duality Lie algebra $\mathfrak{sl}(3,\mathbb{R})\oplus \mathfrak{sl}(3,\mathbb{R})$, whose Tits-Satake diagram is sketched in  \autoref{sl3ctitsatake1}.
Correspondingly, the action of the Cartan involution is non-trivial, and it is given again by (\autoref{CI}).

We start by reporting the Dynkin tree of the $\left( \mathbf{3}^{\prime },\mathbf{3}\right) $ of $\mathfrak{sl}(3,\mathbb{R})\oplus \mathfrak{sl}(3,\mathbb{R})$ in \autoref{3bar3sl3cdynkintree1}. All the weights in this representation have the same length and are real.
\begin{figure}[h]
\centering
\subfloat[][Dynkin tree of the $\mathbf{(3}^{\prime }\mathbf{,3)}$ of $\mathfrak{sl}(3,\mathbb{R})\oplus \mathfrak{sl}(3,\mathbb{R})$.\label{3bar3sl3cdynkintree1}]{
% Generated with LaTeXDraw 2.0.8
% Wed Oct 14 10:12:21 CEST 2015
% \usepackage[usenames,dvipsnames]{pstricks}
% \usepackage{epsfig}
% \usepackage{pst-grad} % For gradients
% \usepackage{pst-plot} % For axes
\scalebox{0.55} % Change this value to rescale the drawing.
{\
\begin{pspicture}(0,-6.0452604)(15.309063,6.0452604)
\usefont{T1}{ppl}{m}{n}
\rput{-40.588333}(4.4377284,3.670598){\rput(7.1453123,-4.1452603){\large $\alpha_{1}$}}
\psline[linewidth=0.02cm,fillcolor=black,dotsize=0.07055555cm 2.0]{*-*}(5.341875,-2.8402605)(8.541875,-5.64026)
\psline[linewidth=0.02cm,fillcolor=black,dotsize=0.07055555cm 2.0]{*-*}(4.341875,-0.040260416)(5.341875,-2.8402605)
\usefont{T1}{ppl}{m}{n}
\rput{-68.655}(4.5652714,3.7956471){\rput(5.0253124,-1.4252604){\large $\alpha_{2}$}}
\psline[linewidth=0.02cm,fillcolor=black,dotsize=0.07055555cm 2.0]{*-*}(11.741875,-2.8402605)(8.541875,-5.64026)
\usefont{T1}{ppl}{m}{n}
\rput{41.345284}(-0.26004365,-7.6007156){\rput(9.905313,-4.1252604){\large $\alpha_{4}$}}
\psline[linewidth=0.02cm,fillcolor=black,dotsize=0.07055555cm 2.0]{*-*}(12.741875,-0.040260416)(11.741875,-2.8402605)
\usefont{T1}{ppl}{m}{n}
\rput{68.5604}(6.351212,-12.087339){\rput(12.005313,-1.3652604){\large $\alpha_{3}$}}
\usefont{T1}{ppl}{m}{n}
\rput{-40.588333}(3.3859015,6.42625){\rput(10.345312,-1.3452604){\large $\alpha_{1}$}}
\psline[linewidth=0.02cm,fillcolor=black,dotsize=0.07055555cm 2.0]{*-*}(8.541875,-0.040260416)(11.741875,-2.8402605)
\psline[linewidth=0.02cm,fillcolor=black,dotsize=0.07055555cm 2.0]{*-*}(7.541875,2.7597396)(8.541875,-0.040260416)
\usefont{T1}{ppl}{m}{n}
\rput{-68.655}(3.9925904,8.5569935){\rput(8.225312,1.3747395){\large $\alpha_{2}$}}
\psline[linewidth=0.02cm,fillcolor=black,dotsize=0.07055555cm 2.0]{*-*}(7.541875,2.7597396)(4.341875,-0.040260416)
\usefont{T1}{ppl}{m}{n}
\rput{41.345284}(2.3924072,-3.4303722){\rput(5.7053127,1.4747396){\large $\alpha_{4}$}}
\psline[linewidth=0.02cm,fillcolor=black,dotsize=0.07055555cm 2.0]{*-*}(9.541875,2.7597396)(8.541875,-0.040260416)
\usefont{T1}{ppl}{m}{n}
\rput{68.5604}(6.9271264,-7.3322253){\rput(8.805312,1.4347396){\large $\alpha_{3}$}}
\usefont{T1}{ppl}{m}{n}
\rput{-40.588333}(1.8047627,7.750539){\rput(11.345312,1.4547396){\large $\alpha_{1}$}}
\psline[linewidth=0.02cm,fillcolor=black,dotsize=0.07055555cm 2.0]{*-*}(9.541875,2.7597396)(12.741875,-0.040260416)
\psline[linewidth=0.02cm,fillcolor=black,dotsize=0.07055555cm 2.0]{*-*}(8.541875,5.5597396)(9.541875,2.7597396)
\usefont{T1}{ppl}{m}{n}
\rput{-68.655}(2.0206718,11.269247){\rput(9.225312,4.1747394){\large $\alpha_{2}$}}
\psline[linewidth=0.02cm,fillcolor=white,dotsize=0.07055555cm 2.0]{*-*}(8.541875,-0.040260416)(5.341875,-2.8402605)
\usefont{T1}{ppl}{m}{n}
\rput{41.345284}(0.7919982,-4.788889){\rput(6.7053127,-1.3252604){\large $\alpha_{4}$}}
\psline[linewidth=0.02cm,fillcolor=white,dotsize=0.07055555cm 2.0]{*-*}(8.541875,5.5597396)(7.541875,2.7597396)
\usefont{T1}{ppl}{m}{n}
\rput{68.5604}(8.898896,-4.6248784){\rput(7.8053126,4.23474){\large $\alpha_{3}$}}
\usefont{T1}{ppl}{m}{n}
\rput(11.765156,-2.8302605){\huge\color{black} \psframebox[linewidth=0.02,fillstyle=solid,fillcolor=white]{-1 0 -1 1}}
\usefont{T1}{ppl}{m}{n}
\rput(5.54375,-2.8302605){\huge\color{black} \psframebox[linewidth=0.02,fillstyle=solid,fillcolor=white]{1 -1 0 -1}}
\usefont{T1}{ppl}{m}{n}
\rput(12.644218,-0.030260416){\huge\color{black} \psframebox[linewidth=0.02,fillstyle=solid,fillcolor=white]{-1 0 1 0}}
\usefont{T1}{ppl}{m}{n}
\rput(8.54375,-0.030260416){\huge\color{black} \psframebox[linewidth=0.02,fillstyle=solid,fillcolor=white]{1 -1 -1 1}}
\usefont{T1}{ppl}{m}{n}
\rput(4.4290624,-0.030260416){\huge\color{black} \psframebox[linewidth=0.02,fillstyle=solid,fillcolor=white]{0 1 0 -1}}
\usefont{T1}{ppl}{m}{n}
\rput(10.022813,2.7697396){\huge\color{black} \psframebox[linewidth=0.02,fillstyle=solid,fillcolor=white]{1 -1 1 0}}
\usefont{T1}{ppl}{m}{n}
\rput(7.2290626,2.7697396){\huge\color{black} \psframebox[linewidth=0.02,fillstyle=solid,fillcolor=white]{0 1 -1 1}}
\usefont{T1}{ppl}{m}{n}
\rput(8.508125,5.56974){\huge\color{black} \psframebox[linewidth=0.02,fillstyle=solid,fillcolor=white]{0 1 1 0}}
\usefont{T1}{ppl}{m}{n}
\rput(8.565156,-5.6302605){\huge\color{black} \psframebox[linewidth=0.02,fillstyle=solid,fillcolor=white]{-1 0 0 -1}}
\usefont{T1}{ppl}{m}{n}
\rput(10.486406,5.56974){\huge $\Lambda_{1}$}
\usefont{T1}{ppl}{m}{n}
\rput(5.076406,2.7697396){\huge $\Sigma_{1}$}
\usefont{T1}{ppl}{m}{n}
\rput(12.176406,2.7697396){\huge $\Sigma_{2}$}
\usefont{T1}{ppl}{m}{n}
\rput(2.0764062,-0.030260416){\huge $\Sigma_{3}$}
\usefont{T1}{ppl}{m}{n}
\rput(3.2764062,-2.8302605){\huge $\Sigma_{5}$}
\usefont{T1}{ppl}{m}{n}
\rput(14.776406,-0.030260416){\huge $\Sigma_{4}$}
\usefont{T1}{ppl}{m}{n}
\rput(10.886406,-5.6302605){\huge $\Lambda_{3}$}
\usefont{T1}{ppl}{m}{n}
\rput(8.686406,-1.0302604){\huge $\Lambda_{2}$}
\usefont{T1}{ppl}{m}{n}
\rput(13.976406,-2.8302605){\huge $\Sigma_{6}$}
\end{pspicture}
}
}\hspace{1cm}
\subfloat[$\mathbf{(\overline{3},3)}$ of $\mathfrak{sl}(3,\mathbb{C})$ orbits][Orbits of the real weights $\Lambda _{1}$, $\Lambda _{2}$ and $\Lambda _{3}$ in the $\mathbf{(3}^{\prime }\mathbf{,3)}$ of $\mathfrak{sl}(3,\mathbb{R})\oplus \mathfrak{sl}(3,\mathbb{R})$. The red circles denote the starting points of the corresponding orbit.\label{realweightsorbits}]{

% Generated with LaTeXDraw 2.0.8
% Sun May 07 18:48:32 CEST 2017
% \usepackage[usenames,dvipsnames]{pstricks}
% \usepackage{epsfig}
% \usepackage{pst-grad} % For gradients
% \usepackage{pst-plot} % For axes
\scalebox{0.55} % Change this value to rescale the drawing.
{
\begin{pspicture}(0,-6.3225)(9.22,6.3025)
 \definecolor{color995b}{rgb}{0.9882352941176471,1.0,0.0}
 \definecolor{yellow07b}{rgb}{0.0,0.36470588235294116,1.0}
 \definecolor{color53b}{rgb}{0.8509803921568627,0.0,0.0}
\definecolor{color1644b}{rgb}{0.8509803921568627,0.0,0.0}
\pspolygon[linewidth=0.02,fillstyle=solid,fillcolor=teal!70!blue,opacity=0.3](3.4,2.6825)(3.6,3.4825)(4.6,0.6825)(5.6,3.4825)(5.8,2.6825)(5.0,0.2825)(8.2,-2.7175)(7.8,-3.1175)(4.6,-0.3175)(1.4,-3.1175)(1.0,-2.7175)(4.2,0.2825)
\pspolygon[linewidth=0.02,fillstyle=solid,fillcolor=darkgray,opacity=0.3](0.0,0.0825)(0.4,0.4825)(1.6,-2.5175)(4.6,-5.1175)(7.6,-2.5175)(8.8,0.4825)(9.2,0.0825)(8.2,-2.7175)(4.6,-6.1175)(1.0,-2.7175)
\pspolygon[linewidth=0.02,fillstyle=solid,fillcolor=lime!30!red,opacity=0.3](0.0,0.0825)(0.4,-0.3175)(3.8,2.6825)(4.6,4.8825)(5.4,2.6825)(8.8,-0.3175)(9.2,0.0825)(5.8,3.0825)(4.6,6.2825)(3.4,3.0825)
\usefont{T1}{ppl}{m}{n}
\rput{-40.588333}(3.4094582,1.1354725){\rput(3.2034376,-4.0225){\large $\alpha_{1}$}}
\psline[linewidth=0.02cm,fillcolor=black,dotsize=0.07055555cm 2.0]{*-*}(1.4,-2.7175)(4.6,-5.5175)
\psline[linewidth=0.02cm,fillcolor=black,dotsize=0.07055555cm 2.0]{*-*}(0.4,0.0825)(1.4,-2.7175)
\usefont{T1}{ppl}{m}{n}
\rput{-68.655}(1.943832,0.20224045){\rput(1.0834374,-1.3025){\large $\alpha_{2}$}}
\psline[linewidth=0.02cm,fillcolor=black,dotsize=0.07055555cm 2.0]{*-*}(7.8,-2.7175)(4.6,-5.5175)
\usefont{T1}{ppl}{m}{n}
\rput{41.345284}(-1.1614916,-4.9661326){\rput(5.9634376,-4.0025){\large $\alpha_{4}$}}
\psline[linewidth=0.02cm,fillcolor=black,dotsize=0.07055555cm 2.0]{*-*}(8.8,0.0825)(7.8,-2.7175)
\usefont{T1}{ppl}{m}{n}
\rput{68.5604}(3.9644382,-8.340341){\rput(8.063437,-1.2425){\large $\alpha_{3}$}}
\usefont{T1}{ppl}{m}{n}
\rput{-40.588333}(2.357631,3.8911245){\rput(6.4034376,-1.2225){\large $\alpha_{1}$}}
\psline[linewidth=0.02cm,fillcolor=black,dotsize=0.07055555cm 2.0]{*-*}(4.6,0.0825)(7.8,-2.7175)
\psline[linewidth=0.02cm,fillcolor=black,dotsize=0.07055555cm 2.0]{*-*}(3.6,2.8825)(4.6,0.0825)
\usefont{T1}{ppl}{m}{n}
\rput{-68.655}(1.371151,4.9635863){\rput(4.2834377,1.4975){\large $\alpha_{2}$}}
\psline[linewidth=0.02cm,fillcolor=black,dotsize=0.07055555cm 2.0]{*-*}(3.6,2.8825)(0.4,0.0825)
\usefont{T1}{ppl}{m}{n}
\rput{41.345284}(1.4909593,-0.79578936){\rput(1.7634375,1.5975){\large $\alpha_{4}$}}
\psline[linewidth=0.02cm,fillcolor=black,dotsize=0.07055555cm 2.0]{*-*}(5.6,2.8825)(4.6,0.0825)
\usefont{T1}{ppl}{m}{n}
\rput{68.5604}(4.540352,-3.5852258){\rput(4.8634377,1.5575){\large $\alpha_{3}$}}
\usefont{T1}{ppl}{m}{n}
\rput{-40.588333}(0.77649236,5.2154136){\rput(7.4034376,1.5775){\large $\alpha_{1}$}}
\psline[linewidth=0.02cm,fillcolor=black,dotsize=0.07055555cm 2.0]{*-*}(5.6,2.8825)(8.8,0.0825)
\psline[linewidth=0.02cm,fillcolor=black,dotsize=0.07055555cm 2.0]{*-*}(4.6,5.6825)(5.6,2.8825)
\usefont{T1}{ppl}{m}{n}
\rput{-68.655}(-0.6007677,7.67584){\rput(5.2834377,4.2975){\large $\alpha_{2}$}}
\psline[linewidth=0.02cm,fillcolor=black,dotsize=0.07055555cm 2.0]{*-*}(4.6,0.0825)(1.4,-2.7175)
\usefont{T1}{ppl}{m}{n}
\rput{41.345284}(-0.10944963,-2.1543062){\rput(2.7634375,-1.2025){\large $\alpha_{4}$}}
\psline[linewidth=0.02cm,fillcolor=black,dotsize=0.07055555cm 2.0]{*-*}(4.6,5.6825)(3.6,2.8825)
\usefont{T1}{ppl}{m}{n}
\rput{68.5604}(6.5121217,-0.8778791){\rput(3.8634374,4.3575){\large $\alpha_{3}$}}
\usefont{T1}{ppl}{m}{n}
\rput(5.744531,5.6925){\huge $\Lambda_{1}$}
\usefont{T1}{ppl}{m}{n}
\rput(5.744531,-5.9075){\huge $\Lambda_{3}$}
\usefont{T1}{ppl}{m}{n}
\rput(4.744531,-0.9075){\huge $\Lambda_{2}$}
\pscircle[linewidth=0.02,dimen=outer,fillstyle=solid,fillcolor=color1644b](4.6,5.6825){0.2}
\pscircle[linewidth=0.02,dimen=outer,fillstyle=solid,fillcolor=color1644b](4.6,0.0825){0.2}
\pscircle[linewidth=0.02,dimen=outer,fillstyle=solid,fillcolor=color1644b](4.6,-5.5175){0.2}
\end{pspicture}
}
}
\end{figure}

The action of the generators on the weights appearing above is described in detail in \autoref{appendix:33primesl3sl3}.

\subsubsection{1-weight}

The 1-weight stabilizers are listed in \autoref{stab33barsl3c}.
\begin{table}[h]
\renewcommand{\arraystretch}{1.6}
\par
\begin{center}
\resizebox{\textwidth}{!}{
\begin{tabular}{|c|c|c|}
\hline
\multicolumn{3}{|c|}{\textcolor{black}{\bf Stabilizer}}\\
\hline
\textcolor{black}{$\Lambda_{1}$}&\textcolor{black}{$\Lambda_{2}$}&\textcolor{black}{$\Lambda_{3}$}\\
\hline\hline
$\alpha_{1}+\alpha_{2}\quad \alpha_{3}+\alpha_{4} $&$\alpha_{1}+\alpha_{2}\quad \alpha_{3}+\alpha_{4} $&\\

$\alpha_{1}\quad \alpha_{2}\quad \alpha_{3}\quad \alpha_{4}$&$\alpha_{1}\quad \alpha_{4}$&
$\alpha_{2}\quad \alpha_{3}$\\

$H_{\alpha_{1}}\quad H_{\alpha_{4}}\quad H_{\alpha_{2}}-H_{\alpha_{3}}$&$H_{\alpha_{1}}+H_{\alpha_{2}}\quad H_{\alpha_{3}}+H_{\alpha_{4}}\quad H_{\alpha_{1}}-H_{\alpha_{4}}$&$H_{\alpha_{2}}\quad H_{\alpha_{3}}\quad H_{\alpha_{1}}-H_{\alpha_{4}}$\\

$-\alpha_{1}\quad -\alpha_{4}$&$-\alpha_{2}\quad -\alpha_{3}$&$-\alpha_{1}\quad -\alpha_{2}\quad -\alpha_{3}\quad -\alpha_{4}$\\
&$-\alpha_{1}-\alpha_{2}\quad -\alpha_{3}-\alpha_{4} $&$-\alpha_{1}-\alpha_{2}\quad -\alpha_{3}-\alpha_{4} $\\
\hline
\end{tabular}
}
\end{center}
\caption{Stabilizers for the weights in the $\mathbf{(3}^{\prime }\mathbf{,3)}$ of $\mathfrak{sl}(3,\mathbb{R})\oplus \mathfrak{sl}(3,\mathbb{R})$.}
\label{stab33barsl3c}
\end{table}
\newline
Correspondingly, the orbit of a real weight reads
\begin{align}
\partial I_{3,mag}=0:&&\frac{SL(3,\mathbb{R})\times SL(3,\mathbb{R})}{\left[SL(2,\mathbb{R})\times SL(2,\mathbb{R})\times SO(1,1)\right] \ltimes \mathbb{R}^{\left( 2,2\right) }},  \label{p-spp-1-II}
\end{align}
where $\mathbb{R}^{\left( 2,2\right) }\simeq \left( \mathbf{2,1}\right)_{1}+\left( \mathbf{1,2}\right) _{-1}$ of the split form $SL(2,\mathbb{R})\times SL(2,\mathbb{R})\times SO(1,1)$. Note that $SL(2,\mathbb{R})\times SL(2,\mathbb{R})$ is the $U$-duality group of the corresponding theory uplifted in $D=6$, and $\left( \mathbf{2,1}\right) _{1}+\left(\mathbf{1,2}\right) _{-1}$ ($\left( \mathbf{2,1}\right) _{-1}+\left(\mathbf{1,2}\right) _{1}$) is the irrep. relevant to non-dyonic asymptotically flat branes (black holes and black $2$-branes, respectively) in $D=6$. We also reported the corresponding $\left( SL(3,\mathbb{R})\times SL(3,\mathbb{R}%
)\right) $-invariant constraint on the cubic (magnetic) invariant $I_{3,mag}$of the $\mathbf{(3}^{\prime }\mathbf{,3)}$; note that the orbit (\autoref{p-spp-1-II}) is the orbit of rank-1 elements of the cubic Jordan algebra $J_{3}^{\mathbb{C}_{s}}$.

\subsubsection{2-weights}

Without loss of generality, in this case a bound state can be obtained as combination of $\Lambda _{1}\pm \Lambda _{2}$. The stabilizers are listed in  \autoref{L1pmL1stabs}; the conjunctions can be easily visualized by looking at the overlaps of the orbits sketched in  \autoref{realweightsorbits}.

\begin{table}[h]
\renewcommand{\arraystretch}{1}
\par
\begin{center}
\begin{tabular}{|c|c|c|}
\hline
\textcolor{black}{\bf Common} & \multicolumn{2}{|c|}{\textcolor{black}{\bf Conjunction}} \\ \hline
\textcolor{black}{$\Lambda_{1},\Lambda_{2}$} & \textcolor{black}{$\Lambda_{1}+\Lambda_{2}$} & \textcolor{black}{$\Lambda_{1}-\Lambda_{2}$} \\
\hline \hline
\begin{tabular}{c}
$E_{\alpha_{1}+\alpha_{2}}$ \\
$E_{\alpha_{3}+\alpha_{4}}$ \\
$E_{\alpha_{1}}$ \\
$E_{\alpha_{4}}$ \\
$H_{\alpha_{1}}-H_{\alpha_{4}}$ \\
$H_{\alpha_{2}}-H_{-\alpha_{3}}$ \\
\end{tabular}
&
\begin{tabular}{c}
$E_{\alpha_{2}}-E_{-\alpha_{3}}$ \\
$E_{\alpha_{3}}-E_{-\alpha_{2}}$ \\
\end{tabular}
&
\begin{tabular}{c}
$E_{\alpha_{2}}+E_{-\alpha_{3}}$ \\
$E_{\alpha_{3}}+E_{-\alpha_{2}}$ \\
\end{tabular}
\\ \hline
\end{tabular}%
\end{center}
\caption[$\Lambda _{1}\pm \Lambda _{2}$ stabilizers]{$\Lambda _{1}\pm
\Lambda _{2}$ stabilizers.}
\label{L1pmL1stabs}
\end{table}

The orbit for both $\Lambda _{1}\pm \Lambda _{2}$ read
\begin{align}
I_{3,mag}=0:&&\frac{SL(3,\mathbb{R})\times SL(3,\mathbb{R})}{\left[ SL(2,\mathbb{R})_{d}\times SO(1,1)\right] \ltimes \mathbb{R}^{(2,2)}},\label{p-spp-2-II}
\end{align}
where $SL(2,\mathbb{R})_{d}$ is diagonally embedded into the $SL(2,\mathbb{R})\times SL(2,\mathbb{R})$ (non-maximal) subgroup of the e.m. duality, and $\mathbb{R}^{(2,2)}\simeq \left( \mathbf{2,2}\right) $denotes the bi-fundamental of $SL(2,\mathbb{R})_{d}\times SO(1,1)$. We also reported the corresponding $\left( SL(3,\mathbb{R})\times SL(3,\mathbb{R})\right) $-invariant constraint on the cubic (magnetic) invariant $I_{3,mag}$ of the $\mathbf{(3}^{\prime }\mathbf{,3)}$; note that the orbit (\autoref%
{p-spp-2-II}) is the orbit of rank-2 elements of the cubic Jordan algebra $J_{3}^{\mathbb{C}_{s}}$.

\subsubsection{3-weights}

For the 3-weights bound states, we consider $\Lambda _{1}+\Lambda _{2}\pm \Lambda _{3}$, with stabilizers reported in \autoref{L1L2pmL3stabs}.
\begin{table}[h]
\renewcommand{\arraystretch}{1}
\par
\begin{center}
\begin{tabular}{|c|c|c|}
\hline
{\bf Common}& \multicolumn{2}{|c|}{\bf Conjunction} \\ \hline
$\Lambda_{1},\Lambda_{2},\Lambda_3$ & \textcolor{black}{$\Lambda_{1}+\Lambda_{2}+\Lambda_3$} & \textcolor{black}{$\Lambda_{1}+\Lambda_{2}-\Lambda_3$} \\ \hline \hline
$H_{\alpha_{1}}-H_{\alpha_{4}}$ & $E_{\alpha_2}-E_{-\alpha_3}$ & $E_{\alpha_2}-E_{-\alpha_3}$ \\
$H_{\alpha_{2}}-H_{\alpha_{3}}$ & $E_{\alpha_3}-E_{-\alpha_2}$ & $E_{\alpha_3}-E_{-\alpha_2}$ \\
& $E_{\alpha_1}-E_{-\alpha_4}$ & $E_{\alpha_1}+E_{-\alpha_4}$ \\
& $E_{\alpha_4}-E_{-\alpha_1}$ & $E_{\alpha_4}+E_{-\alpha_1}$ \\
& $E_{\alpha_1+\alpha_2}-E_{-\alpha_3-\alpha_4}$ & $E_{\alpha_1+\alpha_2}+E_{-\alpha_3-\alpha_4}$ \\
& $E_{\alpha_3+\alpha_4}-E_{-\alpha_1-\alpha_2}$ & $E_{\alpha_3+\alpha_4}+E_{-\alpha_1-\alpha_2}$ \\ \hline
\end{tabular}
\end{center}
\caption{$\Lambda _{1}+\Lambda _{2}\pm \Lambda _{3}$ stabilizers.}
\label{L1L2pmL3stabs}
\end{table}
\newline
Thus, for both $\Lambda _{1}+\Lambda _{2}\pm \Lambda _{3}$, we obtain the following 3-weights bound state orbit :
\begin{align}
I_{3,mag}\neq 0:&&\frac{SL(3,\mathbb{R})\times SL(3,\mathbb{R})}{SL(3,\mathbb{R})_{d}},  \label{p-spp-3-II}
\end{align}
where the stabilizer $SL(3,\mathbb{R})_{d}$ is embedded in the e.m. duality group through the (maximal and symmetric; \textit{cfr. e.g.} \cite{Pasquier}) \textit{diagonal} embedding (with simple roots $\frac{\alpha _{2}-\alpha_{3}}{2}$ and $\frac{\alpha _{1}-\alpha _{4}}{2}$); thus, this orbit is a symmetric manifold. We also reported the corresponding $\left(SL(3,\mathbb{R})\times SL(3,\mathbb{R})\right)$-invariant constraint on the cubic (magnetic) invariant $I_{3,mag}$ of the $\mathbf{(3}^{\prime }\mathbf{,3)}$; note that the orbit (\autoref{p-spp-3-II}) is the orbit of rank-3 elements of the cubic Jordan algebra $J_{3}^{\mathbb{C}_{s}}$.\bigskip

We summarize the stratification of the $\mathbf{(3}^{\prime }\mathbf{,3)}$ of the split form $\mathfrak{sl}(3,\mathbb{R})\oplus \mathfrak{sl}(3,\mathbb{R})$ of the Lie algebra $\mathfrak{a}_{2}\oplus \mathfrak{a}_{2}$ in  \autoref{3bar3summary}.
\begin{table}[h]
\renewcommand{\arraystretch}{1.3}
\par
\begin{center}
\resizebox{\textwidth}{!}{
\begin{tabular}{|c|c|c|c|c|c|c|}
\hline
\multicolumn{2}{|c|}{}&\multicolumn{2}{|c|}{\textcolor{black}{$\theta$}}&&&\\ \cline{3-4}
\multicolumn{2}{|c|}{\multirow{-2}{*}{\textcolor{black}{\bf State}}}&\textcolor{black}{+}&\textcolor{black}{-}&\multirow{-2}{*}{\textcolor{black}{\bf Semisimple Stabilizer}}&\multirow{-2}{*}{\textcolor{black}{\bf Stabilizer}}&\multirow{-2}{*}{\textcolor{black}{\bf Rank}}\\ \hline\hline

1-w&$\Lambda_{1}$&2&5&$SL(2,\mathbb{R})\times SL(2,\mathbb{R})\times SO(1,1)$&
$[SL(2,\mathbb{R})\times SL(2,\mathbb{R})\times SO(1,1)]\ltimes \mathbb{R}^{(2,2)}$&1\\ \hline

2-w&$\Lambda_{1}\pm\Lambda_{2}$&1&3&$SL(2,\mathbb{R})_{d}\times SO(1,1)$&$[SL(2,\mathbb{R})\times SO(1,1)]\ltimes \mathbb{R}^{(2,2)}$&2\\
\hline

3-w&$\Lambda_{1}+\Lambda_{2}\pm\Lambda_{3}$&3&5&$SL(3,\mathbb{R})$&$SL(3,\mathbb{R})_{d}$&3\\
\hline

\hline

\hline
\end{tabular}
}
\end{center}
\caption{Orbit stabilizers in the $\mathbf{(3}^{\prime }\mathbf{,3)}$ of $\mathfrak{sl}(3,\mathbb{R})\oplus \mathfrak{sl}(3,\mathbb{R})$.}
\label{3bar3summary}
\end{table}

\FloatBarrier
\subsection{\label{D=5-Hs}$\mathbb{H}_{s}$}

The $U$-duality group is $SL(6,\mathbb{R})$, and the scalar manifold reads (\textit{cfr.} (\autoref{Cs-D=4}))%
\begin{equation}
\frac{Str_{0}\left( J_{3}^{\mathbb{H}_{s}}\right) }{mcs\left( Str_{0}\left(J_{3}^{\mathbb{H}_{s}}\right) \right) }=\frac{SL(6,\mathbb{R})}{SO(6)}\simeq\frac{Conf\left( J_{3}^{\mathbb{C}_{s}}\right) }{mcs\left( Conf\left( J_{3}^{\mathbb{C}_{s}}\right) \right) }.  \label{sm-Hs-D=5}
\end{equation}

The electric $0$-brane and magnetic $1$-brane irreps. are the rank-$2$ antisymmetric $\mathbf{15}$, and its conjugate (rank-$4$ antisymmetric) $\mathbf{15}^{\prime }$, respectively. They are both characterized by a unique independent cubic invariant polynomial, which we will denote by $I_{3,el}$ and $I_{3,mag}$, respectively \cite{Sato-Kimura, Kac-80,
Garibaldi-2, Garibaldi-2}.

While the non-linear action of $SL(6,\mathbb{R})$ on the scalar manifold (\autoref{sm-Hs-D=5}) is transitive, the linear action of $SL(6,\mathbb{R})$ on the $\mathbf{15}$ and $\mathbf{15}^{\prime }$ determines the stratification into three orbits, classified in terms of invariant constraints on $I_{3}$, or equivalently in terms of the \textit{rank} of the corresponding
Jordan algebra $J_{3}^{\mathbb{H}_{s}}$ \cite{rank-J,rank-FTS}.\bigskip

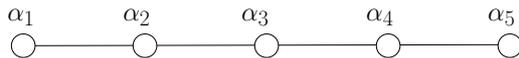
\begin{figure}[h]
\centering
% \usepackage[usenames,dvipsnames]{pstricks}
% \usepackage{epsfig}
% \usepackage{pst-grad} % For gradients
% \usepackage{pst-plot} % For axes
% \usepackage[space]{grffile} % For spaces in paths
% \usepackage{etoolbox} % For spaces in paths
% \makeatletter % For spaces in paths
% \patchcmd\Gread@eps{\@inputcheck#1 }{\@inputcheck"#1"\relax}{}{}
% \makeatother
% % User Packages:
%
%

\psscalebox{0.4} % Change this value to rescale the drawing.
{
\begin{pspicture}(0,-0.77234375)(17.527187,0.77234375)
\psline[linecolor=black, linewidth=0.02](0.39,-0.37165624)(16.426874,-0.33765626)
\pscircle[linecolor=black, linewidth=0.02, fillstyle=solid, dimen=outer](0.4,-0.37234366){0.4}
\pscircle[linecolor=black, linewidth=0.02, fillstyle=solid, dimen=outer](4.4,-0.37234375){0.4}
\rput(12.128906,0.5623438){\huge $\alpha_4$}
\pscircle[linecolor=black, linewidth=0.02, fillstyle=solid, dimen=outer](8.4,-0.37234366){0.4}
\pscircle[linecolor=black, linewidth=0.02, fillstyle=solid, dimen=outer](12.4,-0.37234375){0.4}
\rput(8.028906,0.5623438){\huge $\alpha_3$}
\rput(4.128906,0.5623438){\huge $\alpha_2$}
\rput(0.3289062,0.5623438){\huge $\alpha_1$}
\rput(16.128906,0.5623438){\huge $\alpha_5$}
\pscircle[linecolor=black, linewidth=0.02, fillstyle=solid, dimen=outer](16.4,-0.37234375){0.4}
\end{pspicture}
}
\caption{Tits-Satake diagram of $\mathfrak{sl}(6,\mathbb{R})$.}
\label{fig:sl6titsatake1}
\end{figure}
\begin{figure}[h!]
\subfloat[][{\bf 15} of $SL(6,\mathbb{R})$. All the weights in the representation have the same length and we refer to the weights without label as $\Sigma_i$ with i ranging from 1 to 12 and increasing from the top to the bottom from the left to the rights of the diagram.\label{fig:15sl6dynkintree}]{
\centering
% Generated with LaTeXDraw 2.0.8
% Fri Apr 21 22:56:49 CEST 2017
% \usepackage[usenames,dvipsnames]{pstricks}
% \usepackage{epsfig}
% \usepackage{pst-grad} % For gradients
% \usepackage{pst-plot} % For axes
\scalebox{0.6} % Change this value to rescale the drawing.
{
\begin{pspicture}(0,-9.197396)(13.787239,9.197396)
\usefont{T1}{ppl}{m}{n}
\rput(5.544427,8.8203125){\Large $\Lambda_{1}$}
\usefont{T1}{ppl}{m}{n}
\rput(11.844427,-0.179687){\Large $\Lambda_{2}$}
\usefont{T1}{ppl}{m}{n}
\rput(5.644427,-8.879687){\Large $\Lambda_{3}$}
\usefont{T1}{ppl}{m}{n}
\rput{35.35122}(-1.9191236,-4.999212){\rput(6.854265,-5.494954){$\alpha_{5}$}}
\psline[linewidth=0.02cm,fillcolor=black,dotsize=0.07055555cm 2.0]{*-*}(8.618802,-4.524687)(5.4188023,-6.724687)
\usefont{T1}{ppl}{m}{n}
\rput{35.35122}(1.0514055,-1.1208084){\rput(2.2542653,1.1050457){$\alpha_{5}$}}
\psline[linewidth=0.02cm,fillcolor=black,dotsize=0.07055555cm 2.0]{*-*}(4.018802,2.075313)(0.81880236,-0.124687)
\psline[linewidth=0.02cm,fillcolor=black,dotsize=0.07055555cm 2.0]{*-*}(5.4188023,-6.724687)(4.018802,-8.924687)
\usefont{T1}{ppl}{m}{n}
\rput{54.790813}(-4.547955,-6.987708){\rput(4.4376736,-7.865881){$\alpha_{4}$}}
\usefont{T1}{ppl}{m}{n}
\rput{35.35122}(-0.6462323,-4.5935774){\rput(6.854265,-3.2949543){$\alpha_{5}$}}
\psline[linewidth=0.02cm,fillcolor=black,dotsize=0.07055555cm 2.0]{*-*}(8.618802,-2.324687)(5.4188023,-4.524687)
\psline[linewidth=0.02cm,fillcolor=black,dotsize=0.07055555cm 2.0]{*-*}(0.81880236,-0.124687)(4.018802,-2.324687)
\usefont{T1}{ppl}{m}{n}
\rput{-32.069756}(0.9199811,1.1271031){\rput(2.3908226,-1.0211673){$\alpha_{1}$}}
\psline[linewidth=0.02cm,fillcolor=black,dotsize=0.07055555cm 2.0]{*-*}(5.4188023,4.275313)(4.018802,2.075313)
\usefont{T1}{ppl}{m}{n}
\rput{54.790813}(4.439622,-2.329905){\rput(4.4376736,3.1341188){$\alpha_{4}$}}
\psline[linewidth=0.02cm,fillcolor=black,dotsize=0.07055555cm 2.0]{*-*}(4.018802,-2.3246872)(5.4188023,-4.524687)
\usefont{T1}{ppl}{m}{n}
\rput{-58.967655}(5.244566,2.5710318){\rput(4.8659215,-3.336626){$\alpha_{2}$}}
\psline[linewidth=0.02cm,fillcolor=black,dotsize=0.07055555cm 2.0]{*-*}(4.018802,2.075313)(7.2188025,-0.124687)
\usefont{T1}{ppl}{m}{n}
\rput{-32.069756}(0.24020079,3.1618621){\rput(5.5908227,1.1788328){$\alpha_{1}$}}
\psline[linewidth=0.02cm,fillcolor=black,dotsize=0.07055555cm 2.0]{*-*}(10.018803,-0.1246872)(8.618802,-2.324687)
\usefont{T1}{ppl}{m}{n}
\rput{54.790813}(2.7924001,-7.951468){\rput(9.037673,-1.2658811){$\alpha_{4}$}}
\psline[linewidth=0.02cm,fillcolor=black,dotsize=0.07055555cm 2.0]{*-*}(7.2188025,-0.1246872)(8.618802,-2.324687)
\usefont{T1}{ppl}{m}{n}
\rput{-58.967655}(4.9097676,6.378888){\rput(8.065922,-1.1366261){$\alpha_{2}$}}
\usefont{T1}{ppl}{m}{n}
\rput{35.35122}(0.36852798,-3.377921){\rput(5.454265,-1.0949543){$\alpha_{5}$}}
\psline[linewidth=0.02cm,fillcolor=black,dotsize=0.07055555cm 2.0]{*-*}(7.2188025,-0.124687)(4.018802,-2.324687)
\psline[linewidth=0.02cm,fillcolor=black,dotsize=0.07055555cm 2.0]{*-*}(5.4188023,6.475313)(8.618802,4.275313)
\usefont{T1}{ppl}{m}{n}
\rput{-32.069756}(-1.8823484,4.576624){\rput(6.990823,5.5788326){$\alpha_{1}$}}
\psline[linewidth=0.02cm,fillcolor=black,dotsize=0.07055555cm 2.0]{*-*}(8.618802,2.0753129)(7.2188025,-0.124687)
\usefont{T1}{ppl}{m}{n}
\rput{54.790813}(3.9971042,-5.876034){\rput(7.6376734,0.9341189){$\alpha_{4}$}}
\psline[linewidth=0.02cm,fillcolor=black,dotsize=0.07055555cm 2.0]{*-*}(8.618802,2.0753129)(10.018803,-0.124687)
\usefont{T1}{ppl}{m}{n}
\rput{-58.967655}(3.702909,8.644367){\rput(9.465921,1.0633738){$\alpha_{2}$}}
\psline[linewidth=0.02cm,fillcolor=black,dotsize=0.07055555cm 2.0]{*-*}(5.4188023,4.275313)(8.618802,2.075313)
\usefont{T1}{ppl}{m}{n}
\rput{-32.069756}(-0.71425533,4.240909){\rput(6.990823,3.3788328){$\alpha_{1}$}}
\psline[linewidth=0.02cm,fillcolor=black,dotsize=0.07055555cm 2.0]{*-*}(4.018802,8.675313)(5.4188023,6.475313)
\usefont{T1}{ppl}{m}{n}
\rput{-58.967655}(-4.1810746,7.900291){\rput(4.8659215,7.663374){$\alpha_{2}$}}
\psline[linewidth=0.02cm,fillcolor=black,dotsize=0.07055555cm 2.0]{*-*}(5.4188023,-4.5246863)(5.4188023,-6.724687)
\usefont{T1}{ppl}{m}{n}
\rput{90.38534}(-0.5186829,-11.077504){\rput(5.212285,-5.7805743){$\alpha_{3}$}}
\psline[linewidth=0.02cm,fillcolor=black,dotsize=0.07055555cm 2.0]{*-*}(8.618802,4.2753134)(8.618802,2.0753129)
\usefont{T1}{ppl}{m}{n}
\rput{90.38534}(11.502639,-5.4182487){\rput(8.412285,3.0194259){$\alpha_{3}$}}
\psline[linewidth=0.02cm,fillcolor=black,dotsize=0.07055555cm 2.0]{*-*}(5.4188023,6.4753137)(5.4188023,4.275313)
\usefont{T1}{ppl}{m}{n}
\rput{90.38534}(10.481069,-0.0035248662){\rput(5.212285,5.2194257){$\alpha_{3}$}}
\psline[linewidth=0.02cm,fillcolor=black,dotsize=0.07055555cm 2.0]{*-*}(8.618802,-2.3246865)(8.618802,-4.5246873)
\usefont{T1}{ppl}{m}{n}
\rput{90.38534}(4.902788,-12.062636){\rput(8.412285,-3.5805743){$\alpha_{3}$}}
\usefont{T1}{ppl}{m}{n}
\rput(4.191927,-2.379687){\Large \psframebox[linewidth=0.02,fillstyle=solid]{-1 1 0 0 -1}}
\usefont{T1}{ppl}{m}{n}
\rput(4.0206776,-8.879687){\Large \psframebox[linewidth=0.02,fillstyle=solid]{0 0 0 -1 0}}
\usefont{T1}{ppl}{m}{n}
\rput(5.408646,-6.679687){\Large \psframebox[linewidth=0.02,fillstyle=solid]{0 0 -1 1 -1}}
\usefont{T1}{ppl}{m}{n}
\rput(5.3939586,-4.479687){\Large \psframebox[linewidth=0.02,fillstyle=solid]{0 -1 1 0 -1}}
\usefont{T1}{ppl}{m}{n}
\rput(7.091927,-0.179687){\Large \psframebox[linewidth=0.02,fillstyle=solid]{-1 1 0 -1 1}}
\usefont{T1}{ppl}{m}{n}
\rput(10.118646,-0.179687){\Large \psframebox[linewidth=0.02,fillstyle=solid]{0 -1 0 1 0}}
\usefont{T1}{ppl}{m}{n}
\rput(1.1872395,-0.179687){\Large \psframebox[linewidth=0.02,fillstyle=solid]{1 0 0 0 -1}}
\usefont{T1}{ppl}{m}{n}
\rput(3.8978648,2.120313){\Large \psframebox[linewidth=0.02,fillstyle=solid]{1 0 0 -1 1}}
\usefont{T1}{ppl}{m}{n}
\rput(8.498021,2.120313){\Large \psframebox[linewidth=0.02,fillstyle=solid]{-1 1 -1 1 0}}
\usefont{T1}{ppl}{m}{n}
\rput(5.5039587,4.320313){\Large \psframebox[linewidth=0.02,fillstyle=solid]{1 0 -1 1 0}}
\usefont{T1}{ppl}{m}{n}
\rput(8.514115,4.320313){\Large \psframebox[linewidth=0.02,fillstyle=solid]{-1 0 1 0 0}}
\usefont{T1}{ppl}{m}{n}
\rput(5.3100524,6.620313){\Large \psframebox[linewidth=0.02,fillstyle=solid]{1 -1 1 0 0}}
\usefont{T1}{ppl}{m}{n}
\rput(3.9306777,8.8203125){\Large \psframebox[linewidth=0.02,fillstyle=solid]{0 1 0 0 0}}
\usefont{T1}{ppl}{m}{n}
\rput(8.502552,-2.379687){\Large \psframebox[linewidth=0.02,fillstyle=solid]{0 -1 1 -1 1}}
\usefont{T1}{ppl}{m}{n}
\rput(8.512552,-4.479687){\Large \psframebox[linewidth=0.02,fillstyle=solid]{0 0 -1 0 1}}
\end{pspicture}
}
}
\subfloat[][Orbits in the {\bf 15} of $SL(6,\mathbb{R})$.\label{fig:15sl6orbits}]{
\centering
% Generated with LaTeXDraw 2.0.8
% Sat May 06 18:17:57 CEST 2017
% \usepackage[usenames,dvipsnames]{pstricks}
% \usepackage{epsfig}
% \usepackage{pst-grad} % For gradients
% \usepackage{pst-plot} % For axes
\scalebox{0.6} % Change this value to rescale the drawing.
{
\begin{pspicture}(0,-9.32)(13.505625,9.32)
\definecolor{color395b}{rgb}{1.0,0.792156862745098,0.0}
\definecolor{color336b}{rgb}{0.14901960784313725,0.42745098039215684,0.8313725490196079}
\definecolor{color181b}{rgb}{0.8509803921568627,0.0,0.0}
\definecolor{color519b}{rgb}{0.8352941176470589,0.00392156862745098,0.00392156862745098}
\pspolygon[linewidth=0.02,fillstyle=solid,fillcolor=color395b,opacity=0.3](3.2,-9.0)(3.7,-9.3)(5.5,-6.7)(8.6,-4.5)(8.6,-1.9)(7.1,0.3)(3.8,2.6)(0.0,0.0)(3.5,-2.4)(4.8,-4.6)(4.8,-6.6)
\pspolygon[linewidth=0.02,fillstyle=solid,fillcolor=color519b,opacity=0.3](10.1,-0.1)(8.6,1.8)(8.6,4.5)(4.8,7.2)(4.8,6.7)(4.8,4.0)(8.1,1.7)(9.4,0.0)(8.1,-1.8)(4.8,-4.1)(4.8,-7.2)(8.6,-4.5)(8.6,-1.9)
\pspolygon[linewidth=0.02,fillstyle=solid,fillcolor=color336b,opacity=0.3](3.2,9.0)(3.7,9.3)(5.5,6.7)(8.6,4.5)(8.6,1.9)(7.1,-0.3)(3.8,-2.6)(0.0,0.0)(3.5,2.4)(4.8,4.6)(4.8,6.6)
\usefont{T1}{ppl}{m}{n}
\rput(5.262812,8.922916){\Large $\Lambda_{1}$}
\usefont{T1}{ppl}{m}{n}
\rput(11.562812,-0.077082835){\Large $\Lambda_{2}$}
\usefont{T1}{ppl}{m}{n}
\rput(5.362812,-8.777083){\Large $\Lambda_{3}$}
\usefont{T1}{ppl}{m}{n}
\rput{35.35122}(-1.9116821,-4.817355){\rput(6.5726504,-5.39235){$\alpha_{5}$}}
\psline[linewidth=0.02cm,fillcolor=black,dotsize=0.07055555cm 2.0]{*-*}(8.337188,-4.422083)(5.1371875,-6.6220827)
\usefont{T1}{ppl}{m}{n}
\rput{35.35122}(1.0588471,-0.9389516){\rput(1.9726504,1.20765){$\alpha_{5}$}}
\psline[linewidth=0.02cm,fillcolor=black,dotsize=0.07055555cm 2.0]{*-*}(3.7371874,2.1779172)(0.5371875,-0.022082834)
\psline[linewidth=0.02cm,fillcolor=black,dotsize=0.07055555cm 2.0]{*-*}(5.1371875,-6.622083)(3.7371874,-8.8220825)
\usefont{T1}{ppl}{m}{n}
\rput{54.790813}(-4.583368,-6.7141676){\rput(4.1560583,-7.763277){$\alpha_{4}$}}
\usefont{T1}{ppl}{m}{n}
\rput{35.35122}(-0.63879085,-4.4117208){\rput(6.5726504,-3.1923501){$\alpha_{5}$}}
\psline[linewidth=0.02cm,fillcolor=black,dotsize=0.07055555cm 2.0]{*-*}(8.337188,-2.2220829)(5.1371875,-4.422083)
\psline[linewidth=0.02cm,fillcolor=black,dotsize=0.07055555cm 2.0]{*-*}(0.5371875,-0.022082834)(3.7371874,-2.2220829)
\usefont{T1}{ppl}{m}{n}
\rput{-32.069756}(0.82252955,0.9932365){\rput(2.1092076,-0.91856307){$\alpha_{1}$}}
\psline[linewidth=0.02cm,fillcolor=black,dotsize=0.07055555cm 2.0]{*-*}(5.1371875,4.377917)(3.7371874,2.1779172)
\usefont{T1}{ppl}{m}{n}
\rput{54.790813}(4.404209,-2.0563645){\rput(4.1560583,3.2367232){$\alpha_{4}$}}
\psline[linewidth=0.02cm,fillcolor=black,dotsize=0.07055555cm 2.0]{*-*}(3.7371874,-2.222083)(5.1371875,-4.422083)
\usefont{T1}{ppl}{m}{n}
\rput{-58.967655}(5.0202107,2.3794322){\rput(4.5843067,-3.234022){$\alpha_{2}$}}
\psline[linewidth=0.02cm,fillcolor=black,dotsize=0.07055555cm 2.0]{*-*}(3.7371874,2.1779172)(6.9371877,-0.022082834)
\usefont{T1}{ppl}{m}{n}
\rput{-32.069756}(0.1427492,3.0279956){\rput(5.309208,1.2814369){$\alpha_{1}$}}
\psline[linewidth=0.02cm,fillcolor=black,dotsize=0.07055555cm 2.0]{*-*}(9.737187,-0.022083033)(8.337188,-2.2220829)
\usefont{T1}{ppl}{m}{n}
\rput{54.790813}(2.7569869,-7.677927){\rput(8.756059,-1.1632769){$\alpha_{4}$}}
\psline[linewidth=0.02cm,fillcolor=black,dotsize=0.07055555cm 2.0]{*-*}(6.9371877,-0.022083033)(8.337188,-2.2220829)
\usefont{T1}{ppl}{m}{n}
\rput{-58.967655}(4.6854124,6.1872883){\rput(7.784307,-1.034022){$\alpha_{2}$}}
\usefont{T1}{ppl}{m}{n}
\rput{35.35122}(0.37596944,-3.1960645){\rput(5.1726503,-0.9923501){$\alpha_{5}$}}
\psline[linewidth=0.02cm,fillcolor=black,dotsize=0.07055555cm 2.0]{*-*}(6.9371877,-0.022082834)(3.7371874,-2.2220829)
\psline[linewidth=0.02cm,fillcolor=black,dotsize=0.07055555cm 2.0]{*-*}(5.1371875,6.577917)(8.337188,4.3779173)
\usefont{T1}{ppl}{m}{n}
\rput{-32.069756}(-1.9797999,4.4427576){\rput(6.7092075,5.681437){$\alpha_{1}$}}
\psline[linewidth=0.02cm,fillcolor=black,dotsize=0.07055555cm 2.0]{*-*}(8.337188,2.177917)(6.9371877,-0.022082834)
\usefont{T1}{ppl}{m}{n}
\rput{54.790813}(3.961691,-5.602493){\rput(7.3560586,1.036723){$\alpha_{4}$}}
\psline[linewidth=0.02cm,fillcolor=black,dotsize=0.07055555cm 2.0]{*-*}(8.337188,2.177917)(9.737187,-0.022082834)
\usefont{T1}{ppl}{m}{n}
\rput{-58.967655}(3.4785535,8.452767){\rput(9.184307,1.1659781){$\alpha_{2}$}}
\psline[linewidth=0.02cm,fillcolor=black,dotsize=0.07055555cm 2.0]{*-*}(5.1371875,4.3779173)(8.337188,2.1779172)
\usefont{T1}{ppl}{m}{n}
\rput{-32.069756}(-0.8117069,4.1070423){\rput(6.7092075,3.481437){$\alpha_{1}$}}
\psline[linewidth=0.02cm,fillcolor=black,dotsize=0.07055555cm 2.0]{*-*}(3.7371874,8.777917)(5.1371875,6.577917)
\usefont{T1}{ppl}{m}{n}
\rput{-58.967655}(-4.4054303,7.7086916){\rput(4.5843067,7.765978){$\alpha_{2}$}}
\psline[linewidth=0.02cm,fillcolor=black,dotsize=0.07055555cm 2.0]{*-*}(5.1371875,-4.4220824)(5.1371875,-6.622083)
\usefont{T1}{ppl}{m}{n}
\rput{90.38534}(-0.6995901,-10.692602){\rput(4.9306703,-5.67797){$\alpha_{3}$}}
\psline[linewidth=0.02cm,fillcolor=black,dotsize=0.07055555cm 2.0]{*-*}(8.337188,4.377918)(8.337188,2.177917)
\usefont{T1}{ppl}{m}{n}
\rput{90.38534}(11.3217325,-5.033346){\rput(8.130671,3.12203){$\alpha_{3}$}}
\psline[linewidth=0.02cm,fillcolor=black,dotsize=0.07055555cm 2.0]{*-*}(5.1371875,6.5779176)(5.1371875,4.377917)
\usefont{T1}{ppl}{m}{n}
\rput{90.38534}(10.300161,0.38137752){\rput(4.9306703,5.32203){$\alpha_{3}$}}
\psline[linewidth=0.02cm,fillcolor=black,dotsize=0.07055555cm 2.0]{*-*}(8.337188,-2.2220824)(8.337188,-4.422083)
\usefont{T1}{ppl}{m}{n}
\rput{90.38534}(4.721882,-11.677733){\rput(8.130671,-3.4779701){$\alpha_{3}$}}
\pscircle[linewidth=0.02,dimen=outer,fillstyle=solid,fillcolor=color181b](3.7,-8.8){0.2}
\pscircle[linewidth=0.02,dimen=outer,fillstyle=solid,fillcolor=color181b](9.7,-0.1){0.2}
\pscircle[linewidth=0.02,dimen=outer,fillstyle=solid,fillcolor=color181b](3.7,8.8){0.2}
\end{pspicture}
}
}
\caption{Dynkin tree of the {\bf 15} of $\mathfrak{sl}(6,\mathbb{R})$ and the orbits of the weights $\Lambda_1, \Lambda_2$ and $\Lambda_3$.}\label{fig:15sl6dynkintreeandorbits}
\end{figure}
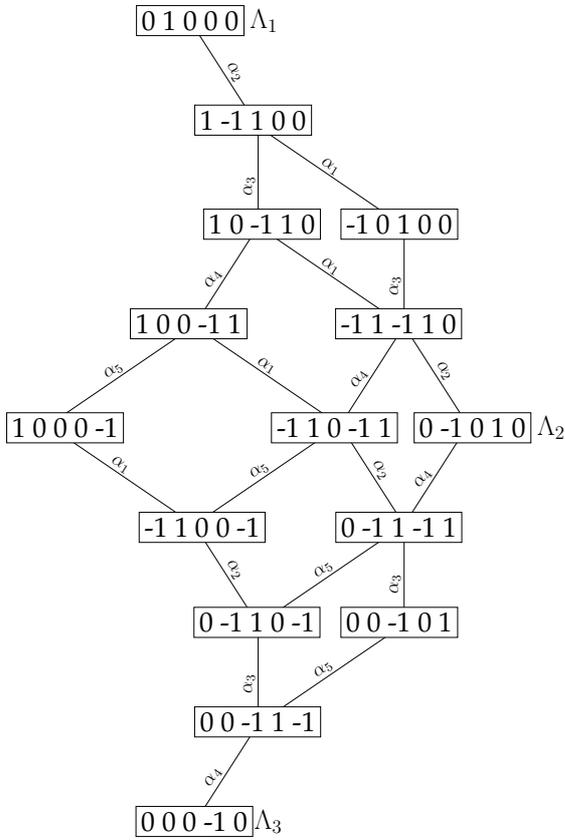
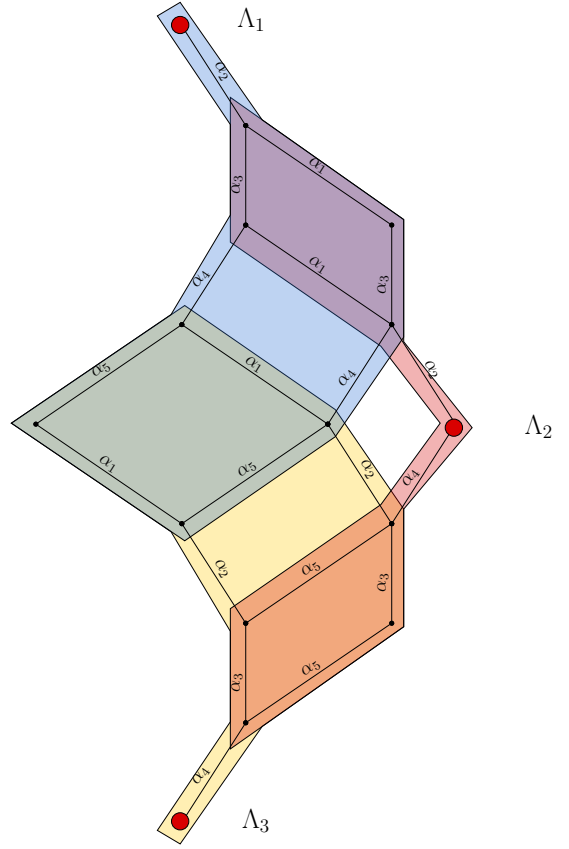
In \autoref{fig:15sl6dynkintreeandorbits} and  we show the Dynkin tree of the {\bf 15} and the orbits of the weights $\Lambda_1, \Lambda_2$ and $\Lambda_3$ we have chosen as our representatives. The simple roots have been denoted as in \autoref{fig:sl6titsatake1}. All the weights in the representation have the same length.

\begin{table}[h!]
\renewcommand{\arraystretch}{1.6}
\begin{center}
\resizebox{\textwidth}{!}{
\begin{tabular}{|c|c|c|}
\hline
\multicolumn{3}{|c|}{\textbf{Stabilizer}}\\
\hline
$\Lambda_{1}$&$\Lambda_{2}$&$\Lambda_{3}$\\ \hline\hline

$\alpha_1+\alpha_2+\alpha_3+\alpha_4+\alpha_5$&$\alpha_1+\alpha_2+\alpha_3+\alpha_4+\alpha_5$&\\

$\alpha_1+\alpha_2+\alpha_3+\alpha_4\quad \alpha_2+\alpha_3+\alpha_4+\alpha_5$&$\alpha_1+\alpha_2+\alpha_3+\alpha_4\quad \alpha_2+\alpha_3+\alpha_4+\alpha_5$&\\

$\alpha_1+\alpha_2+\alpha_3\quad \alpha_2+\alpha_3+\alpha_4\quad \alpha_3+\alpha_4+\alpha_5$&$\alpha_2+\alpha_3+\alpha_4\quad \alpha_3+\alpha_4+\alpha_5$&$\alpha_1+\alpha_2+\alpha_3$\\

$\alpha_1+\alpha_2\quad\alpha_2+\alpha_3\quad \alpha_3+\alpha_4\quad \alpha_4+\alpha_5$&$\alpha_3+\alpha_4\quad \alpha_4+\alpha_5$&$\alpha_1+\alpha_2\quad\alpha_2+\alpha_3$\\

$\alpha_1\quad\alpha_2\quad\alpha_3\quad \alpha_4\quad \alpha_5$&$\alpha_1\quad\alpha_3\quad \alpha_4\quad \alpha_5$&$\alpha_1\quad\alpha_2\quad\alpha_3\quad \alpha_5$\\

$H_{\alpha_1}\quad H_{\alpha_3}\quad H_{\alpha_4}\quad H_{\alpha_5}$&$H_{\alpha_1}\quad H_{\alpha_3}\quad H_{\alpha_2}+ H_{\alpha_4}\quad H_{\alpha_5}$&$H_{\alpha_1}\quad H_{\alpha_2}\quad H_{\alpha_3}\quad  H_{\alpha_5}$\\

$-\alpha_1\quad-\alpha_3\quad -\alpha_4\quad -\alpha_5$&$-\alpha_1\quad-\alpha_2\quad-\alpha_3\quad -\alpha_5$&$-\alpha_1\quad-\alpha_2\quad-\alpha_3\quad -\alpha_4\quad -\alpha_5$\\

$-\alpha_3-\alpha_4\quad -\alpha_4-\alpha_5$&$-\alpha_1-\alpha_2\quad-\alpha_2-\alpha_3$&$-\alpha_1-\alpha_2\quad-\alpha_2-\alpha_3\quad -\alpha_3-\alpha_4\quad -\alpha_4-\alpha_5$\\

$-\alpha_3-\alpha_4-\alpha_5$&$-\alpha_1-\alpha_2-\alpha_3\quad -\alpha_2-\alpha_3-\alpha_4$&$-\alpha_1-\alpha_2-\alpha_3\quad -\alpha_2-\alpha_3-\alpha_4\quad -\alpha_3-\alpha_4-\alpha_5$\\

&$-\alpha_1-\alpha_2-\alpha_3-\alpha_4\quad -\alpha_2-\alpha_3-\alpha_4-\alpha_5$&$-\alpha_1-\alpha_2-\alpha_3-\alpha_4\quad -\alpha_2-\alpha_3-\alpha_4-\alpha_5$\\

&$-\alpha_1-\alpha_2-\alpha_3-\alpha_4-\alpha_5$&$-\alpha_1-\alpha_2-\alpha_3-\alpha_4-\alpha_5$\\

\hline
\end{tabular}
}
\caption{Stabilizers for the weights in the ${\bf 15}$ of $\mathfrak{sl}(6,\mathbb{R})$.}\label{tab:15sl6stabilizers}
\end{center}
\end{table}

\begin{table}[h!]
\renewcommand{\arraystretch}{1.6}
\begin{center}
%\resizebox{24cm}{!}{
\begin{tabular}{|c|c|}
\hline
\multicolumn{2}{|c|}{\textbf{Common Stabilizers}}\\
\hline
$\Lambda_{1}\Lambda_3$&$\Lambda_{1}\Lambda_2\Lambda_{3}$\\ \hline\hline

$\alpha_1+\alpha_2+\alpha_3$&\\
$\alpha_1+\alpha_2\quad \alpha_2+\alpha_3$&\\
$\alpha_1\quad \alpha_2\quad \alpha_3\quad \alpha_5$&$\alpha_1\quad \alpha_3\quad \alpha_5$\\

$H_{\alpha_1}\quad H_{\alpha_3}\quad H_{\alpha_5}$&$H_{\alpha_1}\quad H_{\alpha_3}\quad H_{\alpha_5}$\\

$-\alpha_1\quad -\alpha_3\quad -\alpha_4\quad -\alpha_5$&$-\alpha_1\quad -\alpha_3\quad -\alpha_5$\\
$-\alpha_3-\alpha_4\quad -\alpha_4-\alpha_5$&\\
$-\alpha_3-\alpha_4+\alpha_5$&\\

\hline
\end{tabular}
%}
\caption{Stabilizers for the weights in the ${\bf 15}$ of $\mathfrak{sl}(6,\mathbb{R})$}\label{tab:15sl6common}
\end{center}
\end{table}

%
%
% \begin{table}[h!]
% \renewcommand{\arraystretch}{1.6}
% \begin{center}
% \resizebox{\textwidth}{!}{
% \begin{tabular}{|c|c|c|}
% \hline
% \multicolumn{3}{|c|}{stabilizer}\\
% \hline
% common&\multicolumn{2}{|c|}{conjunction}\\ \hline
% $\Lambda_1,\Lambda_3$&$\Lambda_1\pm\Lambda_3$&$\Lambda_1-\Lambda_3$\\
% \hline
% &$[\alpha_1+\alpha_2+\alpha_3+\alpha_4+\alpha_5]\mp [-\alpha_2-\alpha_3-\alpha_4]$&\\
% &$[\alpha_1+\alpha_2+\alpha_3+\alpha_4]\mp [-\alpha_2-\alpha_3-\alpha_4-\alpha_5]$&\\
% &$[\alpha_1+\alpha_2+\alpha_3+\alpha_4]\mp [-\alpha_2-\alpha_3-\alpha_4-\alpha_5]$&\\
% &$[\alpha_2+\alpha_3+\alpha_4+\alpha_5]\mp [-\alpha_1-\alpha_2-\alpha_3-\alpha_4-\alpha_5]$&\\
%
% \hline
% \end{tabular}
% }
% \caption{Stabilizer for the weights in the ${\bf 15}$ of $\mathfrak{sl}(6,\mathbb{R})$}
% \end{center}
% \end{table}
%
%
% \begin{flalign}
%  \mathcal{O}(\Lambda_1)=\dfrac{SL(6,\mathbb{R})}{[SL(4,\mathbb{R})\times SL(2,\mathbb{R})]\ltimes \mathbb{R}^{4,2}}
% \end{flalign}

\begin{table}[h!]
\renewcommand{\arraystretch}{1.6}
\begin{center}
%\resizebox{\textwidth}{!}{
\begin{tabular}{|c|c|}
\hline
\multicolumn{2}{|c|}{{\bf Conjunction Stabilizers}}\\
\hline\hline
$\Lambda_{1}+a\Lambda_3$&$\Lambda_{2}$\\ \hline

$E_{\alpha_1+\alpha_2+\alpha_3+\alpha_4+\alpha_5}-aE_{-\alpha_2-\alpha_3-\alpha_4}$&-\\
$E_{\alpha_1+\alpha_2+\alpha_3+\alpha_4}-aE_{-\alpha_2-\alpha_3-\alpha_4-\alpha_5}$&-\\
$E_{\alpha_2+\alpha_3+\alpha_4+\alpha_5}-aE_{-\alpha_1-\alpha_2-\alpha_3-\alpha_4}$&-\\
$E_{\alpha_2+\alpha_3+\alpha_4}-aE_{-\alpha_1-\alpha_2-\alpha_3-\alpha_4-\alpha_5}$&-\\  \hline

$\Lambda_{1}+a\Lambda_3$&$a\Lambda_{3}$\\ \hline
$E_{\alpha_1+\alpha_2+\alpha_3}-E_{-\alpha_2}$&-\\
$E_{\alpha_1+\alpha_2}-E_{-\alpha_2-\alpha_3}$&-\\
$E_{\alpha_2+\alpha_3}-E_{-\alpha_1-\alpha_2}$&-\\
$E_{\alpha_2}-E_{-\alpha_1-\alpha_2-\alpha_3}$&-\\ \hline

$\Lambda_{2}+a\Lambda_3$&$\Lambda_{1}$\\ \hline
$E_{\alpha_3+\alpha_4+\alpha_5}-aE_{-\alpha_4}$&-\\
$E_{\alpha_4+\alpha_5}-aE_{-\alpha_3-\alpha_4}$&-\\
$E_{\alpha_3+\alpha_4}-aE_{-\alpha_4-\alpha_5}$&-\\
$E_{\alpha_4}-aE_{-\alpha_3-\alpha_4-\alpha_5}$&-\\ \hline

\hline
\end{tabular}
%}
\caption{Stabilizers for the weights in the ${\bf 15}$ of $\mathfrak{sl}(6,\mathbb{R})$.  For any pair of weights appearing in the combination $\Lambda_{1}+\Lambda_{2}+a\Lambda_{3}$ we have identified the corresponding conjunction stabilizers in the first column. In the second comulmn the action of each of them has been evaluated on the remaining weight of the bound state.}\label{tab:15sl6conjunction}
\end{center}
\end{table}

\begin{description}

\item[Rank-1 (1-weight orbit)] the rank-one orbit could be identified looking at \autoref{tab:15sl6stabilizers} and corresponds to
\begin{align}
\partial I_{3,mag}=0:&&\frac{SL(6,\mathbb{R})}{\left[ SL(4,\mathbb{R})\times SL\left( 2,\mathbb{R}\right) \right] \ltimes \mathbb{R}^{\left( 4,2\right) }},
\end{align}
where $\mathbb{R}^{\left( 4,2\right) }\simeq \left( \mathbf{4,2}\right) $real bi-fundamental of the split form $SL(4,\mathbb{R})\times SL\left( 2,\mathbb{R}\right) $. Note that $SL(4,\mathbb{R})\times SL\left( 2,\mathbb{R}\right) $ is the $U$-duality group of the corresponding theory in $D=6$, and $\left( \mathbf{4,2}\right) $ ($\left( \mathbf{4}^{\prime }\mathbf{,2}\right) $) is the irrep. relevant to non-dyonic asymptotically flat branes (black holes and black $2$-branes, respectively) in $D=6$.

\item[Rank-2 (2-weights orbit)] to study the rank 2 orbit we select the combinations $\Lambda_1\pm\Lambda_3$. Their common stabilizers, appearing in \autoref{tab:15sl6common} together with the conjunction listed in \autoref{tab:15sl6conjunction}  define the full set of generators annihilating these linear combinations. The orbit reads
\begin{flalign}
 &[SL(2,\mathbb{R})\times SL(2,\mathbb{R})]\ltimes \mathbb{R}^{(4,2)}
\end{flalign}
where the $SO(2,3)$ has simple roots
\begin{flalign}
 &\beta_1=\alpha_1\qquad \beta_{2}=\frac{\alpha_5-\alpha_1}{2}
\end{flalign}
and $SL(2,\mathbb{R})$ is associated with the simple roots $\alpha_3$. The orbit is
\begin{align}
I_{3,mag}=0:&&\frac{SL(6,\mathbb{R})}{\left[ Sp(4,\mathbb{R})\times Sp\left( 2,\mathbb{R}\right) \right] \ltimes \mathbb{R}^{\left( 4,2\right) }},
\end{align}
where $\mathbb{R}^{\left( 4,2\right) }\simeq \left( \mathbf{4,2}\right) $denotes the real bi-fundamental of the split form $Sp(4,\mathbb{R})\times Sp\left( 2,\mathbb{R}\right) \simeq SO(3,2)\times SL(2,\mathbb{R})\simeq SO(3,2)\times SO(2,1)$.

\item[Rank-3 (3-weights orbit)] for the rank-three orbit the $[SL(2,\mathbb{R})]^3$ defined by the common stabilizers, \autoref{tab:15sl6common}, is promoted by the twelve conjunction stabilizers appearing in \autoref{tab:15sl6conjunction} to the algebra $\mathfrak{sp}(6,\mathbb{R})$ with simple roots
\begin{flalign}
 &\beta_1=\frac{\alpha_1-\alpha_3}{2}\quad \beta_2=\frac{\alpha_3-\alpha_5}{2}\quad\beta_3=\alpha_5.
\end{flalign}
The orbit is
\begin{align}
I_{3,mag}\neq 0:&&\frac{SL(6,\mathbb{R})}{Sp(6,\mathbb{R})}.
\end{align}
Note that this is a symmetric manifold. We summarize our results for the five dimensional $\mathbb{H}_{s}$ case in \autoref{tab:15sl6summary}.
\end{description}

\begin{table}[h!]
\renewcommand{\arraystretch}{1.8}
\begin{center}
\resizebox{\textwidth}{!}{
\begin{tabular}{|c|c|c|c|c|c|c|c|}
\hline
\multicolumn{2}{|c|}{}&\multicolumn{2}{|c|}{{$\mathbf{\theta}$}}&&&\\ \cline{3-4}
\multicolumn{2}{|c|}{\multirow{-2}{*}{{\textbf{States}}}}&{\bf +}&{\bf -}&\multirow{-2}{*}{{\textbf{semisimple stabilizer}}}&\multirow{-2}{*}{{\textbf{stabilizer}}}&\multirow{-2}{*}{{\bf rank}}\\ \hline\hline

{1-w}&$\Lambda_{1}$&7&11&$SL(4,\mathbb{R})\times SL\left( 2,\mathbb{R}\right)$&$\left[ SL(4,\mathbb{R})\times SL\left( 2,\mathbb{R}\right) \right] \ltimes \mathbb{R}^{\left( 4,2\right) }$&1\\ \hline

{2-w}&$\Lambda_{1}\pm\Lambda_{2}$&5&8& $Sp(4,\mathbb{R})\times Sp(2,\mathbb{R}) $&$[Sp(4,\mathbb{R})\times Sp( 2,\mathbb{R})]\ltimes \mathbb{R}^{(4,2)})$&2\\
\hline

{3-w}&$\Lambda_{1}+\Lambda_{2}\pm\Lambda_{3}$&9&12&$Sp(6,\mathbb{R})$&$Sp(6,\mathbb{R})$&3\\ \hline

\hline
\end{tabular}
}
\caption{Summary of the orbits in the {\bf 15} of $\mathfrak{sl}(6,\mathbb{R})$}\label{tab:15sl6summary}
\end{center}
\end{table}

\FloatBarrier
\section{\label{D=4}$\mathbf{D=4}$}
\FloatBarrier

In $D=4$ dimensions, the unique asymptotically flat branes are black holes ($0$-branes), which can be dyonic.
\FloatBarrier
\subsection{$\mathbb{C}_{s}$}\label{section:20ofsl6}

The $U$-duality group is $SL(6,\mathbb{R})$ (split form of $SU(6)$), and the
scalar manifold reads
\begin{equation}
\frac{Conf\left( J_{3}^{\mathbb{C}_{s}}\right) }{mcs\left( Conf\left( J_{3}^{\mathbb{C}_{s}}\right) \right) }=\frac{SL(6,\mathbb{R})}{SO(6)},\label{Cs-D=4}
\end{equation}%
where $Conf\left( J_{3}^{\mathbb{C}_{s}}\right) \simeq Aut\left( \mathfrak{F}\left( J_{3}^{\mathbb{C}_{s}}\right) \right) $ is the \textit{conformal} group \cite{GNK} of the cubic Jordan algebra $J_{3}^{\mathbb{C}_{s}}$, or equivalently, the automorphism group of the Freudenthal triple system (FTS) $\mathfrak{F}$ over $J_{3}^{\mathbb{C}_{s}}$ \cite{FTS} ($mcs$ stands for \textit{maximal compact subgroup} throughout). The $0$-brane dyonic irrep. is the rank-3 antisymmetric self-dual (real) $\mathbf{20}$, such that the pair $\left( SL(6,\mathbb{R}),\mathbf{20}\right) $ defines a group \textquotedblleft of type $E_{7}$", characterized by a unique independent quartic invariant polynomial $I_{4}$ \cite{Sato-Kimura, Kac-80, Garibaldi-2,Garibaldi-2}.\\

While the non-linear action of $SL(6,\mathbb{R})$ on the scalar manifold (\autoref{Cs-D=4}) is transitive, the linear action of $SL(6,\mathbb{R})$ on the $\mathbf{20}$ representation space determines the stratification into orbits, classified in terms of invariant constraints on $I_{4}$, or equivalently in terms of the \textit{rank} \cite{rank-J,rank-FTS} of the corresponding representative in the Freudenthal triple system $\mathfrak{F}\left( J_{3}^{\mathbb{C}_{s}}\right) $.\\

In order to define the U-duality orbits now we are going to study the stabilizers of bound states of the weights of the $\mathbf{20}$ of the $U$-duality Lie algebra $\mathfrak{sl}(6,\mathbb{R})$, whose Tits-Satake diagram is the same as the one appearing in \autoref{fig:sl6titsatake1}, along the lines defined in the previous sections. First of all we note that the action of the Cartan involution is, as expected for a split real form,
\begin{align}
&\theta(\alpha_{i})=-\alpha_{i}
\end{align}
for any root of the algebra.\\

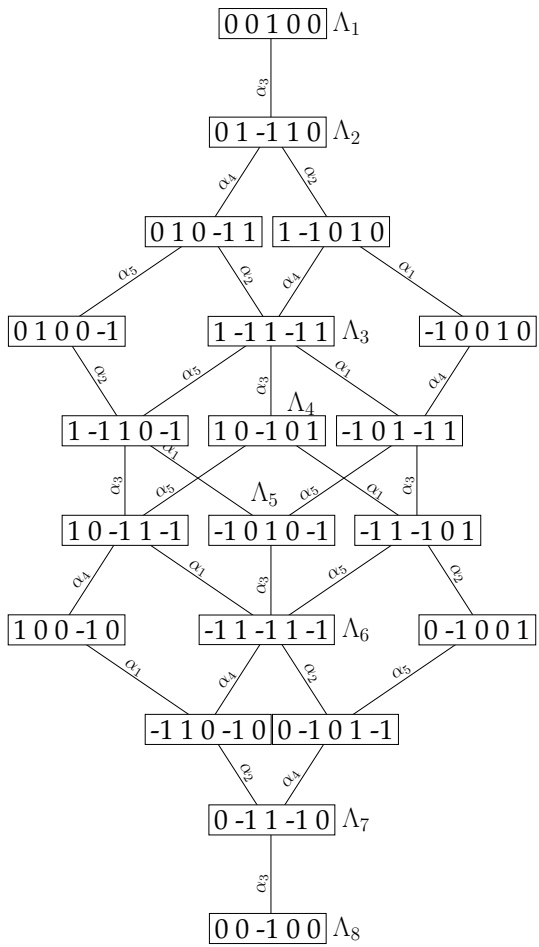
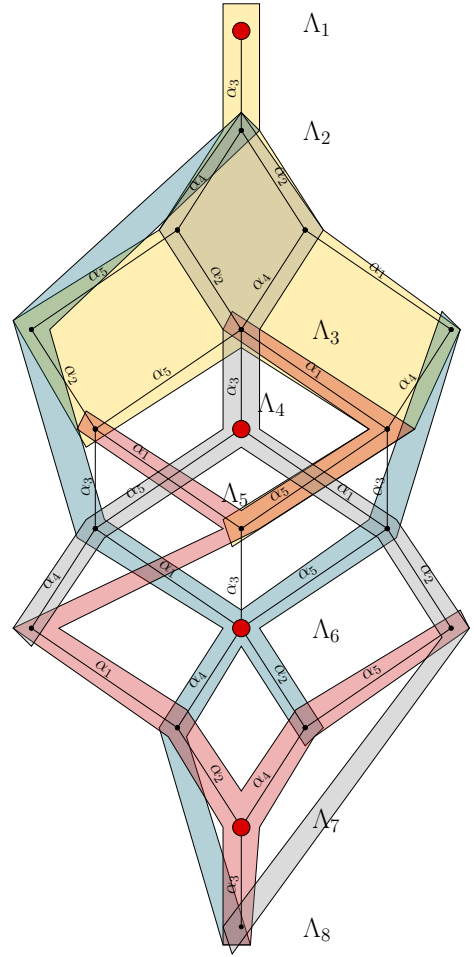
\begin{figure}[h]
\subfloat[][\label{fig:20sl6dynkintree} Dynkin tree of the $\mathbf{20}$ of $\mathfrak{sl}(6,\mathbb{R})$.]{
\centering
% Generated with LaTeXDraw 2.0.8
% Wed Sep 21 10:44:47 CEST 2016
% \usepackage[usenames,dvipsnames]{pstricks}
% \usepackage{epsfig}
% \usepackage{pst-grad} % For gradients
% \usepackage{pst-plot} % For axes
\scalebox{0.6} % Change this value to rescale the drawing.
{\
\begin{pspicture}(0,-10.347396)(15.865625,10.347396)
\usefont{T1}{ppl}{m}{n}
\rput{35.35122}(-0.5468636,-6.9648657){\rput(10.624531,-4.3246875){$\alpha_{5}$}}
\psline[linewidth=0.02cm,fillcolor=black,dotsize=0.07055555cm 2.0]{*-*}(12.36,-3.3746874)(9.16,-5.5746875)
\psline[linewidth=0.02cm,fillcolor=black,dotsize=0.07055555cm 2.0]{*-*}(4.56,-1.1746875)(7.76,-3.3746874)
\usefont{T1}{ppl}{m}{n}
\rput{-32.069756}(2.0574658,2.957849){\rput(6.1445312,-2.0846875){$\alpha_{1}$}}
\usefont{T1}{ppl}{m}{n}
\rput{35.35122}(1.360323,-5.0696654){\rput(8.604531,-0.3846875){$\alpha_{5}$}}
\psline[linewidth=0.02cm,fillcolor=black,dotsize=0.07055555cm 2.0]{*-*}(10.96,1.0253125)(7.76,-1.1746875)
\psline[linewidth=0.02cm,fillcolor=black,dotsize=0.07055555cm 2.0]{*-*}(7.76,1.0253125)(10.96,-1.1746875)
\usefont{T1}{ppl}{m}{n}
\rput{-32.069756}(1.7499796,5.2879753){\rput(10.044531,-0.3846875){$\alpha_{1}$}}
\psline[linewidth=0.02cm,fillcolor=black,dotsize=0.07055555cm 2.0]{*-*}(7.76,-1.1746875)(7.76,-3.3746874)
\usefont{T1}{ppl}{m}{n}
\rput{90.38534}(5.245192,-10.010972){\rput(7.5645313,-2.3846874){$\alpha_{3}$}}
\usefont{T1}{ppl}{m}{n}
\rput{35.35122}(2.4236658,-3.086462){\rput(6.0245314,2.2753124){$\alpha_{5}$}}
\psline[linewidth=0.02cm,fillcolor=black,dotsize=0.07055555cm 2.0]{*-*}(7.76,3.2253125)(4.56,1.0253125)
\psline[linewidth=0.02cm,fillcolor=black,dotsize=0.07055555cm 2.0]{*-*}(9.16,-5.5746875)(7.76,-7.7746873)
\usefont{T1}{ppl}{m}{n}
\rput{54.790813}(-1.9878318,-9.5652685){\rput(8.204532,-6.6846876){$\alpha_{4}$}}
\psline[linewidth=0.02cm,fillcolor=black,dotsize=0.07055555cm 2.0]{*-*}(4.56,1.0253125)(7.76,-1.1746875)
\usefont{T1}{ppl}{m}{n}
\rput{-32.069756}(0.5884857,3.046651){\rput(5.5645313,0.5153125){$\alpha_{1}$}}
\psline[linewidth=0.02cm,fillcolor=black,dotsize=0.07055555cm 2.0]{*-*}(7.76,-3.3746874)(9.16,-5.5746875)
\usefont{T1}{ppl}{m}{n}
\rput{-58.967655}(7.971037,5.2571054){\rput(8.604531,-4.4046874){$\alpha_{2}$}}
\psline[linewidth=0.02cm,fillcolor=black,dotsize=0.07055555cm 2.0]{*-*}(7.76,-7.7746873)(7.76,-9.974688)
\usefont{T1}{ppl}{m}{n}
\rput{90.38534}(-1.354659,-16.65536){\rput(7.5645313,-8.984688){$\alpha_{3}$}}
\usefont{T1}{ppl}{m}{n}
\rput{35.35122}(3.438426,-1.870806){\rput(4.6245313,4.4753127){$\alpha_{5}$}}
\psline[linewidth=0.02cm,fillcolor=black,dotsize=0.07055555cm 2.0]{*-*}(6.36,5.4253125)(3.16,3.2253125)
\psline[linewidth=0.02cm,fillcolor=black,dotsize=0.07055555cm 2.0]{*-*}(7.76,-3.3746874)(6.36,-5.5746875)
\usefont{T1}{ppl}{m}{n}
\rput{54.790813}(-0.7831279,-7.489834){\rput(6.804531,-4.4846873){$\alpha_{4}$}}
\psline[linewidth=0.02cm,fillcolor=black,dotsize=0.07055555cm 2.0]{*-*}(9.16,5.4253125)(12.36,3.2253125)
\usefont{T1}{ppl}{m}{n}
\rput{-32.069756}(-0.74486357,6.4073696){\rput(10.744532,4.5153127){$\alpha_{1}$}}
\psline[linewidth=0.02cm,fillcolor=black,dotsize=0.07055555cm 2.0]{*-*}(6.36,-5.5746875)(7.76,-7.7746873)
\usefont{T1}{ppl}{m}{n}
\rput{-58.967655}(9.177896,2.9916272){\rput(7.204531,-6.6046877){$\alpha_{2}$}}
\psline[linewidth=0.02cm,fillcolor=black,dotsize=0.07055555cm 2.0]{*-*}(10.96,1.0253125)(10.96,-1.1746875)
\usefont{T1}{ppl}{m}{n}
\rput{90.38534}(10.666663,-10.996103){\rput(10.764531,-0.1846875){$\alpha_{3}$}}
\psline[linewidth=0.02cm,fillcolor=black,dotsize=0.07055555cm 2.0]{*-*}(4.56,-1.1746875)(3.16,-3.3746874)
\usefont{T1}{ppl}{m}{n}
\rput{54.790813}(-0.34060982,-3.9437056){\rput(3.6045313,-2.2846875){$\alpha_{4}$}}
\psline[linewidth=0.02cm,fillcolor=black,dotsize=0.07055555cm 2.0]{*-*}(6.36,5.4253125)(7.76,3.2253125)
\usefont{T1}{ppl}{m}{n}
\rput{-58.967655}(-0.24774492,8.320887){\rput(7.204531,4.3953123){$\alpha_{2}$}}
\psline[linewidth=0.02cm,fillcolor=black,dotsize=0.07055555cm 2.0]{*-*}(4.56,1.0253125)(4.56,-1.1746875)
\usefont{T1}{ppl}{m}{n}
\rput{90.38534}(4.223622,-4.5962486){\rput(4.364531,-0.1846875){$\alpha_{3}$}}
\usefont{T1}{ppl}{m}{n}
\rput{35.35122}(0.8123995,-3.2302678){\rput(5.4445314,-0.3246875){$\alpha_{5}$}}
\psline[linewidth=0.02cm,fillcolor=black,dotsize=0.07055555cm 2.0]{*-*}(7.76,1.0253125)(4.56,-1.1746875)
\psline[linewidth=0.02cm,fillcolor=black,dotsize=0.07055555cm 2.0]{*-*}(12.36,3.2253125)(10.96,1.0253125)
\usefont{T1}{ppl}{m}{n}
\rput{54.790813}(6.5572267,-8.453593){\rput(11.4045315,2.1153126){$\alpha_{4}$}}
\psline[linewidth=0.02cm,fillcolor=black,dotsize=0.07055555cm 2.0]{*-*}(3.16,-3.3746874)(6.36,-5.5746875)
\usefont{T1}{ppl}{m}{n}
\rput{-32.069756}(3.0119221,1.8788022){\rput(4.744531,-4.2846875){$\alpha_{1}$}}
\psline[linewidth=0.02cm,fillcolor=black,dotsize=0.07055555cm 2.0]{*-*}(10.96,-1.1746875)(12.36,-3.3746874)
\usefont{T1}{ppl}{m}{n}
\rput{-58.967655}(7.6362386,9.064961){\rput(11.804531,-2.2046876){$\alpha_{2}$}}
\psline[linewidth=0.02cm,fillcolor=black,dotsize=0.07055555cm 2.0]{*-*}(7.76,3.2253125)(7.76,1.0253125)
\usefont{T1}{ppl}{m}{n}
\rput{90.38534}(9.645093,-5.5813804){\rput(7.5645313,2.0153124){$\alpha_{3}$}}
\psline[linewidth=0.02cm,fillcolor=black,dotsize=0.07055555cm 2.0]{*-*}(9.16,5.4253125)(7.76,3.2253125)
\usefont{T1}{ppl}{m}{n}
\rput{54.790813}(6.9997454,-4.907465){\rput(8.204532,4.3153124){$\alpha_{4}$}}
\psline[linewidth=0.02cm,fillcolor=black,dotsize=0.07055555cm 2.0]{*-*}(3.16,3.2253125)(4.56,1.0253125)
\usefont{T1}{ppl}{m}{n}
\rput{-58.967655}(0.08705309,4.51303){\rput(4.0045314,2.1953125){$\alpha_{2}$}}
\psline[linewidth=0.02cm,fillcolor=black,dotsize=0.07055555cm 2.0]{*-*}(7.76,9.825313)(7.76,7.6253123)
\usefont{T1}{ppl}{m}{n}
\rput{90.38534}(16.244944,1.0630065){\rput(7.5645313,8.615313){$\alpha_{3}$}}
\usefont{T1}{ppl}{m}{n}
\rput{35.35122}(0.46789667,-5.7492094){\rput(9.224531,-2.1246874){$\alpha_{5}$}}
\psline[linewidth=0.02cm,fillcolor=black,dotsize=0.07055555cm 2.0]{*-*}(10.96,-1.1746875)(7.76,-3.3746874)
\psline[linewidth=0.02cm,fillcolor=black,dotsize=0.07055555cm 2.0]{*-*}(7.76,7.6253123)(6.36,5.4253125)
\usefont{T1}{ppl}{m}{n}
\rput{54.790813}(8.204449,-2.8320308){\rput(6.804531,6.5153127){$\alpha_{4}$}}
\psline[linewidth=0.02cm,fillcolor=black,dotsize=0.07055555cm 2.0]{*-*}(7.76,3.2253125)(10.96,1.0253125)
\usefont{T1}{ppl}{m}{n}
\rput{-32.069756}(0.20959252,5.328323){\rput(9.344531,2.3153124){$\alpha_{1}$}}
\psline[linewidth=0.02cm,fillcolor=black,dotsize=0.07055555cm 2.0]{*-*}(7.76,7.6253123)(9.16,5.4253125)
\usefont{T1}{ppl}{m}{n}
\rput{-58.967655}(-1.4546037,10.586365){\rput(8.604531,6.5953126){$\alpha_{2}$}}
\usefont{T1}{ppl}{m}{n}
\rput(7.8009377,9.970312){\Large \psframebox[linewidth=0.02,fillstyle=solid]{0 0 1 0 0}}
\usefont{T1}{ppl}{m}{n}
\rput(7.6909375,7.5703125){\Large \psframebox[linewidth=0.02,fillstyle=solid]{0 1 -1 1 0}}
\usefont{T1}{ppl}{m}{n}
\rput(6.284844,5.3703127){\Large \psframebox[linewidth=0.02,fillstyle=solid]{0 1 0 -1 1}}
\usefont{T1}{ppl}{m}{n}
\rput(3.2848437,3.1703124){\Large \psframebox[linewidth=0.02,fillstyle=solid]{0 1 0 0 -1}}
\usefont{T1}{ppl}{m}{n}
\rput(9.080313,5.3703127){\Large \psframebox[linewidth=0.02,fillstyle=solid]{1 -1 0 1 0}}
\usefont{T1}{ppl}{m}{n}
\rput(7.764219,3.1703124){\Large \psframebox[linewidth=0.02,fillstyle=solid]{1 -1 1 -1 1}}
\usefont{T1}{ppl}{m}{n}
\rput(12.295,3.1703124){\Large \psframebox[linewidth=0.02,fillstyle=solid]{-1 0 0 1 0}}
\usefont{T1}{ppl}{m}{n}
\rput(4.5642185,0.9703125){\Large \psframebox[linewidth=0.02,fillstyle=solid]{1 -1 1 0 -1}}
\usefont{T1}{ppl}{m}{n}
\rput(7.6742187,0.9703125){\Large \psframebox[linewidth=0.02,fillstyle=solid]{1 0 -1 0 1}}
\usefont{T1}{ppl}{m}{n}
\rput(10.578906,0.9703125){\Large \psframebox[linewidth=0.02,fillstyle=solid]{-1 0 1 -1 1}}
\usefont{T1}{ppl}{m}{n}
\rput(4.5642185,-1.2296875){\Large \psframebox[linewidth=0.02,fillstyle=solid]{1 0 -1 1 -1}}
\usefont{T1}{ppl}{m}{n}
\rput(7.7789063,-1.2296875){\Large \psframebox[linewidth=0.02,fillstyle=solid]{-1 0 1 0 -1}}
\usefont{T1}{ppl}{m}{n}
\rput(10.978907,-1.2296875){\Large \psframebox[linewidth=0.02,fillstyle=solid]{-1 1 -1 0 1}}
\usefont{T1}{ppl}{m}{n}
\rput(7.6909375,-10.029688){\Large \psframebox[linewidth=0.02,fillstyle=solid]{0 0 -1 0 0}}
\usefont{T1}{ppl}{m}{n}
\rput(7.7809377,-7.6296873){\Large \psframebox[linewidth=0.02,fillstyle=solid]{0 -1 1 -1 0}}
\usefont{T1}{ppl}{m}{n}
\rput(9.174844,-5.6296873){\Large \psframebox[linewidth=0.02,fillstyle=solid]{0 -1 0 1 -1}}
\usefont{T1}{ppl}{m}{n}
\rput(6.385,-5.6296873){\Large \psframebox[linewidth=0.02,fillstyle=solid]{-1 1 0 -1 0}}
\usefont{T1}{ppl}{m}{n}
\rput(3.2803125,-3.4296875){\Large \psframebox[linewidth=0.02,fillstyle=solid]{1 0 0 -1 0}}
\usefont{T1}{ppl}{m}{n}
\rput(12.284843,-3.4296875){\Large \psframebox[linewidth=0.02,fillstyle=solid]{0 -1 0 0 1}}
\usefont{T1}{ppl}{m}{n}
\rput(7.668906,-3.4296875){\Large \psframebox[linewidth=0.02,fillstyle=solid]{-1 1 -1 1 -1}}
\usefont{T1}{ppl}{m}{n}
\rput(9.422812,9.970312){\Large $\Lambda_{1}$}
% \usefont{T1}{ppl}{m}{n}
% \rput(4.8228126,5.3703127){\Large $\Sigma_{1}$}
\usefont{T1}{ppl}{m}{n}
\rput(9.422812,7.5703125){\Large $\Lambda_{2}$}
\usefont{T1}{ppl}{m}{n}
\rput(9.622812,3.1703124){\Large $\Lambda_{3}$}
\usefont{T1}{ppl}{m}{n}
\rput(8.422812,1.5703125){\Large $\Lambda_{4}$}
\usefont{T1}{ppl}{m}{n}
\rput(9.622812,-3.4296875){\Large $\Lambda_{6}$}
\usefont{T1}{ppl}{m}{n}
\rput(9.622812,-7.6296873){\Large $\Lambda_{7}$}
\usefont{T1}{ppl}{m}{n}
\rput(7.6228123,-0.4296875){\Large $\Lambda_{5}$}
\usefont{T1}{ppl}{m}{n}
\rput(9.422812,-10.029688){\Large $\Lambda_{8}$}
\usefont{T1}{ppl}{m}{n}
% \rput(10.622812,5.3703127){\Large $\Sigma_{2}$}
% \usefont{T1}{ppl}{m}{n}
% \rput(1.6228125,3.1703124){\Large $\Sigma_{3}$}
% \usefont{T1}{ppl}{m}{n}
% \rput(2.8228126,0.9703125){\Large $\Sigma_{5}$}
% \usefont{T1}{ppl}{m}{n}
% \rput(13.822812,3.1703124){\Large $\Sigma_{4}$}
% \usefont{T1}{ppl}{m}{n}
% \rput(12.222813,0.9703125){\Large $\Sigma_{6}$}
% \usefont{T1}{ppl}{m}{n}
% \rput(2.8228126,-1.2296875){\Large $\Sigma_{7}$}
% \usefont{T1}{ppl}{m}{n}
% \rput(1.6228125,-3.4296875){\Large $\Sigma_{9}$}
% \usefont{T1}{ppl}{m}{n}
% \rput(12.622812,-1.2296875){\Large $\Sigma_{8}$}
% \usefont{T1}{ppl}{m}{n}
% \rput(13.972813,-3.4296875){\Large $\Sigma_{10}$}
% \usefont{T1}{ppl}{m}{n}
% \rput(4.7728124,-5.6296873){\Large $\Sigma_{11}$}
% \usefont{T1}{ppl}{m}{n}
% \rput(10.972813,-5.6296873){\Large $\Sigma_{12}$}
\end{pspicture}
}
%\caption{Dynkin tree of the $\mathbf{20}$ of $\mathfrak{sl}(6,\mathbb{R})$.}
%\label{20sl6dynkintree}
}
\subfloat[orbits][\label{fig:20sl6dynkintreeorbits}Orbits of the real weights $\Lambda _{1}$, $\Lambda _{4}$, $\Lambda_{6}$ and $\Lambda _{7}$ in the $\mathbf{20}$ of $\mathfrak{sl}(6,\mathbb{R}) $. The red circles denote the starting points of the corresponding orbit]{
\centering
% Generated with LaTeXDraw 2.0.8
% Sun May 07 19:05:53 CEST 2017
% \usepackage[usenames,dvipsnames]{pstricks}
% \usepackage{epsfig}
% \usepackage{pst-grad} % For gradients
% \usepackage{pst-plot} % For axes
\scalebox{0.6} % Change this value to rescale the drawing.
{
\begin{pspicture}(0,-10.52)(10.02,10.52)
\definecolor{color613b}{rgb}{1.0,0.792156862745098,0.0}
\definecolor{color415b}{rgb}{0.00392156862745098,0.403921568627451,0.4823529411764706}
\definecolor{color406b}{rgb}{0.8,0.00392156862745098,0.00392156862745098}
\definecolor{color79b}{rgb}{0.5333333333333333,0.5333333333333333,0.5333333333333333}
\definecolor{color716b}{rgb}{0.8509803921568627,0.0,0.0}
\pspolygon[linewidth=0.02,fillstyle=solid,fillcolor=color613b,opacity=0.3](4.6,10.5)(5.4,10.5)(5.4,7.7)(6.8,5.5)(9.8,3.3)(8.8,1.1)(4.8,-1.5)(4.6,-0.9)(7.8,1.1)(5.0,2.9)(1.6,0.7)(0.0,3.5)(3.2,5.5)(4.6,7.7)
\pspolygon[linewidth=0.02,fillstyle=solid,fillcolor=color415b,opacity=0.3](5.0,8.1)(5.4,7.7)(0.8,3.3)(2.0,-0.9)(5.0,-2.9)(7.8,-0.9)(9.4,3.7)(9.8,3.3)(8.4,-1.3)(5.4,-3.3)(6.8,-5.5)(6.4,-5.9)(5.0,-3.7)(3.8,-5.7)(5.2,-10.3)(4.6,-10.3)(3.2,-5.5)(4.6,-3.3)(1.4,-1.3)(0.0,3.5)
\pspolygon[linewidth=0.02,fillstyle=solid,fillcolor=color406b,opacity=0.3](4.6,-10.3)(5.2,-10.3)(5.4,-7.7)(6.6,-5.7)(10.0,-3.3)(9.8,-2.9)(6.2,-5.1)(5.0,-6.9)(3.8,-5.1)(1.0,-3.3)(5.4,-1.1)(8.8,1.1)(4.8,3.7)(4.6,3.3)(7.8,1.1)(5.0,-0.7)(1.6,1.5)(1.4,1.1)(4.6,-1.1)(0.0,-3.3)(3.4,-5.7)(4.6,-7.7)
\pspolygon[linewidth=0.02,fillstyle=solid,fillcolor=color79b,opacity=0.3](5.0,8.1)(6.8,5.5)(5.4,3.3)(5.4,1.1)(8.4,-0.9)(10.0,-3.3)(4.8,-10.5)(4.6,-9.9)(9.4,-3.5)(8.0,-1.3)(5.0,0.7)(2.0,-1.3)(0.4,-3.7)(0.0,-3.3)(1.6,-0.9)(4.6,1.1)(4.6,3.3)(3.2,5.5)
\usefont{T1}{ppl}{m}{n}
\rput{35.35122}(-1.0125374,-5.3541946){\rput(7.864531,-4.25){$\alpha_{5}$}}
\psline[linewidth=0.02cm,fillcolor=black,dotsize=0.07055555cm 2.0]{*-*}(9.6,-3.3)(6.4,-5.5)
\psline[linewidth=0.02cm,fillcolor=black,dotsize=0.07055555cm 2.0]{*-*}(1.8,-1.1)(5.0,-3.3)
\usefont{T1}{ppl}{m}{n}
\rput{-32.069756}(1.5966407,1.5038204){\rput(3.3845313,-2.01){$\alpha_{1}$}}
\usefont{T1}{ppl}{m}{n}
\rput{35.35122}(0.89464927,-3.4589946){\rput(5.844531,-0.31){$\alpha_{5}$}}
\psline[linewidth=0.02cm,fillcolor=black,dotsize=0.07055555cm 2.0]{*-*}(8.2,1.1)(5.0,-1.1)
\psline[linewidth=0.02cm,fillcolor=black,dotsize=0.07055555cm 2.0]{*-*}(5.0,1.1)(8.2,-1.1)
\usefont{T1}{ppl}{m}{n}
\rput{-32.069756}(1.2891545,3.8339467){\rput(7.284531,-0.31){$\alpha_{1}$}}
\psline[linewidth=0.02cm,fillcolor=black,dotsize=0.07055555cm 2.0]{*-*}(5.0,-1.1)(5.0,-3.3)
\usefont{T1}{ppl}{m}{n}
\rput{90.38534}(2.5413163,-7.175845){\rput(4.804531,-2.31){$\alpha_{3}$}}
\usefont{T1}{ppl}{m}{n}
\rput{35.35122}(1.957992,-1.4757913){\rput(3.2645311,2.35){$\alpha_{5}$}}
\psline[linewidth=0.02cm,fillcolor=black,dotsize=0.07055555cm 2.0]{*-*}(5.0,3.3)(1.8,1.1)
\psline[linewidth=0.02cm,fillcolor=black,dotsize=0.07055555cm 2.0]{*-*}(6.4,-5.5)(5.0,-7.7)
\usefont{T1}{ppl}{m}{n}
\rput{54.790813}(-3.0954936,-7.278578){\rput(5.4445314,-6.61){$\alpha_{4}$}}
\psline[linewidth=0.02cm,fillcolor=black,dotsize=0.07055555cm 2.0]{*-*}(1.8,1.1)(5.0,-1.1)
\usefont{T1}{ppl}{m}{n}
\rput{-32.069756}(0.1276607,1.5926226){\rput(2.8045313,0.59){$\alpha_{1}$}}
\psline[linewidth=0.02cm,fillcolor=black,dotsize=0.07055555cm 2.0]{*-*}(5.0,-3.3)(6.4,-5.5)
\usefont{T1}{ppl}{m}{n}
\rput{-58.967655}(6.5698795,2.9283113){\rput(5.844531,-4.33){$\alpha_{2}$}}
\psline[linewidth=0.02cm,fillcolor=black,dotsize=0.07055555cm 2.0]{*-*}(5.0,-7.7)(5.0,-9.9)
\usefont{T1}{ppl}{m}{n}
\rput{90.38534}(-4.058534,-13.820232){\rput(4.804531,-8.91){$\alpha_{3}$}}
\usefont{T1}{ppl}{m}{n}
\rput{35.35122}(2.9727523,-0.26013497){\rput(1.8645313,4.55){$\alpha_{5}$}}
\psline[linewidth=0.02cm,fillcolor=black,dotsize=0.07055555cm 2.0]{*-*}(3.6,5.5)(0.4,3.3)
\psline[linewidth=0.02cm,fillcolor=black,dotsize=0.07055555cm 2.0]{*-*}(5.0,-3.3)(3.6,-5.5)
\usefont{T1}{ppl}{m}{n}
\rput{54.790813}(-1.8907896,-5.2031436){\rput(4.0445313,-4.41){$\alpha_{4}$}}
\psline[linewidth=0.02cm,fillcolor=black,dotsize=0.07055555cm 2.0]{*-*}(6.4,5.5)(9.6,3.3)
\usefont{T1}{ppl}{m}{n}
\rput{-32.069756}(-1.2056886,4.953341){\rput(7.9845314,4.59){$\alpha_{1}$}}
\psline[linewidth=0.02cm,fillcolor=black,dotsize=0.07055555cm 2.0]{*-*}(3.6,-5.5)(5.0,-7.7)
\usefont{T1}{ppl}{m}{n}
\rput{-58.967655}(7.776738,0.66283244){\rput(4.4445314,-6.53){$\alpha_{2}$}}
\psline[linewidth=0.02cm,fillcolor=black,dotsize=0.07055555cm 2.0]{*-*}(8.2,1.1)(8.2,-1.1)
\usefont{T1}{ppl}{m}{n}
\rput{90.38534}(7.962788,-8.160977){\rput(8.004531,-0.11){$\alpha_{3}$}}
\psline[linewidth=0.02cm,fillcolor=black,dotsize=0.07055555cm 2.0]{*-*}(1.8,-1.1)(0.4,-3.3)
\usefont{T1}{ppl}{m}{n}
\rput{54.790813}(-1.4482714,-1.6570153){\rput(0.84453124,-2.21){$\alpha_{4}$}}
\psline[linewidth=0.02cm,fillcolor=black,dotsize=0.07055555cm 2.0]{*-*}(3.6,5.5)(5.0,3.3)
\usefont{T1}{ppl}{m}{n}
\rput{-58.967655}(-1.6489027,5.992092){\rput(4.4445314,4.47){$\alpha_{2}$}}
\psline[linewidth=0.02cm,fillcolor=black,dotsize=0.07055555cm 2.0]{*-*}(1.8,1.1)(1.8,-1.1)
\usefont{T1}{ppl}{m}{n}
\rput{90.38534}(1.5197456,-1.7611215){\rput(1.6045313,-0.11){$\alpha_{3}$}}
\usefont{T1}{ppl}{m}{n}
\rput{35.35122}(0.34672582,-1.619597){\rput(2.6845312,-0.25){$\alpha_{5}$}}
\psline[linewidth=0.02cm,fillcolor=black,dotsize=0.07055555cm 2.0]{*-*}(5.0,1.1)(1.8,-1.1)
\psline[linewidth=0.02cm,fillcolor=black,dotsize=0.07055555cm 2.0]{*-*}(9.6,3.3)(8.2,1.1)
\usefont{T1}{ppl}{m}{n}
\rput{54.790813}(5.449565,-6.166903){\rput(8.644531,2.19){$\alpha_{4}$}}
\psline[linewidth=0.02cm,fillcolor=black,dotsize=0.07055555cm 2.0]{*-*}(0.4,-3.3)(3.6,-5.5)
\usefont{T1}{ppl}{m}{n}
\rput{-32.069756}(2.5510972,0.4247735){\rput(1.9845313,-4.21){$\alpha_{1}$}}
\psline[linewidth=0.02cm,fillcolor=black,dotsize=0.07055555cm 2.0]{*-*}(8.2,-1.1)(9.6,-3.3)
\usefont{T1}{ppl}{m}{n}
\rput{-58.967655}(6.2350817,6.736168){\rput(9.044531,-2.13){$\alpha_{2}$}}
\psline[linewidth=0.02cm,fillcolor=black,dotsize=0.07055555cm 2.0]{*-*}(5.0,3.3)(5.0,1.1)
\usefont{T1}{ppl}{m}{n}
\rput{90.38534}(6.941217,-2.7462535){\rput(4.804531,2.09){$\alpha_{3}$}}
\psline[linewidth=0.02cm,fillcolor=black,dotsize=0.07055555cm 2.0]{*-*}(6.4,5.5)(5.0,3.3)
\usefont{T1}{ppl}{m}{n}
\rput{54.790813}(5.8920836,-2.6207745){\rput(5.4445314,4.39){$\alpha_{4}$}}
\psline[linewidth=0.02cm,fillcolor=black,dotsize=0.07055555cm 2.0]{*-*}(0.4,3.3)(1.8,1.1)
\usefont{T1}{ppl}{m}{n}
\rput{-58.967655}(-1.3141044,2.1842356){\rput(1.2445313,2.27){$\alpha_{2}$}}
\psline[linewidth=0.02cm,fillcolor=black,dotsize=0.07055555cm 2.0]{*-*}(5.0,9.9)(5.0,7.7)
\usefont{T1}{ppl}{m}{n}
\rput{90.38534}(13.541068,3.898134){\rput(4.804531,8.69){$\alpha_{3}$}}
\usefont{T1}{ppl}{m}{n}
\rput{35.35122}(0.0022230004,-4.1385384){\rput(6.4645314,-2.05){$\alpha_{5}$}}
\psline[linewidth=0.02cm,fillcolor=black,dotsize=0.07055555cm 2.0]{*-*}(8.2,-1.1)(5.0,-3.3)
\psline[linewidth=0.02cm,fillcolor=black,dotsize=0.07055555cm 2.0]{*-*}(5.0,7.7)(3.6,5.5)
\usefont{T1}{ppl}{m}{n}
\rput{54.790813}(7.0967875,-0.5453404){\rput(4.0445313,6.59){$\alpha_{4}$}}
\psline[linewidth=0.02cm,fillcolor=black,dotsize=0.07055555cm 2.0]{*-*}(5.0,3.3)(8.2,1.1)
\usefont{T1}{ppl}{m}{n}
\rput{-32.069756}(-0.25123248,3.8742943){\rput(6.5845313,2.39){$\alpha_{1}$}}
\psline[linewidth=0.02cm,fillcolor=black,dotsize=0.07055555cm 2.0]{*-*}(5.0,7.7)(6.4,5.5)
\usefont{T1}{ppl}{m}{n}
\rput{-58.967655}(-2.855761,8.257571){\rput(5.844531,6.67){$\alpha_{2}$}}
\usefont{T1}{ppl}{m}{n}
\rput(6.6628127,10.045){\Large $\Lambda_{1}$}
\usefont{T1}{ppl}{m}{n}
\rput(6.6628127,7.645){\Large $\Lambda_{2}$}
\usefont{T1}{ppl}{m}{n}
\rput(6.8628125,3.245){\Large $\Lambda_{3}$}
\usefont{T1}{ppl}{m}{n}
\rput(5.6628127,1.645){\Large $\Lambda_{4}$}
\usefont{T1}{ppl}{m}{n}
\rput(6.8628125,-3.355){\Large $\Lambda_{6}$}
\usefont{T1}{ppl}{m}{n}
\rput(6.8628125,-7.555){\Large $\Lambda_{7}$}
\usefont{T1}{ppl}{m}{n}
\rput(4.8628125,-0.355){\Large $\Lambda_{5}$}
\usefont{T1}{ppl}{m}{n}
\rput(6.6628127,-9.955){\Large $\Lambda_{8}$}
\pscircle[linewidth=0.02,dimen=outer,fillstyle=solid,fillcolor=color716b](5.0,-7.7){0.2}
\pscircle[linewidth=0.02,dimen=outer,fillstyle=solid,fillcolor=color716b](5.0,-3.3){0.2}
\pscircle[linewidth=0.02,dimen=outer,fillstyle=solid,fillcolor=color716b](5.0,1.1){0.2}
\pscircle[linewidth=0.02,dimen=outer,fillstyle=solid,fillcolor=color716b](5.0,9.9){0.2}
\end{pspicture}
}

}
\caption{In the figure are sketched the structure of representation {\bf 20} of SL(6,$\mathbb{R}$) and the orbits of four different weights.}\label{fig:20sl6dynkintreeandorbits}
\end{figure}

The Dynkin tree of the irrep $\mathbf{20}$ of $\mathfrak{sl}
(6,\mathbb{R})$ is depicted in \autoref{fig:20sl6dynkintreeandorbits}, while in \autoref{fig:20sl6dynkintree} we sketchily represent the orbits of the four weights $\Lambda_{1}$, $\Lambda_{4}$, $\Lambda_{6}$, $\Lambda_{7}$  and their overlaps.

\subsubsection{1-weight}
To determine the 1-weight stabilizer, one simply can look at the Dynkin tree in \autoref{fig:20sl6dynkintree}, and the results listed in \autoref{stabs20sl6} immediately follow.
\begin{sidewaystable}[h]
\renewcommand{\arraystretch}{1.8}
\begin{center}
\resizebox{\textwidth}{!}{
\begin{tabular}{|c|c|c|c|}
\hline
\multicolumn{4}{|c|}{\textbf{Stabilizer}}\\
\hline
\textcolor{black}{$\Lambda_{1}$}&\textcolor{black}{$\Lambda_{4}$}&\textcolor{black}{$\Lambda_{6}$}&$\Lambda_{7}$\\
\hline\hline
$\alpha_{1}+\alpha_{2}+\alpha_{3}+\alpha_{4}+\alpha_{5}$&$\alpha_{1}+\alpha_{2}+\alpha_{3}+\alpha_{4}+\alpha_{5}$&&\\

$\alpha_{1}+\alpha_{2}+\alpha_{3}+\alpha_{4}\quad \alpha_{2}+\alpha_{3}+\alpha_{4}+\alpha_{5}$&$\alpha_{1}+\alpha_{2}+\alpha_{3}+\alpha_{4}\quad \alpha_{2}+\alpha_{3}+\alpha_{4}+\alpha_{5}$&$\alpha_{1}+\alpha_{2}+\alpha_{3}+\alpha_{4}\quad \alpha_{2}+\alpha_{3}+\alpha_{4}+\alpha_{5}$&\\

$\alpha_{1}+\alpha_{2}+\alpha_{3}\quad \alpha_{2}+\alpha_{3}+\alpha_{4}\quad \alpha_{3}+\alpha_{4}+\alpha_{5}$&$\alpha_{1}+\alpha_{2}+\alpha_{3}\quad \alpha_{3}+\alpha_{4}+\alpha_{5}$&$\alpha_{2}+\alpha_{3}+\alpha_{4}$&$\alpha_{1}+\alpha_{2}+\alpha_{3}\quad \alpha_{3}+\alpha_{4}+\alpha_{5}$\\

$\alpha_{1}+\alpha_{2}\quad \alpha_{2}+\alpha_{3}\quad \alpha_{3}+\alpha_{4}\quad \alpha_{4}+\alpha_{5}$
&$\alpha_{1}+\alpha_{2}\quad \alpha_{4}+\alpha_{5}$&
$\alpha_{1}+\alpha_{2}\quad \alpha_{2}+\alpha_{3}\quad \alpha_{3}+\alpha_{4}\quad \alpha_{4}+\alpha_{5}$&$\alpha_{2}+\alpha_{3}\quad \alpha_{3}+\alpha_{4}$\\

$\alpha_{1}\quad \alpha_{2}\quad \alpha_{3}\quad \alpha_{4}\quad \alpha_{5}$&
$\alpha_{1}\quad \alpha_{2}\quad \alpha_{4}\quad \alpha_{5}$&
$\alpha_{2}\quad \alpha_{4}$&$\alpha_{1}\quad \alpha_{3}\quad \alpha_{5}$\\

$H_{\alpha_{1}}\quad H_{\alpha_{2}}\quad H_{\alpha_{4}}\quad H_{\alpha_{5}}$&$H_{\alpha_{2}}\quad H_{\alpha_{4}}\quad H_{\alpha_{1}}+H_{\alpha_{3}}\quad H_{\alpha_{3}}+H_{\alpha_{5}}$&$H_{\alpha_{1}}+H_{\alpha_{2}}\quad H_{\alpha_{2}}+H_{\alpha_{3}}\quad H_{\alpha_{3}}+H_{\alpha_{4}}\quad H_{\alpha_{4}}+H_{\alpha_{5}}$&$H_{\alpha_{1}}\quad H_{\alpha_{5}}\quad H_{\alpha_{2}}+H_{\alpha_{3}}\quad H_{\alpha_{3}}+H_{\alpha_{4}}$\\

$-\alpha_{1}\quad -\alpha_{2}\quad -\alpha_{4}\quad -\alpha_{5}$&
$-\alpha_{2}\quad -\alpha_{3}\quad -\alpha_{4}$&
$-\alpha_{1}\quad -\alpha_{3}\quad -\alpha_{5}$&$-\alpha_{1}\quad -\alpha_{2}\quad -\alpha_{4}\quad -\alpha_{5}$\\

$-\alpha_{1}-\alpha_{2}\quad -\alpha_{4}-\alpha_{5}$&$-\alpha_{2}-\alpha_{3}\quad -\alpha_{3}-\alpha_{4}$
&$-\alpha_{1}-\alpha_{2}\quad -\alpha_{2}-\alpha_{3}\quad -\alpha_{3}-\alpha_{4}\quad -\alpha_{4}-\alpha_{5}$
&$-\alpha_{1}-\alpha_{2}\quad -\alpha_{2}-\alpha_{3}\quad -\alpha_{3}-\alpha_{4}\quad -\alpha_{4}-\alpha_{5}$\\

&$-\alpha_{1}-\alpha_{2}-\alpha_{3}\quad -\alpha_{2}-\alpha_{3}-\alpha_{4}\quad -\alpha_{3}-\alpha_{4}-\alpha_{5}$&
$-\alpha_{1}-\alpha_{2}-\alpha_{3}\quad -\alpha_{3}-\alpha_{4}-\alpha_{5}$&$-\alpha_{1}-\alpha_{2}-\alpha_{3}\quad -\alpha_{2}-\alpha_{3}-\alpha_{4}\quad -\alpha_{3}-\alpha_{4}-\alpha_{5}$\\

&$-\alpha_{1}-\alpha_{2}-\alpha_{3}-\alpha_{4}\quad \alpha_{2}-\alpha_{3}-\alpha_{4}-\alpha_{5}$&$-\alpha_{1}-\alpha_{2}-\alpha_{3}-\alpha_{4}\quad \alpha_{2}-\alpha_{3}-\alpha_{4}-\alpha_{5}$&$-\alpha_{1}-\alpha_{2}-\alpha_{3}-\alpha_{4}\quad \alpha_{2}-\alpha_{3}-\alpha_{4}-\alpha_{5}$\\

&&$-\alpha_{1}-\alpha_{2}-\alpha_{3}-\alpha_{4}-\alpha_{5}$&$-\alpha_{1}-\alpha_{2}-\alpha_{3}-\alpha_{4}-\alpha_{5}$\\
\hline
\end{tabular}
}
\end{center}
\caption{Stabilizers for the weights in the $\mathbf{20}$ of $\mathfrak{sl}(6,\mathbb{R})$.}
\label{stabs20sl6}
\end{sidewaystable}

Consequently, the orbit of a real weight is
\begin{align}
\left. \partial ^{2}I_{4}\right\vert _{\mathbf{35}}=0\text{ }:&&\frac{SL(6,\mathbb{R})}{\left[ SL(3,\mathbb{R})\times SL(3,\mathbb{R})\right] \ltimes\mathbb{R}^{\left( 3,3^{\prime }\right) }},  \label{p-sp-1-II}
\end{align}
where $\mathbb{R}^{\left( 3,3^{\prime }\right) }\simeq \left( \mathbf{3},\mathbf{3}^{\prime }\right) $ of $SL(3,\mathbb{R})\times SL(3,\mathbb{R})$, which is the $U$-duality group in $D=5$; furthermore, $\left( \mathbf{3},\mathbf{3}^{\prime }\right) $ ($\left( \mathbf{3}^{\prime },\mathbf{3}\right) $) is the irrep. relevant to asymptotically flat branes (black holes and black strings, respectively) in $D=5$. We also reported the corresponding $SL(6,\mathbb{R})$-invariant constraint on the quartic invariant $I_{4}$ of the $\mathbf{20}$ (with $\mathbf{35}$ denoting the adjoint of $SL(6,\mathbb{R})$ itself); note that the orbit (\autoref{p-sp-1-II}) is the orbit of rank-1 elements of the FTS $\mathfrak{F}$ over $J_{3}^{\mathbb{C}_{s}}$.

\subsubsection{2-weights}
We consider the 2-weights bound states $\Lambda _{1}\pm \Lambda_{4}$. The corresponding stabilizers are listed in  \autoref{stabL1L4}. The conjunctions can be visualized in  \autoref{fig:20sl6dynkintreeorbits}, in which the orbits of the real weights $\Lambda _{1}$, $\Lambda _{4}$, $\Lambda _{6}$ and $\Lambda _{7}$ are drawn according to the notation of \cite{Marrani-Riccioni-Romano}.

\begin{table}[h]
\renewcommand{\arraystretch}{1}
\par
\begin{center}
\begin{tabular}{|c|c|c|}
\hline
\textcolor{black}{\textbf{Common}} & \multicolumn{2}{|c|}{\textcolor{black}{\textbf{Conjunction}}} \\ \hline
\textcolor{black}{$\Lambda_{1},\Lambda_{4}$} & \textcolor{black}{$\Lambda_{1}+\Lambda_{4}$} & \textcolor{black}{$\Lambda_{1}-\Lambda_{4}$} \\
\hline\hline
$E_{\alpha_{1}+\alpha_{2}+\alpha_{3}+\alpha_{4}+\alpha_{5}}$ & $E_{\alpha_{2}+\alpha_{3}+\alpha_{4}}-E_{-\alpha_{3}}$ & $E_{\alpha_{2}+\alpha_{3}+\alpha_{4}}+E_{-\alpha_{3}}$ \\
$E_{\alpha_{1}+\alpha_{2}+\alpha_{3}+\alpha_{4}}\quad E_{\alpha_{2}+\alpha_{3}+\alpha_{4}+\alpha_{5}}$ & $E_{\alpha_{3}}-E_{-
\alpha_{2}-\alpha_{3}-\alpha_{4}}$ & $E_{\alpha_{3}}+E_{-\alpha_{2}-\alpha_{3}-\alpha_{4}}$ \\
$E_{\alpha_{1}+\alpha_{2}+\alpha_{3}}\quad E_{\alpha_{3}+\alpha_{4}+\alpha_{5}}$ & $E_{\alpha_{2}+\alpha_{3}}-E_{-\alpha_{3}-\alpha_{4}}$ & $E_{\alpha_{2}+\alpha_{3}}+E_{-\alpha_{3}-\alpha_{4}}$ \\
$E_{\alpha_{1}+\alpha_{2}}\quad E_{\alpha_{4}+\alpha_{5}}$ & $E_{\alpha_{3}+\alpha_{4}}-E_{-\alpha_{2}-\alpha_{3}}$ & $E_{\alpha_{3}+\alpha_{4}}+E_{-\alpha_{2}-\alpha_{3}}$ \\
$E_{\alpha_{1}}\quad E_{\alpha_{2}}\quad E_{\alpha_{4}}\quad E_{\alpha_{5}}$
&  &  \\
$H_{\alpha_{2}}\quad H_{\alpha_{4}}\quad H_{\alpha_{1}}-H_{\alpha_{5}}$ &  &
\\
$E_{-\alpha_{2}}\quad E_{-\alpha_{4}}$ &  &  \\ \hline
\end{tabular}%
\end{center}
\caption[Stabilizers of $\Lambda _{1}+\Lambda _{4}$]{Stabilizers of $\Lambda
_{1}\pm \Lambda _{4}$.}
\label{stabL1L4}
\end{table}

%The semisimple part of the stabilizer of $\Lambda _{1}\pm \Lambda _{4}$ can
%be recognized to be generated by the following generators :
%\begin{align}
%\begin{array}{l}
%E_{\alpha _{2}},~E_{\alpha _{4}}, \\
%H_{\alpha _{2}},~H_{\alpha _{4}},~H_{\alpha 1}-H_{\alpha _{5}}, \\
%E_{-\alpha _{2}},~E_{-\alpha _{4}}, \\
%E_{\alpha _{2}+\alpha _{3}+\alpha _{4}}-E_{-\alpha _{3}}&\rightarrow E_{\frac{1}{2}\left( \alpha _{2}+\alpha _{4}\right) }, \\
%\mp E_{-\left( \alpha _{2}+\alpha _{3}+\alpha _{4}\right) }+E_{\alpha
%_{3}}&\rightarrow E_{-\frac{1}{2}\left( \alpha _{2}+\alpha _{4}\right) }, \\
%E_{\alpha _{2}+\alpha _{3}}\mp E_{-\left( \alpha _{3}+\alpha _{4}\right)
%}&\rightarrow E_{\frac{1}{2}\left( \alpha _{2}-\alpha _{4}\right) }, \\
%\mp E_{-\left( \alpha _{2}+\alpha _{3}\right) }+E_{\alpha _{3}+\alpha
%_{4}}&\rightarrow E_{-\frac{1}{2}\left( \alpha _{2}-\alpha _{4}\right) },
%\end{array}
%\label{gens-1}
%\end{align}
%where the arrows denote the association to root vectors.
The generators in \autoref{stabL1L4} give rise to an algebra $\mathfrak{so}(2,3)\oplus \mathfrak{so}(1,1)$ in both the \textquotedblleft $\pm $" branches, with the $\mathfrak{so}(1,1)$ generated by the Cartan $H_{\alpha _{1}}-H_{\alpha _{5}}$. The simple roots of such a stabilizing algebra are $\alpha _{4}$ and$\ \frac{\alpha _{2}-\alpha _{4}}{2}$. Correspondingly, the resulting 2-weights orbit
reads
\begin{align}
\partial I_{4}=0\text{ }:&&\frac{SL(6,\mathbb{R})}{\left[ Sp(4,\mathbb{R})\times SO(1,1)\right] \ltimes \left( \mathbb{R}^{\left( 4,2\right) }\times \mathbb{R}\right) },  \label{p-sp-2-II}
\end{align}%
where $\mathbb{R}^{\left( 4,2\right) }\simeq \left( \mathbf{4,2}\right)$ denotes the real bi-fundamental\footnote{The real fundamental irrep. of $Sp(4,\mathbb{R})$ is the real spinor of $SO(3,2)$.} of the split form $Sp(4,\mathbb{R})\times SO(1,1)\simeq SO(3,2)\times SO(1,1)$. We also reported the corresponding $SL(6,\mathbb{R})$-invariant constraint on the quartic invariant $I_{4}$ of the $\mathbf{20}$; note that the orbit (\autoref{p-sp-2-II}) is the orbit of rank-2 elements \cite{FTS} of the FTS $\mathfrak{F}$ over $J_{3}^{\mathbb{C}_{s}}$.\medskip

There is another non-isomorphic 2-weights orbit in the $\mathbf{20}$ of $SL(6,\mathbb{R})$, named \textit{dyonic orbit}. Without loss of generality, it can be realized as the orbit of the bound state $\Lambda _{1}+\Lambda_{8} $; in this case, there are only common stabilizers, and these are
listed in \autoref{dyonicstab}.
\begin{table}[h]
\renewcommand{\arraystretch}{1}
\par
\begin{center}
\begin{tabular}{|c|}
\hline
\textcolor{black}{\bf Common} \\ \hline
\textcolor{black}{$\Lambda_1,\Lambda_{8}$} \\ \hline\hline
$E_{\alpha_{1}+\alpha_{2}}\quad E_{\alpha_{4}+\alpha_{5}}$ \\
$E_{\alpha_{1}}\quad E_{\alpha_{2}}\quad E_{\alpha_{4}}\quad E_{\alpha_{5}}$
\\
$H_{\alpha_{1}}\quad H_{\alpha_{2}}\quad H_{\alpha_{4}}\quad H_{\alpha_{5}}$
\\
$E_{-\alpha_{1}}\quad E_{-\alpha_{2}}\quad E_{-\alpha_{4}}\quad
E_{-\alpha_{5}}$ \\
$E_{-\alpha_{1}-\alpha_{2}}\quad E_{-\alpha_{4}-\alpha_{5}}$ \\ \hline
\end{tabular}%
\end{center}
\caption[Stabilizers of $\Lambda _{1}+\Lambda _{8}$]{Stabilizers of $\Lambda
_{1}+\Lambda _{8}$.}
\label{dyonicstab}
\end{table}
\newline
The resulting stabilizing algebra is $\mathfrak{sl}(3,\mathbb{R})\oplus\mathfrak{sl}(3,\mathbb{R})$, which - as mentioned before - is nothing but the $U$-duality Lie algebra of the corresponding theory uplifted to $D=5$. The dyonic orbit reads
\begin{align}
I_{4}<0\text{ }:&&\frac{SL(6,\mathbb{R})}{SL(3,\mathbb{R})\times SL(3,\mathbb{R})},  \label{p-sp-3-II}
\end{align}
where we also reported the corresponding $SL(6,\mathbb{R})$-invariant constraint on the quartic invariant $I_{4}$ of the $\mathbf{20}$; consequently, the orbit (\autoref{p-sp-3-II}) is the orbit of rank-4 elements with $I_{4}<0$ of the FTS $\mathfrak{F}$ over $J_{3}^{\mathbb{C}_{s}}$. We note that the dyonic orbit is realized as a bound states of two weights but it corresponds to rank-4 elements of the FTS $\mathfrak{F}$ over $J_{3}^{\mathbb{C}_{s}}$. The characterization of these weights $\Lambda_{1}$ and $\Lambda_{8}$ with respect to other possible choices is that their single orbits do not overlap. Furthermore fixed one weight in the representation then there is a unique choice of a second weight with this property.\\

No other choice of 2-weights bound states yields other, non-isomorphic orbits.

\subsubsection{3-weights}
With no loss of generality, we consider the 3-weights bound states $\Lambda_{1}+\Lambda _{4}\pm \Lambda _{6}$.
\begin{table}[h]
\renewcommand{\arraystretch}{1}
\par
\begin{center}
\resizebox{\textwidth}{!}{
\begin{tabular}{|c|c|c|}
\hline
\textcolor{black}{\bf Common}&\multicolumn{2}{|c|}{\textcolor{black}{\bf Conjunction}}\\
\hline
\textcolor{black}{$\Lambda_1,\Lambda_4, \Lambda_6$}&\textcolor{black}{$\Lambda_1+\Lambda_4+ \Lambda_6$}&\textcolor{black}{$\Lambda_1+\Lambda_4-\Lambda_6$}\\
\hline\hline
$E_{\alpha_{1}+\alpha_{2}+\alpha_{3}+\alpha_{4}}$&$E_{\alpha_{2}+\alpha_{3}+\alpha_{4}}-E_{-\alpha_{3}}$&$E_{\alpha_{2}+\alpha_{3}+\alpha_{4}}-E_{-\alpha_{3}}$\\
$E_{\alpha_{2}+\alpha_{3}+\alpha_{4}+\alpha_{5}}$&$E_{\alpha_{3}}-E_{-\alpha_{2}-\alpha_{3}-\alpha_{4}}$&$E_{\alpha_{3}}-E_{-\alpha_{2}-\alpha_{3}-\alpha_{4}}$\\
$E_{\alpha_{1}+\alpha_{2}}\quad E_{\alpha_{4}+\alpha_{5}}$&$E_{\alpha_{3}+\alpha_{4}}-E_{-\alpha_{2}-\alpha_{3}}$&$E_{\alpha_{3}+\alpha_{4}}-E_{-\alpha_{2}-\alpha_{3}}$\\
$E_{\alpha_{2}}\quad E_{\alpha_{4}}$&$E_{\alpha_{1}+\alpha_{2}+\alpha_{3}+\alpha_{4}+\alpha_{5}}-E_{-\alpha_{3}}$&$E_{\alpha_{1}+\alpha_{2}+\alpha_{3}+\alpha_{4}+\alpha_{5}}+E_{-\alpha_{3}}$\\
$H_{\alpha_{2}}-H_{\alpha_{4}}\quad H_{\alpha_{1}}-H_{\alpha_{5}}$&$E_{\alpha_{1}+\alpha_{2}+\alpha_{3}}-E_{-\alpha_{3}-\alpha_{4}-\alpha_{5}}$&$E_{\alpha_{1}+\alpha_{2}+\alpha_{3}}+E_{-\alpha_{3}-\alpha_{4}-\alpha_{5}}$\\
&$E_{\alpha_{3}+\alpha_{4}+\alpha_{5}}-E_{-\alpha_{1}-\alpha_{2}-\alpha_{3}}$&$E_{\alpha_{3}+\alpha_{4}+\alpha_{5}}+E_{-\alpha_{1}-\alpha_{2}-\alpha_{3}}$\\
&$E_{\alpha_{2}+\alpha_{3}+\alpha_{4}}-E_{\alpha_{1}+\alpha_{2}+\alpha_{3}+\alpha_{4}+\alpha_{5}}$&$E_{\alpha_{2}+\alpha_{3}+\alpha_{4}}+E_{\alpha_{1}+\alpha_{2}+\alpha_{3}+\alpha_{4}+\alpha_{5}}$\\
&$E_{\alpha_{5}}-E_{-\alpha_{1}}$&$E_{\alpha_{5}}+E_{-\alpha_{1}}$\\
&$E_{\alpha_{1}}-E_{-\alpha_{5}}$&$E_{\alpha_{1}}+E_{-\alpha_{5}}$\\
\hline
\end{tabular}
}
\end{center}
\caption[Stabilizers of $\Lambda _{1}+\Lambda _{4}\pm \Lambda _{6}$]{Stabilizers of $\Lambda _{1}+\Lambda _{4}\pm \Lambda _{6}$.}
\label{L1L4L6stab}
\end{table}
For such bound states, we identify the the stabilizers reported in \autoref{L1L4L6stab}, and the corresponding orbit reads%
\begin{align}
I_{4}=0\text{ }:&&\frac{SL(6,\mathbb{R})}{SL(3,\mathbb{R})\ltimes \mathbb{R}^{8}},  \label{p-sp-4-II}
\end{align}%
where $\mathbb{R}^{8}\simeq \mathbf{8}$ denotes the adjoint of $SL(3,\mathbb{R})$. We also reported the corresponding $SL(6,\mathbb{R})$-invariant constraint on the quartic invariant $I_{4}$ of the $\mathbf{20}$; note that the orbit (\autoref{p-sp-4-II}) is the orbit of rank-3 elements of the FTS $\mathfrak{F}$ over $J_{3}^{\mathbb{C}_{s}}$. In \autoref{appendix:20sl6} we describe explicitly how the stabilizers generates the semisimple part of the full stabilizer.

\subsubsection{4-weights}
Finally, the analysis of the orbit stratification in the $\mathbf{20}$ is completed by considering the orbits of 4-weights bound states. Once chosen $\Lambda _{1}$, $\Lambda _{4}$ and $\Lambda _{6}$, these orbits can be built adding to them $\Lambda _{7}$, which is the only weight disconnected from such weights. The independent combinations one can construct are $\Lambda_{1}+\Lambda _{4}\pm \Lambda _{6}\pm \Lambda _{7}$. Their stabilizers are listed in \autoref{L1L4L6L7stab},in which the various combinations are parametrized by the constants $a=\pm 1$ and $b=\pm 1$. There are two common
stabilizers and fourteen conjunctions.
\begin{table}[h]
\renewcommand{\arraystretch}{1.4}
\par
\begin{center}
%\resizebox{\textwidth}{!}{
\begin{tabular}{|c|c|}
\hline
\textcolor{black}{\bf Common} & \multicolumn{1}{|c|}{\textcolor{black}{\bf Conjunction}} \\ \hline
\textcolor{black}{$\Lambda_1,\Lambda_4, \Lambda_6,\Lambda_7$} & \textcolor{black}{$\Lambda_1+\Lambda_4+a \Lambda_6+b\Lambda_7$} \\ \hline\hline
& $E_{\alpha_{2}+\alpha_{3}}-E_{-\alpha_{3}-\alpha_{4}}$ \\
& $E_{\alpha_{3}+\alpha_{4}}-E_{-\alpha_{2}-\alpha_{3}}$ \\
& $E_{\alpha_{1}+\alpha_{2}+\alpha_{3}}-aE_{-\alpha_{3}-\alpha_{4}-\alpha_{5}}$ \\
& $E_{\alpha_{3}+\alpha_{4}+\alpha_{5}}-aE_{-\alpha_{1}-\alpha_{2}-\alpha_{3}}$ \\
& $E_{\alpha_{5}}-aE_{-\alpha_{1}}$ \\
& $E_{\alpha_{1}}-aE_{-\alpha_{5}}$ \\
& $E_{\alpha_{1}+\alpha_{2}+\alpha_{3}+\alpha_{4}}-bE_{-\alpha_{2}-\alpha_{3}-\alpha_{4}-\alpha_{5}}$ \\
& $E_{\alpha_{2}+\alpha_{3}+\alpha_{4}+\alpha_{5}}-bE_{-\alpha_{1}-\alpha_{2}-\alpha_{3}-\alpha_{4}}$ \\
& $E_{\alpha_{1}+\alpha_{2}}-bE_{-\alpha_{4}-\alpha_{5}}$ \\
& $E_{\alpha_{4}+\alpha_{5}}-bE_{-\alpha_{1}-\alpha_{2}}$ \\
& $E_{\alpha_{2}}-abE_{-\alpha_{4}}$ \\
& $E_{\alpha_{4}}-abE_{-\alpha_{2}}$ \\
& $F_{\alpha_{3}}^{-ab}+abF_{\alpha_{2}+\alpha_{3}+\alpha_{4}}^{-ab}$ \\
\multirow{-14}{*}{ \begin{tabular}{c} $H_{\alpha_{2}}-H_{\alpha_{4}}$\\
$H_{\alpha_{1}}-H_{\alpha_{5}}$\\ \end{tabular} } & $F_{\alpha_{3}}^{-ab}+bF_{\alpha_{1}+\alpha_{2}+\alpha_{3}+\alpha_{4}+\alpha_{5}}^{-ab}$ \\ \hline
\end{tabular}
%}
\end{center}
\caption[Stabilizers of $\Lambda _{1}+\Lambda _{4}+a\Lambda _{6}+b\Lambda_{7}$]{Stabilizers of $\Lambda _{1}+\Lambda _{4}+a\Lambda _{6}+b\Lambda _{7}$.}
\label{L1L4L6L7stab}
\end{table}
\newline
Changing the values of $a$ and $b$, one obtains two different real forms for the stabilizers, namely\footnote{The subscript \textquotedblleft $\mathbb{R}$" denotes the Lie algebra to be considered as an algebra over the reals.}
$\mathfrak{sl}(3,\mathbb{C})_{\mathbb{R}}$ and $\mathfrak{sl}(3,\mathbb{R})\oplus \mathfrak{sl}(3,\mathbb{R})$
\begin{table}[h]
\renewcommand{\arraystretch}{1.4}
\begin{center}
%\resizebox{}{!}{
\begin{tabular}{cc}
\scalebox{0.3} {
\begin{pspicture}(0,-2.7)(16.309063,2.7)
\definecolor{color114b}{rgb}{0.996078431372549,0.996078431372549,0.996078431372549}
\psline[linewidth=0.02cm](9.597343,1.5340624)(13.797344,1.5340624)
\psline[linewidth=0.02cm](1.7973439,1.5340624)(5.7973437,1.5340624)
\pscircle[linewidth=0.02,dimen=outer,fillstyle=solid](1.7973439,1.5340624){0.4}
\pscircle[linewidth=0.02,dimen=outer,fillstyle=solid,fillcolor=color114b](5.7973437,1.5340624){0.4}
\pscircle[linewidth=0.02,dimen=outer,fillstyle=solid](9.797344,1.5340624){0.4}
\pscircle[linewidth=0.02,dimen=outer,fillstyle=solid](13.797344,1.5340624){0.4}
\usefont{T1}{ppl}{m}{n}
\rput(2.0764062,2.3440626){\huge $\beta_{1}$}
\psbezier[linewidth=0.06,arrowsize=0.05291667cm 2.0,arrowlength=1.2,arrowinset=0.0]{<->}(1.7973437,0.7340624)(1.7973437,-2.6659374)(13.797344,-2.6659374)(13.797344,0.73406243)
\psbezier[linewidth=0.06,arrowsize=0.05291667cm 2.0,arrowlength=1.2,arrowinset=0.0]{<->}(5.7973437,0.93406236)(5.7973437,-1.0659376)(9.797344,-1.0659375)(9.797344,0.9340624)
\usefont{T1}{ppl}{m}{n}
\rput(5.876406,2.3440626){\huge $\beta_{2}$}
\usefont{T1}{ppl}{m}{n}
\rput(9.876407,2.3440626){\huge $\beta_{3}$}
\usefont{T1}{ppl}{m}{n}
\rput(14.0764065,2.3440626){\huge $\beta_{4}$}
\end{pspicture}
}
&
\scalebox{0.3} {
\begin{pspicture}(0,-3.5)(16.309063,1.9)
\definecolor{color3b}{rgb}{0.996078431372549,0.996078431372549,0.996078431372549}
\psline[linewidth=0.02cm](9.597343,0.7340625)(13.797344,0.7340625)
\psline[linewidth=0.02cm](1.7973439,0.7340625)(5.7973437,0.7340625)
\pscircle[linewidth=0.02,dimen=outer,fillstyle=solid](1.7973439,0.7340625){0.4}
\pscircle[linewidth=0.02,dimen=outer,fillstyle=solid,fillcolor=color3b](5.7973437,0.7340625){0.4}
\pscircle[linewidth=0.02,dimen=outer,fillstyle=solid](9.797344,0.7340625){0.4}
\pscircle[linewidth=0.02,dimen=outer,fillstyle=solid](13.797344,0.7340625){0.4}
\usefont{T1}{ppl}{m}{n}
\rput(2.0764062,1.5440626){\huge $\beta_{1}$}
\usefont{T1}{ppl}{m}{n}
\rput(5.876406,1.5440626){\huge $\beta_{2}$}
\usefont{T1}{ppl}{m}{n}
\rput(9.876407,1.5440626){\huge $\beta_{3}$}
\usefont{T1}{ppl}{m}{n}
\rput(14.0764065,1.5440626){\huge $\beta_{4}$}
\end{pspicture}
}\\ [0.3cm]
\textcolor{gray}{$a=\pm 1\qquad b=\pm 1$}&\textcolor{gray}{$a=\pm 1\qquad b=\mp 1$}\\
$(8|8)$&$(8|8)$\\ [0.1cm]
$\mathfrak{sl}(3,\mathbb{C})_{\mathbb{R}}$&$\mathfrak{sl}(3,\mathbb{R})\oplus \mathfrak{sl}(3,\mathbb{R})$,\\
\end{tabular}
%}
\label{}
\end{center}
\end{table}
\newline
where signatures, stabilizing algebras and their Tits-Satake diagrams are shown. For the explicit realization of the stabilizing algebra we refer to \autoref{appendix:20sl6}. Summarizing, the 4-weights bound states orbits are given by
\begin{align}
I_{4}>0:&&\frac{SL(6,\mathbb{R})}{SL(3,\mathbb{C})_{\mathbb{R}}}
\label{p-sp-5-II}
\end{align}
and by another orbit isomorphic to the dyonic orbit (\autoref{p-sp-3-II}). Consequently, the orbits (\autoref{p-sp-3-II}) and (\autoref{p-sp-5-II}) are the orbits of rank-4 elements of the FTS $\mathfrak{F}$ over $J_{3}^{\mathbb{C}_{s}}$.\bigskip

We summarize the stratification of the $\mathbf{20}$ of the split form $\mathfrak{sl}(6,\mathbb{R})$ of the Lie algebra $\mathfrak{a}_{5}$ in  \autoref{20sl6summary}.

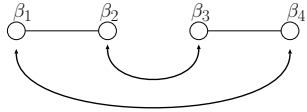
\begin{table}[h]
\renewcommand{\arraystretch}{1.8}
\par
\begin{center}
\resizebox{\textwidth}{!}{
\begin{tabular}{|c|c|c|c|c|c|c|}
\hline
\multicolumn{2}{|c|}{}&\multicolumn{2}{|c|}{\textcolor{black}{$\mathbf{\theta}$}}&&&\\ \cline{3-4}
\multicolumn{2}{|c|}{\multirow{-2}{*}{\textcolor{black}{\bf States}}}&\textcolor{black}{+}&\textcolor{black}{-}&\multirow{-2}{*}{\textcolor{black}{\bf Semisimple Stabilizer}}&\multirow{-2}{*}{\textcolor{black}{\bf Stabilizer}}&\multirow{-2}{*}{\textcolor{black}{\bf Rank}}\\ \hline\hline
\multirow{-1}{*}{1-w}&$\Lambda_{1}$&6&10&$SL(3,\mathbb{R})\times SL(3,\mathbb{R})$&$[SL(3,\mathbb{R})\times SL(3,\mathbb{R})]\ltimes \mathbb{R}^{(3,3')}$&1\\ \hline

&$\Lambda_{1}\pm\Lambda_{4}$&4&7&$SO(2,3)\times SO(1,1)$&$[SO(2,3)\times SO(1,1)]\ltimes (\mathbb{R}\times\mathbb{R}^{(4,2)})$&2\\
\multirow{-2}{*}{2-w}&$\Lambda_{1}+\Lambda_{8}$&6&10&$SL(3,\mathbb{R})\times SL(3,\mathbb{R})$&$SL(3,\mathbb{R})\times SL(3,\mathbb{R})$&dyonic\\
\hline

\multirow{-1}{*}{3-w}&$\Lambda_{1}+\Lambda_{4}\pm\Lambda_{6}$&3&5&$SL(3,\mathbb{R})$&$SL(3,\mathbb{R})\ltimes \mathbb{R}^{8}$&3\\ \hline

&$\Lambda_{1}+\Lambda_{4}\pm\Lambda_{6}\pm\Lambda_{7}$&8&8&$SL(3,\mathbb{C})$&$SL(3,\mathbb{C})$&4\\
\multirow{-2}{*}{4-w}&$\Lambda_{1}+\Lambda_{4}\pm\Lambda_{6}\mp\Lambda_{7}$&6&10&$SL(3,\mathbb{R})\times SL(3,\mathbb{R})$&$SL(3,\mathbb{R})\times SL(3,\mathbb{R})$&4\\

\hline
\end{tabular}
}
\end{center}
\caption{Orbit stabilizers in the $\mathbf{20}$ of $\mathfrak{sl}(6,\mathbb{R})$. }
\label{20sl6summary}
\end{table}

\FloatBarrier
\subsection{$\mathbb{H}_{s}$}\label{section:32ofso(6,6)}

The $U$-duality group is $SO(6,6)$ (split form of $SO(12)$), and the scalar manifold reads
\begin{equation}
\frac{Conf\left( J_{3}^{\mathbb{H}_{s}}\right) }{mcs\left( Conf\left( J_{3}^{\mathbb{H}_{s}}\right) \right) }=\frac{SO(6,6)}{SO(6)\times SO(6)},\label{sm-Hs-D=4}
\end{equation}%
Note that the coset \autoref{sm-Hs-D=4} also characterizes $6$ self-dual and $6 $ anti-self-dual $3$-form field strengths in $D=6$ (see \textit{e.g.} \cite{F-D=6}, and Refs. therein).\\

The $0$-brane dyonic irrep. is the Majorana-Weyl spinor $\mathbf{32}$ (or its conjugate $\mathbf{32}^{\prime }$), such that the pair $\left( SO(6,6),\mathbf{32}\right) $ defines a group \textquotedblleft of type $E_{7}$" \cite{Groups-type-E7}, characterized by a unique independent quartic invariant polynomial\footnote{It should be remarked that the orbit classification done on this Section can
be regarded as the real split form, worked out with completely different
methods, of the classification made by Igusa in \cite{Igusa}, with whom ours
agrees.} $I_{4}$ \cite{Sato-Kimura, Kac-80, Garibaldi-1, Garibaldi-2}.\\

While the non-linear action of $SO(6,6)$ on the scalar manifold (\autoref{sm-Hs-D=4}) is transitive, the linear action of $SO(6,6)$ on the $\mathbf{32}$ determines the stratification into five orbits, classified in terms of invariant constraints on $I_{4}$, or equivalently in terms of the \textit{rank} \cite{rank-J,rank-FTS} of the corresponding representative in the Freudenthal triple system $\mathfrak{F}\left( J_{3}^{\mathbb{H}_{s}}\right) $.
\begin{figure}[h!]
\centering
% \usepackage[usenames,dvipsnames]{pstricks}
% \usepackage{epsfig}
% \usepackage{pst-grad} % For gradients
% \usepackage{pst-plot} % For axes
% \usepackage[space]{grffile} % For spaces in paths
% \usepackage{etoolbox} % For spaces in paths
% \makeatletter % For spaces in paths
% \patchcmd\Gread@eps{\@inputcheck#1 }{\@inputcheck"#1"\relax}{}{}
% \makeatother
% % User Packages:
%
%

\psscalebox{0.4} % Change this value to rescale the drawing.
{
\begin{pspicture}(0,-2.6723437)(16.8,2.6723437)
\psline[linecolor=black, linewidth=0.02](0.39,-2.2716563)(16.426874,-2.2376564)
\pscircle[linecolor=black, linewidth=0.02, fillstyle=solid, dimen=outer](0.4,-2.2723436){0.4}
\pscircle[linecolor=black, linewidth=0.02, fillstyle=solid, dimen=outer](4.4,-2.2723439){0.4}
\rput(11.828906,-1.3376563){\huge $\alpha_4$}
\pscircle[linecolor=black, linewidth=0.02, fillstyle=solid, dimen=outer](8.4,-2.2723436){0.4}
\pscircle[linecolor=black, linewidth=0.02, fillstyle=solid, dimen=outer](12.4,-2.2723439){0.4}
\rput(8.028906,-1.3376563){\huge $\alpha_3$}
\rput(4.128906,-1.3376563){\huge $\alpha_2$}
\rput(0.3289062,-1.3376563){\huge $\alpha_1$}
\rput(16.128906,-1.3376563){\huge $\alpha_5$}
\pscircle[linecolor=black, linewidth=0.02, fillstyle=solid, dimen=outer](16.4,-2.2723439){0.4}
\psline[linecolor=black, linewidth=0.02](12.414873,1.7376014)(12.421088,-2.1994145)
\rput{-90.43946}(10.7817745,14.154607){\pscircle[linecolor=black, linewidth=0.02, fillstyle=solid, dimen=outer](12.414115,1.7276064){0.4}}
\rput{-90.43946}(14.753111,10.0957985){\pscircle[linecolor=black, linewidth=0.02, fillstyle=solid, dimen=outer](12.385885,-2.2722938){0.4}}
\rput(11.728907,2.4623437){\huge $\alpha_6$}
\end{pspicture}
}
\caption{Tits-Satake diagram for $\mathfrak{so}(6,6)$.}\label{fig:titssatakediagramso(6,6)}
\end{figure}
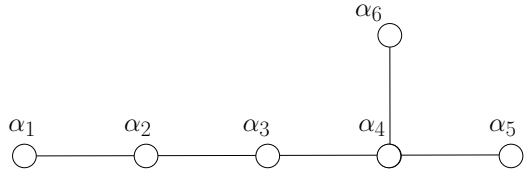
We consider the representation $\bf{32}$ of SO(6,6). In \autoref{fig:titssatakediagramso(6,6)} we report the Tits-Satake diagram for $\mathfrak{so}(6,6)$, fixing the conventions for the simple roots. In this section we only sketch the the Dynkin tree and the orbits with their overlaps in \autoref{fig:32ofso(6,6)dynkinlabels} and \autoref{fig:32ofso(6,6)orbits} respectively, referring to \autoref{appendix:32ofso(6,6)} for the detailed analysis of the orbits.\\

The weights in \autoref{fig:32ofso(6,6)dynkinlabels} with no explicit labels are denoted with $\Sigma_{i}$ with $i$ that goes from 1 to 28 increasing from the top to the bottom from the left to the right. The weights $\Lambda_{i}$ with $i=1,2,3,4$ are the ones we choose to study the orbits. These are not connected by the action of any generator.\\
\FloatBarrier
\begin{figure}[h!]
\centering
% Generated with LaTeXDraw 2.0.8
% Fri May 05 12:09:00 CEST 2017
% \usepackage[usenames,dvipsnames]{pstricks}
% \usepackage{epsfig}
% \usepackage{pst-grad} % For gradients
% \usepackage{pst-plot} % For axes
\scalebox{0.5} % Change this value to rescale the drawing.
{
\begin{pspicture}(0,-16.997396)(24.77521,16.997396)
\usefont{T1}{ppl}{m}{n}
\rput(21.722813,16.620312){\Large $\Lambda_{1}$}
\psline[linewidth=0.02cm,fillcolor=black,dotsize=0.07055555cm 2.0]{*-*}(20.060001,-1.2026042)(14.06,-3.402604)
\usefont{T1}{ppl}{m}{n}
\rput{19.282917}(0.3177689,-5.864427){\rput(17.389208,-1.9811672){$\alpha_{6}$}}
\psline[linewidth=0.02cm,fillcolor=black,dotsize=0.07055555cm 2.0]{*-*}(13.96,0.9973958)(7.9600005,-1.2026042)
\usefont{T1}{ppl}{m}{n}
\rput{19.282917}(0.7020681,-3.7265847){\rput(11.289208,0.21883276){$\alpha_{6}$}}
\psline[linewidth=0.02cm,fillcolor=black,dotsize=0.07055555cm 2.0]{*-*}(12.56,3.1973958)(6.5600004,0.9973958)
\usefont{T1}{ppl}{m}{n}
\rput{19.282917}(1.35004,-3.1408374){\rput(9.889208,2.4188328){$\alpha_{6}$}}
\usefont{T1}{ppl}{m}{n}
\rput{-58.967655}(15.912867,6.239614){\rput(13.444307,-10.936626){$\alpha_{3}$}}
\psline[linewidth=0.02cm,fillcolor=black,dotsize=0.07055555cm 2.0]{*-*}(12.660001,-10.0026045)(14.06,-12.202604)
\psline[linewidth=0.02cm,fillcolor=black,dotsize=0.07055555cm 2.0]{*-*}(9.360001,5.3973956)(3.3600006,3.1973958)
\usefont{T1}{ppl}{m}{n}
\rput{19.282917}(1.8970307,-1.9606705){\rput(6.689208,4.6188326){$\alpha_{6}$}}
\usefont{T1}{ppl}{m}{n}
\rput{-58.967655}(15.578068,10.04747){\rput(16.644308,-8.736627){$\alpha_{3}$}}
\psline[linewidth=0.02cm,fillcolor=black,dotsize=0.07055555cm 2.0]{*-*}(15.860001,-7.802604)(17.26,-10.0026045)
\psline[linewidth=0.02cm,fillcolor=black,dotsize=0.07055555cm 2.0]{*-*}(23.26,-3.402604)(17.26,-5.6026044)
\usefont{T1}{ppl}{m}{n}
\rput{19.282917}(-0.229222,-7.0445933){\rput(20.589209,-4.181167){$\alpha_{6}$}}
\usefont{T1}{ppl}{m}{n}
\rput{-33.691383}(6.7753377,4.742595){\rput(11.189208,-8.8011675){$\alpha_{2}$}}
\psline[linewidth=0.02cm,fillcolor=black,dotsize=0.07055555cm 2.0]{*-*}(9.46,-7.802604)(12.660001,-10.0026045)
\psline[linewidth=0.02cm,fillcolor=black,dotsize=0.07055555cm 2.0]{*-*}(18.66,0.9973958)(12.660001,-1.2026042)
\usefont{T1}{ppl}{m}{n}
\rput{19.282917}(0.96574074,-5.27868){\rput(15.989208,0.21883276){$\alpha_{6}$}}
\psline[linewidth=0.02cm,fillcolor=black,dotsize=0.07055555cm 2.0]{*-*}(12.56,3.1973958)(18.66,0.9973958)
\usefont{T1}{ppl}{m}{n}
\rput{-19.798964}(0.23459768,5.550359){\rput(15.989208,2.1188328){$\alpha_{1}$}}
\usefont{T1}{ppl}{m}{n}
\rput{35.35122}(-3.5012374,-11.034499){\rput(15.532651,-10.994954){$\alpha_{5}$}}
\psline[linewidth=0.02cm,fillcolor=black,dotsize=0.07055555cm 2.0]{*-*}(17.26,-10.0026045)(14.06,-12.202604)
\usefont{T1}{ppl}{m}{n}
\rput{54.790813}(-5.1612425,-16.404325){\rput(13.216059,-13.165881){$\alpha_{4}$}}
\psline[linewidth=0.02cm,fillcolor=black,dotsize=0.07055555cm 2.0]{*-*}(14.06,-12.202604)(12.660001,-14.402604)
\usefont{T1}{ppl}{m}{n}
\rput{-33.691383}(6.092435,6.8872147){\rput(14.389208,-6.601167){$\alpha_{2}$}}
\psline[linewidth=0.02cm,fillcolor=black,dotsize=0.07055555cm 2.0]{*-*}(12.660001,-5.6026044)(15.860001,-7.802604)
\usefont{T1}{ppl}{m}{n}
\rput{-58.967655}(8.372354,10.503021){\rput(13.444307,-2.1366262){$\alpha_{3}$}}
\psline[linewidth=0.02cm,fillcolor=black,dotsize=0.07055555cm 2.0]{*-*}(12.660001,-1.2026042)(14.06,-3.402604)
\psline[linewidth=0.02cm,fillcolor=black,dotsize=0.07055555cm 2.0]{*-*}(6.5600004,0.9973958)(12.660001,-1.2026042)
\usefont{T1}{ppl}{m}{n}
\rput{-19.798964}(0.62510496,3.3879843){\rput(9.989208,-0.081167236){$\alpha_{1}$}}
\usefont{T1}{ppl}{m}{n}
\rput{35.35122}(-2.4864771,-9.818843){\rput(14.132651,-8.794954){$\alpha_{5}$}}
\psline[linewidth=0.02cm,fillcolor=black,dotsize=0.07055555cm 2.0]{*-*}(15.860001,-7.802604)(12.660001,-10.0026045)
\usefont{T1}{ppl}{m}{n}
\rput{54.790813}(1.5863008,-16.224209){\rput(16.41606,-6.5658813){$\alpha_{4}$}}
\psline[linewidth=0.02cm,fillcolor=black,dotsize=0.07055555cm 2.0]{*-*}(17.26,-5.6026044)(15.860001,-7.802604)
\usefont{T1}{ppl}{m}{n}
\rput{-33.691383}(4.8945923,11.731173){\rput(21.789207,-2.2011673){$\alpha_{2}$}}
\psline[linewidth=0.02cm,fillcolor=black,dotsize=0.07055555cm 2.0]{*-*}(20.060001,-1.2026042)(23.26,-3.402604)
\psline[linewidth=0.02cm,fillcolor=black,dotsize=0.07055555cm 2.0]{*-*}(7.9600005,-1.2026042)(14.06,-3.402604)
\usefont{T1}{ppl}{m}{n}
\rput{-19.798964}(1.4530493,3.7321448){\rput(11.389208,-2.2811673){$\alpha_{1}$}}
\usefont{T1}{ppl}{m}{n}
\rput{35.35122}(6.6634555,-7.7315645){\rput(15.432651,6.605046){$\alpha_{5}$}}
\psline[linewidth=0.02cm,fillcolor=black,dotsize=0.07055555cm 2.0]{*-*}(17.16,7.597396)(13.96,5.3973956)
\usefont{T1}{ppl}{m}{n}
\rput{54.790813}(2.0288193,-12.6780815){\rput(13.216059,-4.365881){$\alpha_{4}$}}
\psline[linewidth=0.02cm,fillcolor=black,dotsize=0.07055555cm 2.0]{*-*}(14.06,-3.402604)(12.660001,-5.6026044)
\usefont{T1}{ppl}{m}{n}
\rput{-33.691383}(-0.5637537,6.9042277){\rput(11.089209,4.398833){$\alpha_{2}$}}
\psline[linewidth=0.02cm,fillcolor=black,dotsize=0.07055555cm 2.0]{*-*}(9.360001,5.3973956)(12.56,3.1973958)
\usefont{T1}{ppl}{m}{n}
\rput{-58.967655}(4.5536504,12.549038){\rput(13.344307,2.2633739){$\alpha_{3}$}}
\psline[linewidth=0.02cm,fillcolor=black,dotsize=0.07055555cm 2.0]{*-*}(12.56,3.1973958)(13.96,0.9973958)
\psline[linewidth=0.02cm,fillcolor=black,dotsize=0.07055555cm 2.0]{*-*}(17.16,7.597396)(23.26,5.3973956)
\usefont{T1}{ppl}{m}{n}
\rput{-19.798964}(-0.98385406,7.368572){\rput(20.589209,6.5188327){$\alpha_{1}$}}
\usefont{T1}{ppl}{m}{n}
\rput{35.35122}(-1.8035994,-7.5617304){\rput(10.932651,-6.5949545){$\alpha_{5}$}}
\psline[linewidth=0.02cm,fillcolor=black,dotsize=0.07055555cm 2.0]{*-*}(12.660001,-5.6026044)(9.46,-7.802604)
\usefont{T1}{ppl}{m}{n}
\rput{54.790813}(9.1765375,-8.870133){\rput(13.116059,4.4341187){$\alpha_{4}$}}
\psline[linewidth=0.02cm,fillcolor=black,dotsize=0.07055555cm 2.0]{*-*}(13.96,5.3973956)(12.56,3.1973958)
\usefont{T1}{ppl}{m}{n}
\rput{-33.691383}(5.1071997,8.0333395){\rput(15.789208,-4.4011674){$\alpha_{2}$}}
\psline[linewidth=0.02cm,fillcolor=black,dotsize=0.07055555cm 2.0]{*-*}(14.06,-3.402604)(17.26,-5.6026044)
\usefont{T1}{ppl}{m}{n}
\rput{-58.967655}(3.5319097,6.341928){\rput(7.3443074,0.06337385){$\alpha_{3}$}}
\psline[linewidth=0.02cm,fillcolor=black,dotsize=0.07055555cm 2.0]{*-*}(6.5600004,0.9973958)(7.9600005,-1.2026042)
\psline[linewidth=0.02cm,fillcolor=black,dotsize=0.07055555cm 2.0]{*-*}(13.96,5.3973956)(20.060001,3.1973958)
\usefont{T1}{ppl}{m}{n}
\rput{-19.798964}(-0.42782998,6.154617){\rput(17.389208,4.318833){$\alpha_{1}$}}
\usefont{T1}{ppl}{m}{n}
\rput{35.35122}(6.5152783,-11.666579){\rput(21.53265,4.4050455){$\alpha_{5}$}}
\psline[linewidth=0.02cm,fillcolor=black,dotsize=0.07055555cm 2.0]{*-*}(23.26,5.3973956)(20.060001,3.1973958)
\usefont{T1}{ppl}{m}{n}
\rput{54.790813}(9.619055,-5.324005){\rput(9.9160595,6.634119){$\alpha_{4}$}}
\psline[linewidth=0.02cm,fillcolor=black,dotsize=0.07055555cm 2.0]{*-*}(10.76,7.597396)(9.360001,5.3973956)
\usefont{T1}{ppl}{m}{n}
\rput{-33.691383}(-1.548989,8.050352){\rput(12.489208,6.5988326){$\alpha_{2}$}}
\psline[linewidth=0.02cm,fillcolor=black,dotsize=0.07055555cm 2.0]{*-*}(10.76,7.597396)(13.96,5.3973956)
\usefont{T1}{ppl}{m}{n}
\rput{-58.967655}(-2.9868617,16.812445){\rput(13.344307,11.063374){$\alpha_{3}$}}
\psline[linewidth=0.02cm,fillcolor=black,dotsize=0.07055555cm 2.0]{*-*}(12.56,11.9973955)(13.96,9.797396)
\psline[linewidth=0.02cm,fillcolor=black,dotsize=0.07055555cm 2.0]{*-*}(13.96,0.9973958)(20.060001,-1.2026042)
\usefont{T1}{ppl}{m}{n}
\rput{-19.798964}(1.062542,5.894519){\rput(17.389208,-0.081167236){$\alpha_{1}$}}
\usefont{T1}{ppl}{m}{n}
\rput{35.35122}(7.3463335,-5.4744515){\rput(12.232651,8.805046){$\alpha_{5}$}}
\psline[linewidth=0.02cm,fillcolor=black,dotsize=0.07055555cm 2.0]{*-*}(13.96,9.797396)(10.76,7.597396)
\usefont{T1}{ppl}{m}{n}
\rput{54.790813}(9.961986,-14.785715){\rput(19.216059,2.234119){$\alpha_{4}$}}
\psline[linewidth=0.02cm,fillcolor=black,dotsize=0.07055555cm 2.0]{*-*}(20.060001,3.1973958)(18.66,0.9973958)
\usefont{T1}{ppl}{m}{n}
\rput{-33.691383}(-2.2318916,10.194971){\rput(15.689208,8.798833){$\alpha_{2}$}}
\psline[linewidth=0.02cm,fillcolor=black,dotsize=0.07055555cm 2.0]{*-*}(13.96,9.797396)(17.16,7.597396)
\usefont{T1}{ppl}{m}{n}
\rput{-58.967655}(9.394095,16.710133){\rput(19.444307,0.06337385){$\alpha_{3}$}}
\psline[linewidth=0.02cm,fillcolor=black,dotsize=0.07055555cm 2.0]{*-*}(18.66,0.9973958)(20.060001,-1.2026042)
\psline[linewidth=0.02cm,fillcolor=black,dotsize=0.07055555cm 2.0]{*-*}(19.960001,16.397396)(13.96,14.197396)
\usefont{T1}{ppl}{m}{n}
\rput{19.282917}(6.1242595,-4.844034){\rput(17.289207,15.618833){$\alpha_{6}$}}
\usefont{T1}{ppl}{m}{n}
\rput{35.35122}(8.3610935,-4.2587953){\rput(10.832651,11.005046){$\alpha_{5}$}}
\psline[linewidth=0.02cm,fillcolor=black,dotsize=0.07055555cm 2.0]{*-*}(12.56,11.9973955)(9.360001,9.797396)
\usefont{T1}{ppl}{m}{n}
\rput{54.790813}(16.366598,-5.143891){\rput(13.116059,13.234118){$\alpha_{4}$}}
\psline[linewidth=0.02cm,fillcolor=black,dotsize=0.07055555cm 2.0]{*-*}(13.96,14.197396)(12.56,11.9973955)
\usefont{T1}{ppl}{m}{n}
\rput{-33.691383}(-0.35114607,3.2063944){\rput(5.089208,2.1988328){$\alpha_{2}$}}
\psline[linewidth=0.02cm,fillcolor=black,dotsize=0.07055555cm 2.0]{*-*}(3.3600006,3.1973958)(6.5600004,0.9973958)
\usefont{T1}{ppl}{m}{n}
\rput{-58.967655}(-2.6520636,13.004589){\rput(10.144307,8.863374){$\alpha_{3}$}}
\psline[linewidth=0.02cm,fillcolor=black,dotsize=0.07055555cm 2.0]{*-*}(9.360001,9.797396)(10.76,7.597396)
\usefont{T1}{ppl}{m}{n}
\rput(23.133125,5.3203125){\Large \psframebox[linewidth=0.02,fillstyle=solid]{-1 0 0 0 1 0}}
\usefont{T1}{ppl}{m}{n}
\rput(9.322969,5.3203125){\Large \psframebox[linewidth=0.02,fillstyle=solid]{0 1 0 -1 0 1}}
\usefont{T1}{ppl}{m}{n}
\rput(17.118439,7.6203127){\Large \psframebox[linewidth=0.02,fillstyle=solid]{1 -1 0 0 1 0}}
\usefont{T1}{ppl}{m}{n}
\rput(10.619063,7.6203127){\Large \psframebox[linewidth=0.02,fillstyle=solid]{0 1 -1 1 -1 0}}
\usefont{T1}{ppl}{m}{n}
\rput(13.929063,9.920313){\Large \psframebox[linewidth=0.02,fillstyle=solid]{0 1 -1 0 1 0}}
\usefont{T1}{ppl}{m}{n}
\rput(9.629064,9.920313){\Large \psframebox[linewidth=0.02,fillstyle=solid]{0 0 1 0 -1 0}}
\usefont{T1}{ppl}{m}{n}
\rput(12.729063,12.020312){\Large \psframebox[linewidth=0.02,fillstyle=solid]{0 0 1 -1 1 0}}
\usefont{T1}{ppl}{m}{n}
\rput(19.93297,16.620312){\Large \psframebox[linewidth=0.02,fillstyle=solid]{0 0 0 0 0 1}}
\usefont{T1}{ppl}{m}{n}
\rput(14.022969,14.3203125){\Large \psframebox[linewidth=0.02,fillstyle=solid]{0 0 0 1 0 -1}}
\usefont{T1}{ppl}{m}{n}
\rput(18.917032,1.0203125){\Large \psframebox[linewidth=0.02,fillstyle=solid]{-1 0 1 -1 0 1}}
\usefont{T1}{ppl}{m}{n}
\rput(13.812345,1.0203125){\Large \psframebox[linewidth=0.02,fillstyle=solid]{1 0 -1 0 0 1}}
\usefont{T1}{ppl}{m}{n}
\rput(6.902344,1.0203125){\Large \psframebox[linewidth=0.02,fillstyle=solid]{1 -1 1 0 0 -1}}
\usefont{T1}{ppl}{m}{n}
\rput(20.423126,3.1203125){\Large \psframebox[linewidth=0.02,fillstyle=solid]{-1 0 0 1 -1 0}}
\usefont{T1}{ppl}{m}{n}
\rput(12.702345,3.3203125){\Large \psframebox[linewidth=0.02,fillstyle=solid]{1 -1 1 -1 0 1}}
\usefont{T1}{ppl}{m}{n}
\rput(3.6229694,3.2203126){\Large \psframebox[linewidth=0.02,fillstyle=solid]{0 1 0 0 0 -1}}
\usefont{T1}{ppl}{m}{n}
\rput(14.008438,5.3203125){\Large \psframebox[linewidth=0.02,fillstyle=solid]{1 -1 0 1 -1 0}}
\usefont{T1}{ppl}{m}{n}
\rput(14.119063,-12.279688){\Large \psframebox[linewidth=0.02,fillstyle=solid]{0 0 -1 1 -1 0}}
\usefont{T1}{ppl}{m}{n}
\rput(17.229063,-10.079687){\Large \psframebox[linewidth=0.02,fillstyle=solid]{0 0 -1 0 1 0}}
\usefont{T1}{ppl}{m}{n}
\rput(12.719063,-10.079687){\Large \psframebox[linewidth=0.02,fillstyle=solid]{0 -1 1 0 -1 0}}
\usefont{T1}{ppl}{m}{n}
\rput(16.019064,-7.8796873){\Large \psframebox[linewidth=0.02,fillstyle=solid]{0 -1 1 -1 1 0}}
\usefont{T1}{ppl}{m}{n}
\rput(17.01297,-5.6796875){\Large \psframebox[linewidth=0.02,fillstyle=solid]{0 -1 0 1 0 -1}}
\usefont{T1}{ppl}{m}{n}
\rput(23.22297,-3.4796875){\Large \psframebox[linewidth=0.02,fillstyle=solid]{0 -1 0 0 0 1}}
\usefont{T1}{ppl}{m}{n}
\rput(14.007032,-3.4796875){\Large \psframebox[linewidth=0.02,fillstyle=solid]{-1 1 -1 1 0 -1}}
\usefont{T1}{ppl}{m}{n}
\rput(19.917032,-1.3796875){\Large \psframebox[linewidth=0.02,fillstyle=solid]{-1 1 -1 0 0 1}}
\usefont{T1}{ppl}{m}{n}
\rput(12.817032,-1.4796875){\Large \psframebox[linewidth=0.02,fillstyle=solid]{-1 0 1 0 0 -1}}
\psline[linewidth=0.02cm,fillcolor=black,dotsize=0.07055555cm 2.0]{*-*}(12.660001,-14.402604)(6.6600003,-16.602604)
\usefont{T1}{ppl}{m}{n}
\rput{19.282917}(-4.4564505,-4.16123){\rput(9.989208,-15.181168){$\alpha_{6}$}}
\usefont{T1}{ppl}{m}{n}
\rput(12.62297,-14.379687){\Large \psframebox[linewidth=0.02,fillstyle=solid]{0 0 0 -1 0 1}}
\usefont{T1}{ppl}{m}{n}
\rput(6.5229692,-16.679688){\Large \psframebox[linewidth=0.02,fillstyle=solid]{0 0 0 0 0 -1}}
\psline[linewidth=0.02cm,fillcolor=black,dotsize=0.07055555cm 2.0]{*-*}(6.5600004,-3.402604)(12.660001,-5.6026044)
\usefont{T1}{ppl}{m}{n}
\rput{-19.798964}(2.115477,3.1278865){\rput(9.989208,-4.4811673){$\alpha_{1}$}}
\usefont{T1}{ppl}{m}{n}
\rput{35.35122}(-1.655422,-3.626715){\rput(4.832651,-4.394954){$\alpha_{5}$}}
\psline[linewidth=0.02cm,fillcolor=black,dotsize=0.07055555cm 2.0]{*-*}(6.5600004,-3.402604)(3.3600006,-5.6026044)
\usefont{T1}{ppl}{m}{n}
\rput{54.790813}(1.2433709,-6.762501){\rput(7.116059,-2.1658812){$\alpha_{4}$}}
\psline[linewidth=0.02cm,fillcolor=black,dotsize=0.07055555cm 2.0]{*-*}(7.9600005,-1.2026042)(6.5600004,-3.402604)
\psline[linewidth=0.02cm,fillcolor=black,dotsize=0.07055555cm 2.0]{*-*}(3.3600006,-5.6026044)(9.46,-7.802604)
\usefont{T1}{ppl}{m}{n}
\rput{-19.798964}(2.671501,1.9139309){\rput(6.9892084,-6.681167){$\alpha_{1}$}}
\usefont{T1}{ppl}{m}{n}
\rput(8.002344,-1.4796875){\Large \psframebox[linewidth=0.02,fillstyle=solid]{1 0 -1 1 0 -1}}
\usefont{T1}{ppl}{m}{n}
\rput(6.718438,-3.3796875){\Large \psframebox[linewidth=0.02,fillstyle=solid]{1 0 0 -1 1 0}}
\usefont{T1}{ppl}{m}{n}
\rput(3.6184382,-5.6796875){\Large \psframebox[linewidth=0.02,fillstyle=solid]{1 0 0 0 -1 0}}
\usefont{T1}{ppl}{m}{n}
\rput(12.723125,-5.6796875){\Large \psframebox[linewidth=0.02,fillstyle=solid]{-1 1 0 -1 1 0}}
\usefont{T1}{ppl}{m}{n}
\rput(9.623126,-7.7796874){\Large \psframebox[linewidth=0.02,fillstyle=solid]{-1 1 0 0 -1 0}}
\usefont{T1}{ppl}{m}{n}
\rput(19.222812,-5.6796875){\Large $\Lambda_{3}$}
\usefont{T1}{ppl}{m}{n}
\rput(14.722813,-14.479688){\Large $\Lambda_{4}$}
\usefont{T1}{ppl}{m}{n}
\rput(1.6228126,3.2203126){\Large $\Lambda_{2}$}
%\usefont{T1}{ppl}{m}{n}
%\rput(8.722813,-16.779688){\Large $\Lambda_{5}$}
\end{pspicture}
}

\caption{Dynkin tree of the representation {\bf 32} of $so(6,6)$. All the weights have the same length. We denote with $\Lambda_{i}$, with i=1,2,3,4 four weights not connected by algebra transformations. We choose this weights to study the orbits and we refer to the other weight as $\Sigma_{i}$ with i ranging from 1 to 28 and increasing, in the diagram from the top to the bottom from the left to the right.}\label{fig:32ofso(6,6)dynkinlabels}

\end{figure}
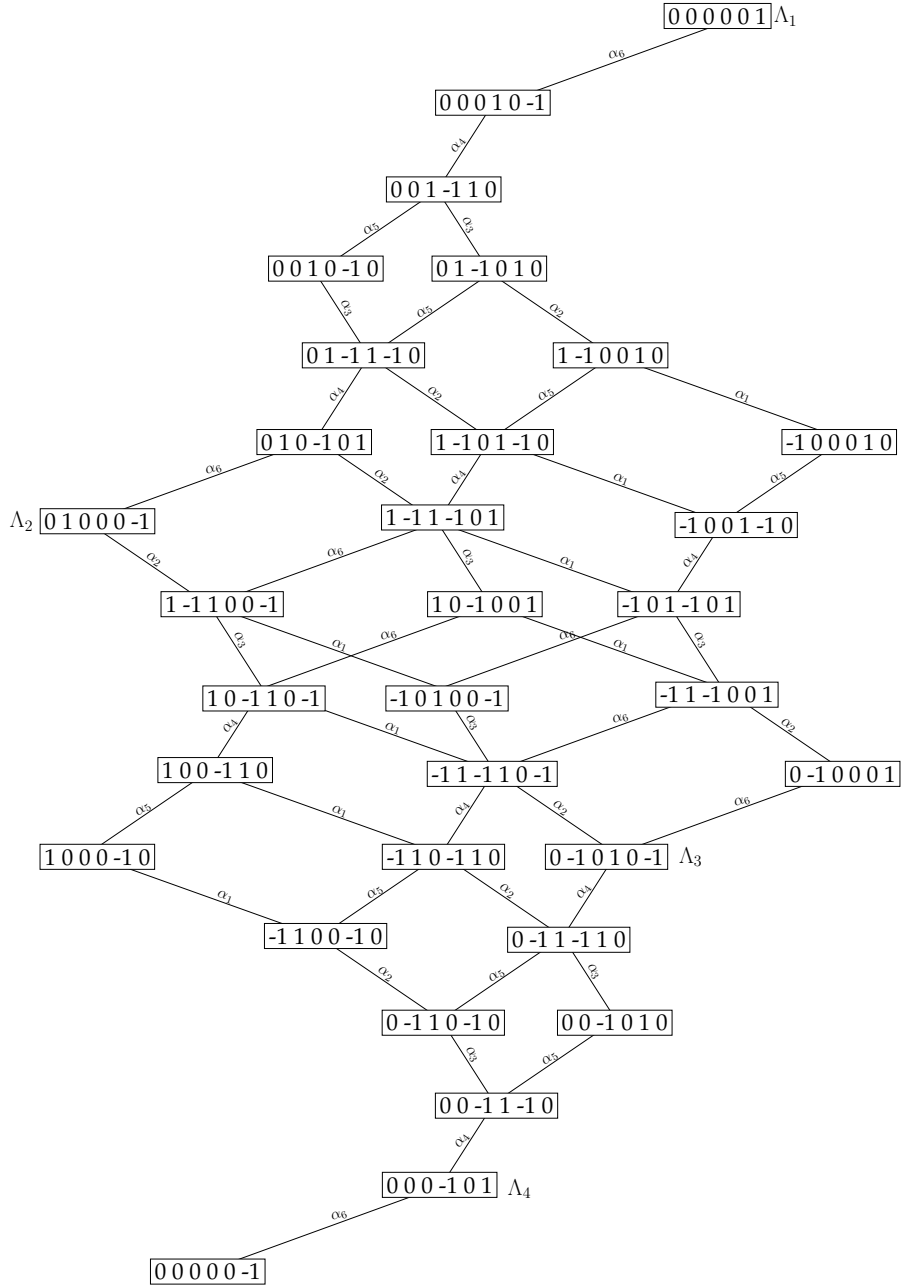

\begin{figure}[h!]
\centering
% Generated with LaTeXDraw 2.0.8
% Sat May 06 11:21:41 CEST 2017
% \usepackage[usenames,dvipsnames]{pstricks}
% \usepackage{epsfig}
% \usepackage{pst-grad} % For gradients
% \usepackage{pst-plot} % For axes
\scalebox{0.5} % Change this value to rescale the drawing.
{
\begin{pspicture}(0,-17.07)(22.02,17.07)
\definecolor{color3313b}{rgb}{1.0,0.8627450980392157,0.0}
\definecolor{color3314b}{rgb}{0.06666666666666667,0.596078431372549,0.6941176470588235}
\definecolor{color3312b}{rgb}{0.5490196078431373,0.5490196078431373,0.5490196078431373}
\definecolor{color3315b}{rgb}{0.8509803921568627,0.00784313725490196,0.00784313725490196}
\definecolor{color3556b}{rgb}{0.8509803921568627,0.0,0.0}
\pspolygon[linewidth=0.04,fillstyle=solid,fillcolor=color3313b,opacity=0.3](0.3,3.45)(6.3,5.55)(7.9,7.75)(6.7,10.05)(9.7,12.15)(11.5,14.85)(12.2,13.95)(11.0,12.25)(12.4,10.15)(12.2,9.45)(9.1,7.25)(7.6,5.05)(2.3,3.05)(4.9,1.35)(11.4,-0.85)(12.8,-3.15)(10.9,-5.95)(7.8,-8.25)(4.7,-17.05)(4.0,-16.55)(6.6,-8.25)(0.0,-5.65)(3.8,-2.95)(5.0,-0.95)(4.1,0.75)
\pspolygon[linewidth=0.04,fillstyle=solid,fillcolor=color3314b,opacity=0.3](15.3,-5.95)(21.3,-3.75)(21.3,-2.85)(15.0,-4.95)(12.8,-3.15)(11.4,-0.85)(17.1,3.05)(18.3,2.95)(22.0,5.65)(16.0,7.85)(11.9,14.95)(11.5,14.75)(11.1,14.25)(14.8,8.35)(10.6,5.35)(16.4,3.35)(10.5,-0.55)(5.0,1.35)(3.7,1.45)(5.3,-1.45)(12.1,-4.15)(13.0,-7.45)(10.0,-9.55)(4.0,-16.55)(4.7,-17.05)(10.2,-10.75)(11.6,-12.75)(15.7,-9.95)(14.2,-7.45)
\pspolygon[linewidth=0.04,fillstyle=solid,fillcolor=color3312b,opacity=0.3](18.0,17.05)(11.4,14.75)(9.7,12.15)(6.7,10.05)(7.9,7.75)(6.3,5.55)(9.8,3.05)(11.5,0.45)(17.3,-1.45)(21.3,-3.75)(21.3,-2.85)(18.2,-0.85)(17.1,1.25)(18.3,2.95)(22.0,5.65)(15.4,8.05)(12.4,10.15)(11.0,12.25)(12.1,13.85)(18.4,16.15)
\pspolygon[linewidth=0.04,fillstyle=solid,fillcolor=color3315b,opacity=0.3](4.7,-17.05)(10.7,-14.75)(12.2,-12.35)(15.7,-9.95)(14.2,-7.45)(14.2,-5.05)(21.3,-3.75)(21.3,-2.85)(18.2,-0.85)(16.9,1.65)(10.6,3.65)(6.8,6.25)(6.3,5.55)(9.8,3.05)(11.5,0.45)(13.7,-2.65)(9.3,-4.95)(3.9,-2.85)(0.0,-5.65)(6.6,-8.25)(9.9,-10.35)(11.2,-12.15)(10.1,-13.95)(4.0,-16.55)
\psline[linewidth=0.02cm,fillcolor=black,dotsize=0.07055555cm 2.0]{*-*}(17.8,-1.15)(11.8,-3.35)
\usefont{T1}{ppl}{m}{n}
\rput{19.282917}(0.20835322,-5.115149){\rput(15.129208,-1.9285631){$\alpha_{6}$}}
\psline[linewidth=0.02cm,fillcolor=black,dotsize=0.07055555cm 2.0]{*-*}(11.7,1.05)(5.7,-1.15)
\usefont{T1}{ppl}{m}{n}
\rput{19.282917}(0.59265244,-2.9773068){\rput(9.029208,0.27143693){$\alpha_{6}$}}
\psline[linewidth=0.02cm,fillcolor=black,dotsize=0.07055555cm 2.0]{*-*}(10.3,3.25)(4.3,1.05)
\usefont{T1}{ppl}{m}{n}
\rput{19.282917}(1.2406242,-2.3915594){\rput(7.6292076,2.471437){$\alpha_{6}$}}
\usefont{T1}{ppl}{m}{n}
\rput{-58.967655}(14.77287,4.328559){\rput(11.184307,-10.884022){$\alpha_{3}$}}
\psline[linewidth=0.02cm,fillcolor=black,dotsize=0.07055555cm 2.0]{*-*}(10.4,-9.95)(11.8,-12.15)
\psline[linewidth=0.02cm,fillcolor=black,dotsize=0.07055555cm 2.0]{*-*}(7.1,5.45)(1.1,3.25)
\usefont{T1}{ppl}{m}{n}
\rput{19.282917}(1.7876152,-1.2113925){\rput(4.429208,4.671437){$\alpha_{6}$}}
\usefont{T1}{ppl}{m}{n}
\rput{-58.967655}(14.438072,8.1364155){\rput(14.384307,-8.684022){$\alpha_{3}$}}
\psline[linewidth=0.02cm,fillcolor=black,dotsize=0.07055555cm 2.0]{*-*}(13.6,-7.75)(15.0,-9.95)
\psline[linewidth=0.02cm,fillcolor=black,dotsize=0.07055555cm 2.0]{*-*}(21.0,-3.35)(15.0,-5.55)
\usefont{T1}{ppl}{m}{n}
\rput{19.282917}(-0.33863762,-6.2953157){\rput(18.329208,-4.128563){$\alpha_{6}$}}
\usefont{T1}{ppl}{m}{n}
\rput{-33.691383}(6.366562,3.4977648){\rput(8.929208,-8.748563){$\alpha_{2}$}}
\psline[linewidth=0.02cm,fillcolor=black,dotsize=0.07055555cm 2.0]{*-*}(7.2,-7.75)(10.4,-9.95)
\psline[linewidth=0.02cm,fillcolor=black,dotsize=0.07055555cm 2.0]{*-*}(16.4,1.05)(10.4,-1.15)
\usefont{T1}{ppl}{m}{n}
\rput{19.282917}(0.85632503,-4.529402){\rput(13.729208,0.27143693){$\alpha_{6}$}}
\psline[linewidth=0.02cm,fillcolor=black,dotsize=0.07055555cm 2.0]{*-*}(10.3,3.25)(16.4,1.05)
\usefont{T1}{ppl}{m}{n}
\rput{-19.798964}(0.083183855,4.7879586){\rput(13.729208,2.171437){$\alpha_{1}$}}
\usefont{T1}{ppl}{m}{n}
\rput{35.35122}(-3.8874989,-9.717194){\rput(13.272651,-10.94235){$\alpha_{5}$}}
\psline[linewidth=0.02cm,fillcolor=black,dotsize=0.07055555cm 2.0]{*-*}(15.0,-9.95)(11.8,-12.15)
\usefont{T1}{ppl}{m}{n}
\rput{54.790813}(-6.075229,-14.53551){\rput(10.9560585,-13.1132765){$\alpha_{4}$}}
\psline[linewidth=0.02cm,fillcolor=black,dotsize=0.07055555cm 2.0]{*-*}(11.8,-12.15)(10.4,-14.35)
\usefont{T1}{ppl}{m}{n}
\rput{-33.691383}(5.683659,5.642384){\rput(12.129208,-6.548563){$\alpha_{2}$}}
\psline[linewidth=0.02cm,fillcolor=black,dotsize=0.07055555cm 2.0]{*-*}(10.4,-5.55)(13.6,-7.75)
\usefont{T1}{ppl}{m}{n}
\rput{-58.967655}(7.232358,8.591967){\rput(11.184307,-2.084022){$\alpha_{3}$}}
\psline[linewidth=0.02cm,fillcolor=black,dotsize=0.07055555cm 2.0]{*-*}(10.4,-1.15)(11.8,-3.35)
\psline[linewidth=0.02cm,fillcolor=black,dotsize=0.07055555cm 2.0]{*-*}(4.3,1.05)(10.4,-1.15)
\usefont{T1}{ppl}{m}{n}
\rput{-19.798964}(0.4736912,2.6255846){\rput(7.7292075,-0.028563065){$\alpha_{1}$}}
\usefont{T1}{ppl}{m}{n}
\rput{35.35122}(-2.8727381,-8.501537){\rput(11.87265,-8.742351){$\alpha_{5}$}}
\psline[linewidth=0.02cm,fillcolor=black,dotsize=0.07055555cm 2.0]{*-*}(13.6,-7.75)(10.4,-9.95)
\usefont{T1}{ppl}{m}{n}
\rput{54.790813}(0.67231464,-14.355397){\rput(14.156058,-6.513277){$\alpha_{4}$}}
\psline[linewidth=0.02cm,fillcolor=black,dotsize=0.07055555cm 2.0]{*-*}(15.0,-5.55)(13.6,-7.75)
\usefont{T1}{ppl}{m}{n}
\rput{-33.691383}(4.4858165,10.486342){\rput(19.529207,-2.1485631){$\alpha_{2}$}}
\psline[linewidth=0.02cm,fillcolor=black,dotsize=0.07055555cm 2.0]{*-*}(17.8,-1.15)(21.0,-3.35)
\psline[linewidth=0.02cm,fillcolor=black,dotsize=0.07055555cm 2.0]{*-*}(5.7,-1.15)(11.8,-3.35)
\usefont{T1}{ppl}{m}{n}
\rput{-19.798964}(1.3016355,2.9697447){\rput(9.129208,-2.228563){$\alpha_{1}$}}
\usefont{T1}{ppl}{m}{n}
\rput{35.35122}(6.2771945,-6.4142585){\rput(13.17265,6.65765){$\alpha_{5}$}}
\psline[linewidth=0.02cm,fillcolor=black,dotsize=0.07055555cm 2.0]{*-*}(14.9,7.65)(11.7,5.45)
\usefont{T1}{ppl}{m}{n}
\rput{54.790813}(1.1148323,-10.809268){\rput(10.9560585,-4.313277){$\alpha_{4}$}}
\psline[linewidth=0.02cm,fillcolor=black,dotsize=0.07055555cm 2.0]{*-*}(11.8,-3.35)(10.4,-5.55)
\usefont{T1}{ppl}{m}{n}
\rput{-33.691383}(-0.9725295,5.6593976){\rput(8.829207,4.451437){$\alpha_{2}$}}
\psline[linewidth=0.02cm,fillcolor=black,dotsize=0.07055555cm 2.0]{*-*}(7.1,5.45)(10.3,3.25)
\usefont{T1}{ppl}{m}{n}
\rput{-58.967655}(3.413654,10.637982){\rput(11.084307,2.315978){$\alpha_{3}$}}
\psline[linewidth=0.02cm,fillcolor=black,dotsize=0.07055555cm 2.0]{*-*}(10.3,3.25)(11.7,1.05)
\psline[linewidth=0.02cm,fillcolor=black,dotsize=0.07055555cm 2.0]{*-*}(14.9,7.65)(21.0,5.45)
\usefont{T1}{ppl}{m}{n}
\rput{-19.798964}(-1.1352679,6.606172){\rput(18.329208,6.571437){$\alpha_{1}$}}
\usefont{T1}{ppl}{m}{n}
\rput{35.35122}(-2.1898608,-6.244424){\rput(8.67265,-6.5423503){$\alpha_{5}$}}
\psline[linewidth=0.02cm,fillcolor=black,dotsize=0.07055555cm 2.0]{*-*}(10.4,-5.55)(7.2,-7.75)
\usefont{T1}{ppl}{m}{n}
\rput{54.790813}(8.26255,-7.0013204){\rput(10.856058,4.486723){$\alpha_{4}$}}
\psline[linewidth=0.02cm,fillcolor=black,dotsize=0.07055555cm 2.0]{*-*}(11.7,5.45)(10.3,3.25)
\usefont{T1}{ppl}{m}{n}
\rput{-33.691383}(4.698424,6.788509){\rput(13.529208,-4.348563){$\alpha_{2}$}}
\psline[linewidth=0.02cm,fillcolor=black,dotsize=0.07055555cm 2.0]{*-*}(11.8,-3.35)(15.0,-5.55)
\usefont{T1}{ppl}{m}{n}
\rput{-58.967655}(2.3919134,4.430872){\rput(5.0843067,0.11597802){$\alpha_{3}$}}
\psline[linewidth=0.02cm,fillcolor=black,dotsize=0.07055555cm 2.0]{*-*}(4.3,1.05)(5.7,-1.15)
\psline[linewidth=0.02cm,fillcolor=black,dotsize=0.07055555cm 2.0]{*-*}(11.7,5.45)(17.8,3.25)
\usefont{T1}{ppl}{m}{n}
\rput{-19.798964}(-0.57924384,5.3922167){\rput(15.129208,4.371437){$\alpha_{1}$}}
\usefont{T1}{ppl}{m}{n}
\rput{35.35122}(6.129017,-10.349274){\rput(19.27265,4.4576497){$\alpha_{5}$}}
\psline[linewidth=0.02cm,fillcolor=black,dotsize=0.07055555cm 2.0]{*-*}(21.0,5.45)(17.8,3.25)
\usefont{T1}{ppl}{m}{n}
\rput{54.790813}(8.705069,-3.4551916){\rput(7.6560583,6.686723){$\alpha_{4}$}}
\psline[linewidth=0.02cm,fillcolor=black,dotsize=0.07055555cm 2.0]{*-*}(8.5,7.65)(7.1,5.45)
\usefont{T1}{ppl}{m}{n}
\rput{-33.691383}(-1.9577647,6.805522){\rput(10.229208,6.651437){$\alpha_{2}$}}
\psline[linewidth=0.02cm,fillcolor=black,dotsize=0.07055555cm 2.0]{*-*}(8.5,7.65)(11.7,5.45)
\usefont{T1}{ppl}{m}{n}
\rput{-58.967655}(-4.1268587,14.901389){\rput(11.084307,11.115978){$\alpha_{3}$}}
\psline[linewidth=0.02cm,fillcolor=black,dotsize=0.07055555cm 2.0]{*-*}(10.3,12.05)(11.7,9.85)
\psline[linewidth=0.02cm,fillcolor=black,dotsize=0.07055555cm 2.0]{*-*}(11.7,1.05)(17.8,-1.15)
\usefont{T1}{ppl}{m}{n}
\rput{-19.798964}(0.91112816,5.132119){\rput(15.129208,-0.028563065){$\alpha_{1}$}}
\usefont{T1}{ppl}{m}{n}
\rput{35.35122}(6.960072,-4.1571455){\rput(9.972651,8.85765){$\alpha_{5}$}}
\psline[linewidth=0.02cm,fillcolor=black,dotsize=0.07055555cm 2.0]{*-*}(11.7,9.85)(8.5,7.65)
\usefont{T1}{ppl}{m}{n}
\rput{54.790813}(9.047998,-12.916902){\rput(16.956059,2.2867231){$\alpha_{4}$}}
\psline[linewidth=0.02cm,fillcolor=black,dotsize=0.07055555cm 2.0]{*-*}(17.8,3.25)(16.4,1.05)
\usefont{T1}{ppl}{m}{n}
\rput{-33.691383}(-2.6406672,8.950141){\rput(13.429208,8.851437){$\alpha_{2}$}}
\psline[linewidth=0.02cm,fillcolor=black,dotsize=0.07055555cm 2.0]{*-*}(11.7,9.85)(14.9,7.65)
\usefont{T1}{ppl}{m}{n}
\rput{-58.967655}(8.254099,14.799077){\rput(17.184307,0.11597802){$\alpha_{3}$}}
\psline[linewidth=0.02cm,fillcolor=black,dotsize=0.07055555cm 2.0]{*-*}(16.4,1.05)(17.8,-1.15)
\psline[linewidth=0.02cm,fillcolor=black,dotsize=0.07055555cm 2.0]{*-*}(17.7,16.45)(11.7,14.25)
\usefont{T1}{ppl}{m}{n}
\rput{19.282917}(6.014844,-4.094756){\rput(15.029208,15.671437){$\alpha_{6}$}}
\usefont{T1}{ppl}{m}{n}
\rput{35.35122}(7.974832,-2.9414897){\rput(8.572651,11.05765){$\alpha_{5}$}}
\psline[linewidth=0.02cm,fillcolor=black,dotsize=0.07055555cm 2.0]{*-*}(10.3,12.05)(7.1,9.85)
\usefont{T1}{ppl}{m}{n}
\rput{54.790813}(15.452612,-3.2750778){\rput(10.856058,13.286723){$\alpha_{4}$}}
\psline[linewidth=0.02cm,fillcolor=black,dotsize=0.07055555cm 2.0]{*-*}(11.7,14.25)(10.3,12.05)
\usefont{T1}{ppl}{m}{n}
\rput{-33.691383}(-0.7599219,1.961564){\rput(2.8292077,2.251437){$\alpha_{2}$}}
\psline[linewidth=0.02cm,fillcolor=black,dotsize=0.07055555cm 2.0]{*-*}(1.1,3.25)(4.3,1.05)
\usefont{T1}{ppl}{m}{n}
\rput{-58.967655}(-3.7920604,11.0935335){\rput(7.884307,8.915978){$\alpha_{3}$}}
\psline[linewidth=0.02cm,fillcolor=black,dotsize=0.07055555cm 2.0]{*-*}(7.1,9.85)(8.5,7.65)
\psline[linewidth=0.02cm,fillcolor=black,dotsize=0.07055555cm 2.0]{*-*}(10.4,-14.35)(4.4,-16.55)
\usefont{T1}{ppl}{m}{n}
\rput{19.282917}(-4.5658665,-3.4119523){\rput(7.7292075,-15.128563){$\alpha_{6}$}}
\psline[linewidth=0.02cm,fillcolor=black,dotsize=0.07055555cm 2.0]{*-*}(4.3,-3.35)(10.4,-5.55)
\usefont{T1}{ppl}{m}{n}
\rput{-19.798964}(1.9640632,2.3654869){\rput(7.7292075,-4.428563){$\alpha_{1}$}}
\usefont{T1}{ppl}{m}{n}
\rput{35.35122}(-2.0416832,-2.309409){\rput(2.5726504,-4.34235){$\alpha_{5}$}}
\psline[linewidth=0.02cm,fillcolor=black,dotsize=0.07055555cm 2.0]{*-*}(4.3,-3.35)(1.1,-5.55)
\usefont{T1}{ppl}{m}{n}
\rput{54.790813}(0.3293841,-4.8936872){\rput(4.8560586,-2.113277){$\alpha_{4}$}}
\psline[linewidth=0.02cm,fillcolor=black,dotsize=0.07055555cm 2.0]{*-*}(5.7,-1.15)(4.3,-3.35)
\psline[linewidth=0.02cm,fillcolor=black,dotsize=0.07055555cm 2.0]{*-*}(1.1,-5.55)(7.2,-7.75)
\usefont{T1}{ppl}{m}{n}
\rput{-19.798964}(2.5200872,1.1515312){\rput(4.5292077,-6.628563){$\alpha_{1}$}}
\pscircle[linewidth=0.02,dimen=outer,fillstyle=solid,fillcolor=color3556b](1.15,3.3){0.15}
\pscircle[linewidth=0.02,dimen=outer,fillstyle=solid,fillcolor=color3556b](10.45,-14.3){0.15}
\pscircle[linewidth=0.02,dimen=outer,fillstyle=solid,fillcolor=color3556b](17.65,16.4){0.15}
\pscircle[linewidth=0.02,dimen=outer,fillstyle=solid,fillcolor=color3556b](15.05,-5.5){0.15}
\end{pspicture}
}

\caption{Diagram of the orbits in the {\bf 32} of the four weights picked as starting points for our analysis. It is possible to appreciate the overlaps of different orbits. There is a triple overlap on four weights and any two orbits overlap on exactly six weights. These will define the conjunction stabilizers.}\label{fig:32ofso(6,6)orbits}
\end{figure}
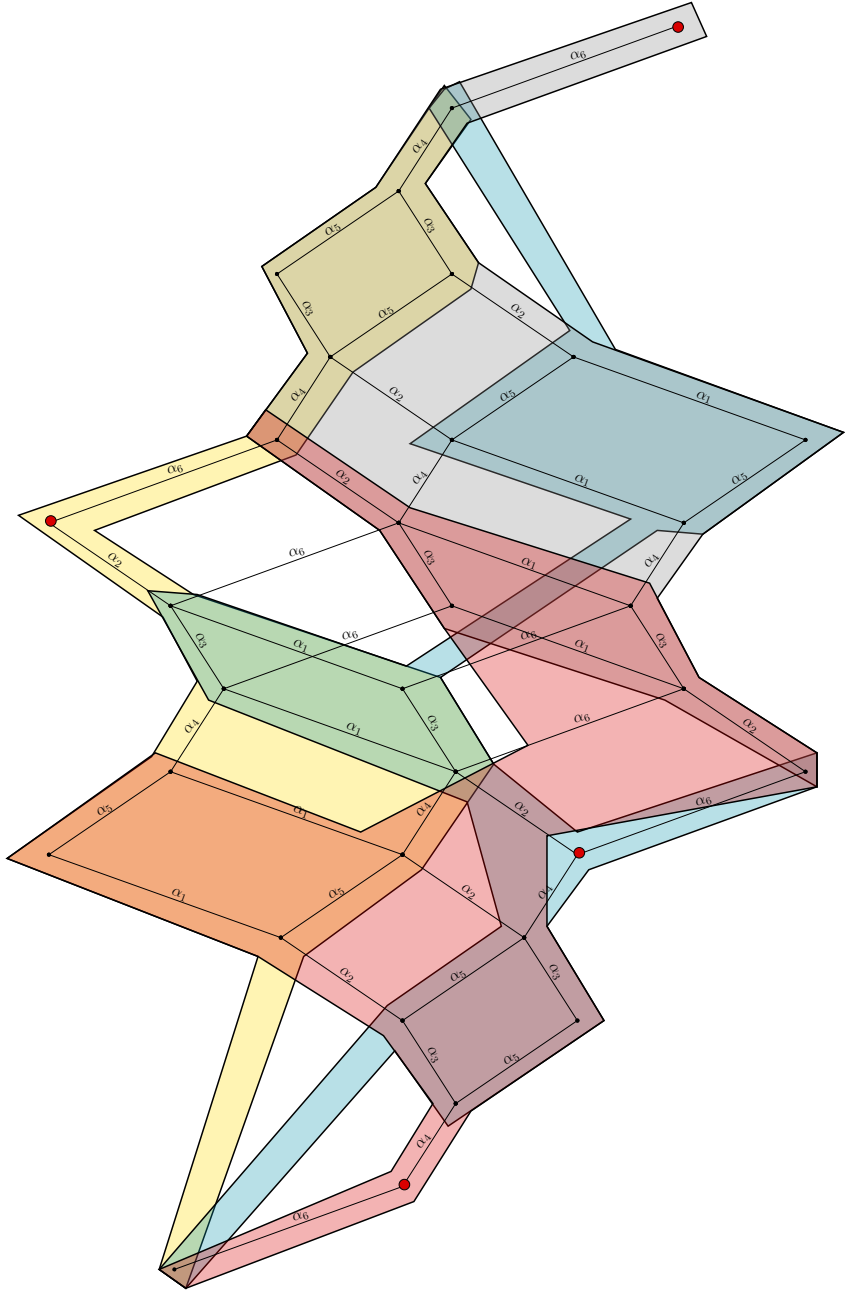

By exploiting the same methods used in the analysis of the $\mathbb{C}_{s}$-based theory in the previous subsection, in this case, the following stratification is determined :

\begin{description}

\item [The rank-1 (1-weight orbit)] orbit could be easily deduced by the stabilizers of $\Lambda_1$ : %in \autoref{tab:commonstabilizer-32so12}:
\begin{align}
\left. \partial ^{2}I_{4}\right\vert _{\mathbf{66}}=0:&&\frac{SO(6,6)}{SL(6,\mathbb{R})\ltimes \mathbb{R}^{15}},  \label{p-sp-1}
\end{align}%
where $\mathbb{R}^{15}\simeq \mathbf{15}$ and $\mathbf{66}$ respectively denote the rank-$2$ antisymmetric irrep. of $SL(6,\mathbb{R})$ and the rank-$2$ antisymmetric (adjoint) irrep. of $SO(6,6)$. The orbit \autoref{p-sp-1} is the \textit{pure spinor} orbit of $SO(6,6)$ \cite{Pure-Spinors}. Note that $SL(6,\mathbb{R})$ is the $U$-duality group in $D=5$ (as well as the $U$-duality group of the $D=4$ theory on $\mathbb{C}_{s}$; see (\autoref{Cs-D=4})),and $\mathbf{15}$ ($\mathbf{15}^{\prime }$) is the irrep. relevant to asymptotically flat branes (black holes and black strings, respectively) in $D=5$.

\item [Rank-2 (2-weights non-dyonic orbit)] this is the orbits of the combinations $\Lambda_{1}\pm \Lambda_{2}$. These bound states have stabilizer made by common and conjunction stabilizers.  The two bound state have the same rank-2 orbit
\begin{align}
\partial I_{4}=0\text{ }&&:\frac{SO(6,6)}{\left[ SO(4,3)\times SL(2,\mathbb{R})\right] \ltimes \left( \mathbb{R}^{\left( 8,2\right) }\times \mathbb{R}\right) },
\end{align}
where $\mathbb{R}^{\left( 8,2\right) }\simeq \left( \mathbf{8,2}\right) $ denotes the real bi-spinor of the split form $SO(4,3)\times SL(2,\mathbb{R})$. The $SO(3,4)$ has simple roots
\begin{flalign}
 &\beta_1=\alpha_4\qquad \beta_{2}=\alpha_3\quad \beta_{3}=\frac{\alpha_5-\alpha_3}{2},
\end{flalign}
while $SL(2,\mathbb{R})$ has simple root $\alpha_1$.

\item [Rank-3 (3-weights orbit)] the rank three orbit, corresponding to the combinations $\Lambda_1+\Lambda_2\pm\Lambda_3$. This bound states has stabilizer of the common and conjunction types.  In particular these lasts  act on two of the three states stabilizing also the remaining one. The semisimple part of the stabilizer is an $Sp(6,\mathbb{R})$ with simple roots recognized as
\begin{flalign}
 &\beta_1=\frac{\alpha_1-\alpha_3}{2}\quad \beta_2=\frac{\alpha_3-\alpha_5}{2}\quad\beta_3=\alpha_5
\end{flalign}
The orbit reads
\begin{align}
I_{4}=0\text{ }:&&\frac{SO(6,6)}{Sp(6,\mathbb{R})\ltimes \mathbb{R}^{14}},
\end{align}%
where $\mathbb{R}^{14}\simeq \mathbf{14}$ denotes the rank-$2$ antisymmetric irrep. of $Sp(6,\mathbb{R})$.

\item [Rank-4 (4-weights and the 2-weights dyonic orbits)] ($I_{4}\neq 0$, these comprise the 4-weights orbits, and the 2-weights dyonic orbit; this latter is isomorphic to the $I_{4}<0$ one). The Dyonic orbit could be realized as the orbit of the highest weight $\Lambda_1$ and the lowest weight $\Sigma_{28}$. These two weights have two not overlapping orbits thus there are no conjunction stabilizers. The common stabilizers correspond to all the generators of $SO(6,6)$ except the one containing $\alpha_6$. The result is a stabilizer $SL(6,\mathbb{R})$. The rank-4 orbits could be seen as the orbits of the combinations $\Lambda_1+\Lambda_2+a\Lambda_3+b\Lambda_4$ with $a,b$ that can take values $\pm 1$. The common stabilizers  define an algebra $[SL(2,\mathbb{R})]^3$ that enlarged by the twenty-four conjunction stabilizers. In particular twenty-two of them have the same compactness properties despite of $a,b$ while there are two of them (for further details see \autoref{appendix:32ofso(6,6)}),
\begin{subequations}
 \begin{flalign}
  &F_{\alpha_6}^{ab}-abF^{ab}_{\alpha_3+2\alpha_4+\alpha_5+\alpha_6}\\
  &F_{\alpha_6}^{ab}-bF^{ab}_{\alpha_1+2\alpha_2+2\alpha_3+2\alpha_4+\alpha_5+\alpha_6},
\end{flalign}
\end{subequations}
where
\begin{align}
&F^{\pm}_{\alpha}=E_{\alpha}\pm E_{-\alpha},
\end{align}
that are compact if $ab=-1$, non-compact if $ab=+1$. This means there are two independent orbits
\begin{subequations}
 \begin{flalign}
 &\mathcal{O}(\Lambda_1+\Lambda_{2}\pm\Lambda_3\pm\Lambda_4)=\frac{SO(6,6)}{SL(6,\mathbb{R})}\\
 &\mathcal{O}(\Lambda_1+\Lambda_{2}\pm\Lambda_3\mp\Lambda_4)=\frac{SO(6,6)}{SU(3,3)}
\end{flalign}
\end{subequations}
corresponding to
\begin{subequations}
\begin{align}
I_{4} >0\text{ }:&&\frac{SO(6,6)}{SU(3,3)}; \\
I_{4} <0\text{ }:&&\frac{SO(6,6)}{SL(6,\mathbb{R})}.
\end{align}
\end{subequations}

\end{description}
We summarize our results in \autoref{tab:32so(6,6)summary}.

\begin{table}[h!]
\renewcommand{\arraystretch}{1.8}
\begin{center}
\resizebox{\textwidth}{!}{
\begin{tabular}{|c|c|c|c|c|c|c|c|}
\hline
\multicolumn{2}{|c|}{}&\multicolumn{2}{|c|}{{$\mathbf{\theta}$}}&&&\\ \cline{3-4}
\multicolumn{2}{|c|}{\multirow{-2}{*}{{\textbf{States}}}}&{\bf +}&{\bf -}&\multirow{-2}{*}{{\textbf{semisimple stabilizer}}}&\multirow{-2}{*}{{\textbf{stabilizer}}}&\multirow{-2}{*}{{\bf rank}}\\ \hline\hline

{1-w}&$\Lambda_{1}$&15&20&$SL(6,\mathbb{R})$&$SL(6,\mathbb{R})\ltimes \mathbb{R}^{15}$&1\\ \hline

{2-w}&$\Lambda_{1}\pm\Lambda_{2}$&10&14& $SO(3,4)\times SL(2,\mathbb{R})$&$[SO(3,4)\times SL(2,\mathbb{R})]\ltimes (\mathbb{R}\times\mathbb{R}^{(8,2)})$&2\\
\hline

{3-w}&$\Lambda_{1}+\Lambda_{2}\pm\Lambda_{3}$&9&12&$Sp(6,\mathbb{R})$&$Sp(6,\mathbb{R})\ltimes (\mathbb{R}^{14}\times \mathbb{R})$&3\\ \hline
&$\Lambda_{1}+\Lambda_{2}+a\Lambda_{3}-a\Lambda_{4}$&17&18&$SU(3,3)$&$SU(3,3)$&4\\

\multirow{-2}{*}{{4-w}}&$\Lambda_{1}+\Lambda_{2}+a\Lambda_{3}+a\Lambda_{4}$&15&20&$SL(6,\mathbb{R})$&$SL(6,\mathbb{R})$&4\\

\hline
\end{tabular}
}
\caption{Summary of the orbits in the {\bf 32} of $SO(6,6)$}\label{tab:32so(6,6)summary}
\end{center}
\end{table}
\FloatBarrier

\section*{Acknowledgments}

This work and LR have been supported in part by the MINECO/FEDER, UE grant PGC2018-
095205-B-I00 and by the Spanish Research Agency (Agencia Estatal de Investigaci´on) through
the grant IFT Centro de Excelencia Severo Ochoa SEV-2016-0597

\FloatBarrier
\appendix
\section{Orbits in Detail}
In this appendix we collect some relevant details on the results obtained for the U-duality orbits in four and five dimensions
\FloatBarrier
\subsection{$(\mathbf{3},\mathbf{3}^{\prime })$ of $SL(3,\mathbb{R})\times SL(3,\mathbb{R})$ in $D=5$}
\label{appendix:33primesl3sl3}
\FloatBarrier

We study the action of the generators of $\mathfrak{sl}(3,\mathbb{R})\oplus \mathfrak{sl}(3,\mathbb{R}) $  on the $(\mathbf{3}, \mathbf{3}^{\prime })$ irrep.. In particular we could consider the two components separately. We describe the action of $\mathfrak{sl}(3,\mathbb{R})$ on the $\mathbf{3}^{\prime }$ in \autoref{3sl3action1}.  Analogously, with another suitable choice of the structure constants, the action on the $\mathbf{3}$ irrep. can be simplified to the form appearing in the table, namely each raising and lowering generators acts on any weights annihilating it or connecting it to another weights without changing sign. This implies that there is a(t least one) choice of the structure constants for $\mathfrak{sl}(3,\mathbb{R})\oplus \mathfrak{sl}(3,\mathbb{R})$ such that its action on the $\mathbf{(3}^{\prime }\mathbf{,3)}$ is analogous to the one described above.
\begin{table}[h]
\renewcommand{\arraystretch}{1.2}
\par
\begin{center}
%\resizebox{\textwidth}{!}{
\begin{tabular}{|c|c|c|c|}
\hline
& \textcolor{black}{$\Lambda_{1}$} & \textcolor{black}{$\Sigma_{2}$} &\textcolor{black}{$\Sigma_4$} \\ \hline\hline $E_{\alpha_{1}+\alpha_{2}}$ &&& $\Lambda_1$ \\ \hline
$E_{\alpha_{2}}$ & & $\Lambda_1$ & \\ \hline
$E_{\alpha_{1}}$ & & & $\Sigma_2$ \\ \hline
$H_{\alpha_{1}}$ & 1 & -1 & \\ \hline
$H_{\alpha_{2}}$ & & 1 & -1 \\ \hline
$E_{-\alpha_{2}}$ & \textcolor{blue}{$\Sigma_2$} & & \\ \hline
$E_{-\alpha_{1}}$ & & $\Sigma_4$ & \\ \hline
$E_{-\alpha_{1}-\alpha_{2}}$ & \textcolor{blue}{$\Sigma_4$} & & \\ \hline
\end{tabular}
%}
\end{center}
\caption{Action of the generators of $\mathfrak{sl}(3,\mathbb{R})$ on the {$\mathbf{3}^{\prime }$}. The blue entries contains an ambiguity since the action is defined up to a sign that could be arbitrarily fixed. Once these ambiguities are fixed the action on the all the other weights is uniquely fixed.}
\label{3sl3action1}
\end{table}

\FloatBarrier
\subsection{$\mathbf{20}$ of $SL(6,\mathbb{R})$ in $D=4$}\label{appendix:20sl6}
\FloatBarrier
In order to compute the orbits of one or bound states of weights the first thing we have to do is to fix the ambiguities in the structure constants of the algebra. In particular we do this as follows. We define the signs of the \textit{extraspecial pairs}\footnote{Extraspecial pairs are pairs of root vectors $(\alpha, \beta)$ for which, in the Chevalley basis, the signs of the structure constats appearing in commutators $[E_{\alpha}, E_{\beta}]=N_{\alpha\beta}E_{\alpha+\beta}$ could be chosen arbitrarily. } (\textit{cfr. e.g.} \cite{Marrani-Riccioni-Romano}, and Refs. therein) in such a way to have $\theta E_{\alpha}=-E_{\theta \alpha }$ for all the raising and lowering operators and this relation holds only if the structure constants satisfy
\begin{equation}
N_{\alpha ,\beta }=-N_{\theta \alpha ,\theta \beta }.
\end{equation}
In the split case (as it is the case for $\mathfrak{sl}(6,\mathbb{R})$) the previous relation automatically holds, since from $\theta \alpha =-\alpha $ we get
\begin{equation}
N_{\alpha ,\beta }=-N_{-\alpha ,-\beta },
\end{equation}%
that is the basic relation satisfied by the structure constants.\\

The structure constants for $\mathfrak{sl}(6,\mathbb{R})$ are listed in \autoref{sl6structconst}. In order to study the orbits, one must define the action of the generators on this representation. We note from  \autoref{20sl6dynkintree} that all loops are made of simple roots such that their sum is never a root; this means that one can set all the arbitrariness associated to simple root vectors to one, and the others follow directly from the structure constants.
\begin{table}[h]
\renewcommand{\arraystretch}{1.8}
\par
\begin{center}
\resizebox{\textwidth}{!}{
  \begin{tabular}{|c|c|c|c|c|c|c|c|c|c|c|c|c|c|c|c|c|c|c|c|c|c|c|c|c|c|c|c|c|c|c|c|} \hline
\centering\textcolor{black}{$N_{\alpha,\beta}$}&\multicolumn{31}{|c|}{\textcolor{black}{$\alpha$}}\\ \cline{1-32}
& &\rotatebox{90}{( 1 0 0 0 0 )} &\rotatebox{90}{( 0 1 0 0 0 )} &\rotatebox{90}{( 0 0 1 0 0 )} &\rotatebox{90}{( 0 0 0 1 0 )} &\rotatebox{90}{( 0 0 0 0 1 )} &\rotatebox{90}{( 1 1 0 0 0 )} &\rotatebox{90}{( 0 1 1 0 0 )} &\rotatebox{90}{( 0 0 1 1 0 )} &\rotatebox{90}{( 0 0 0 1 1 )} &\rotatebox{90}{( 1 1 1 0 0 )} &\rotatebox{90}{( 0 1 1 1 0 )} &\rotatebox{90}{( 0 0 1 1 1 )} &\rotatebox{90}{( 1 1 1 1 0 )} &\rotatebox{90}{( 0 1 1 1 1 )} &\rotatebox{90}{( 1 1 1 1 1 )} &\rotatebox{90}{( -1 0 0 0 0 )} &\rotatebox{90}{( 0 -1 0 0 0 )} &\rotatebox{90}{( 0 0 -1 0 0 )} &\rotatebox{90}{( 0 0 0 -1 0 )} &\rotatebox{90}{( 0 0 0 0 -1 )} &\rotatebox{90}{( -1 -1 0 0 0 )} &\rotatebox{90}{( 0 -1 -1 0 0 )} &\rotatebox{90}{( 0 0 -1 -1 0 )} &\rotatebox{90}{( 0 0 0 -1 -1 )} &\rotatebox{90}{( -1 -1 -1 0 0 )} &\rotatebox{90}{( 0 -1 -1 -1 0 )} &\rotatebox{90}{( 0 0 -1 -1 -1 )} &\rotatebox{90}{( -1 -1 -1 -1 0 )} &\rotatebox{90}{( 0 -1 -1 -1 -1 )} &\rotatebox{90}{( -1 -1 -1 -1 -1 )}\tabularnewline  \cline{2-32}
&( 1 0 0 0 0 )&\cellcolor{black}&1 &0 &0 &0 &0 &-1 &0 &0 &0 &-1 &0 &0 &-1 &0 &\cellcolor{black}&0 &0 &0 &0 &-1 &0 &0 &0 &1 &0 &0 &1 &0 &1 \tabularnewline \cline{2-32}

&( 0 1 0 0 0 )&\cellcolor{gray!40}-1 &\cellcolor{black}&1 &0 &0 &0 &0 &-1 &0 &0 &0 &-1 &0 &0 &0 &0 &\cellcolor{black}&0 &0 &0 &1 &-1 &0 &0 &0 &1 &0 &0 &1 &0 \tabularnewline \cline{2-32}

&( 0 0 1 0 0 )&0 &\cellcolor{gray!40}-1 &\cellcolor{black}&-1 &0 &\cellcolor{gray!40}1 &0 &0 &1 &0 &0 &0 &0 &0 &0 &0 &0 &\cellcolor{black}&0 &0 &0 &1 &1 &0 &-1 &0 &-1 &0 &0 &0 \tabularnewline \cline{2-32}

&( 0 0 0 1 0 )&0 &0 &\cellcolor{gray!40}1 &\cellcolor{black}&-1 &0 &\cellcolor{gray!40}-1 &0 &0 &\cellcolor{gray!40}-1 &0 &0 &0 &0 &0 &0 &0 &0 &\cellcolor{black}&0 &0 &0 &-1 &1 &0 &1 &0 &1 &0 &0 \tabularnewline \cline{2-32}

&( 0 0 0 0 1 )&0 &0 &0 &\cellcolor{gray!40}1 &\cellcolor{black}&0 &0 &\cellcolor{gray!40}-1 &0 &0 &\cellcolor{gray!40}-1 &0 &\cellcolor{gray!40}-1 &0 &0 &0 &0 &0 &0 &\cellcolor{black}&0 &0 &0 &-1 &0 &0 &1 &0 &1 &1 \tabularnewline \cline{2-32}

&( 1 1 0 0 0 )&0 &0 &-1 &0 &0 &\cellcolor{black}&0 &1 &0 &0 &0 &1 &0 &0 &0 &-1 &1 &0 &0 &0 &\cellcolor{black}&0 &0 &0 &1 &0 &0 &-1 &0 &-1 \tabularnewline \cline{2-32}

&( 0 1 1 0 0 )&1 &0 &0 &1 &0 &0 &\cellcolor{black}&0 &-1 &0 &0 &0 &0 &0 &0 &0 &-1 &1 &0 &0 &0 &\cellcolor{black}&0 &0 &-1 &-1 &0 &0 &1 &0 \tabularnewline \cline{2-32}

&( 0 0 1 1 0 )&0 &1 &0 &0 &1 &-1 &0 &\cellcolor{black}&0 &0 &0 &0 &0 &0 &0 &0 &0 &1 &-1 &0 &0 &0 &\cellcolor{black}&0 &0 &-1 &-1 &1 &0 &0 \tabularnewline \cline{2-32}

&( 0 0 0 1 1 )&0 &0 &-1 &0 &0 &0 &1 &0 &\cellcolor{black}&1 &0 &0 &0 &0 &0 &0 &0 &0 &1 &-1 &0 &0 &0 &\cellcolor{black}&0 &0 &1 &0 &-1 &-1 \tabularnewline \cline{2-32}

&( 1 1 1 0 0 )&0 &0 &0 &1 &0 &0 &0 &0 &-1 &\cellcolor{black}&0 &0 &0 &0 &0 &1 &0 &-1 &0 &0 &1 &-1 &0 &0 &\cellcolor{black}&0 &0 &-1 &0 &1 \tabularnewline \cline{2-32}

&( 0 1 1 1 0 )&1 &0 &0 &0 &1 &0 &0 &0 &0 &0 &\cellcolor{black}&0 &0 &0 &0 &0 &1 &0 &1 &0 &0 &-1 &-1 &0 &0 &\cellcolor{black}&0 &-1 &-1 &0 \tabularnewline \cline{2-32}

&( 0 0 1 1 1 )&0 &1 &0 &0 &0 &-1 &0 &0 &0 &0 &0 &\cellcolor{black}&0 &0 &0 &0 &0 &-1 &0 &1 &0 &0 &-1 &1 &0 &0 &\cellcolor{black}&0 &-1 &1 \tabularnewline \cline{2-32}

&( 1 1 1 1 0 )&0 &0 &0 &0 &1 &0 &0 &0 &0 &0 &0 &0 &\cellcolor{black}&0 &0 &1 &0 &0 &1 &0 &-1 &0 &1 &0 &-1 &-1 &0 &\cellcolor{black}&0 &-1 \tabularnewline \cline{2-32}

&( 0 1 1 1 1 )&1 &0 &0 &0 &0 &0 &0 &0 &0 &0 &0 &0 &0 &\cellcolor{black}&0 &0 &1 &0 &0 &1 &0 &1 &0 &-1 &0 &-1 &-1 &0 &\cellcolor{black}&-1 \tabularnewline \cline{2-32}

&( 1 1 1 1 1 )&0 &0 &0 &0 &0 &0 &0 &0 &0 &0 &0 &0 &0 &0 &\cellcolor{black}&1 &0 &0 &0 &1 &-1 &0 &0 &-1 &1 &0 &1 &-1 &-1 &\cellcolor{black}\tabularnewline \cline{2-32}

&( -1 0 0 0 0 )&\cellcolor{black}&0 &0 &0 &0 &1 &0 &0 &0 &-1 &0 &0 &-1 &0 &-1 &\cellcolor{black}&-1 &0 &0 &0 &0 &1 &0 &0 &0 &1 &0 &0 &1 &0 \tabularnewline \cline{2-32}

&( 0 -1 0 0 0 )&0 &\cellcolor{black}&0 &0 &0 &-1 &1 &0 &0 &0 &-1 &0 &0 &-1 &0 &1 &\cellcolor{black}&-1 &0 &0 &0 &0 &1 &0 &0 &0 &1 &0 &0 &0 \tabularnewline \cline{2-32}

&( 0 0 -1 0 0 )&0 &0 &\cellcolor{black}&0 &0 &0 &-1 &-1 &0 &1 &0 &1 &0 &0 &0 &0 &1 &\cellcolor{black}&1 &0 &-1 &0 &0 &-1 &0 &0 &0 &0 &0 &0 \tabularnewline \cline{2-32}

&( 0 0 0 -1 0 )&0 &0 &0 &\cellcolor{black}&0 &0 &0 &1 &-1 &0 &-1 &0 &-1 &0 &0 &0 &0 &-1 &\cellcolor{black}&1 &0 &1 &0 &0 &1 &0 &0 &0 &0 &0 \tabularnewline \cline{2-32}

&( 0 0 0 0 -1 )&0 &0 &0 &0 &\cellcolor{black}&0 &0 &0 &1 &0 &0 &-1 &0 &-1 &-1 &0 &0 &0 &-1 &\cellcolor{black}&0 &0 &1 &0 &0 &1 &0 &1 &0 &0 \tabularnewline \cline{2-32}

&( -1 -1 0 0 0 )&1 &-1 &0 &0 &0 &\cellcolor{black}&0 &0 &0 &-1 &0 &0 &1 &0 &1 &0 &0 &1 &0 &0 &\cellcolor{black}&0 &-1 &0 &0 &0 &-1 &0 &0 &0 \tabularnewline \cline{2-32}

&( 0 -1 -1 0 0 )&0 &1 &-1 &0 &0 &0 &\cellcolor{black}&0 &0 &1 &1 &0 &0 &-1 &0 &-1 &0 &0 &-1 &0 &0 &\cellcolor{black}&0 &1 &0 &0 &0 &0 &0 &0 \tabularnewline \cline{2-32}

&( 0 0 -1 -1 0 )&0 &0 &-1 &1 &0 &0 &0 &\cellcolor{black}&0 &0 &1 &1 &-1 &0 &0 &0 &-1 &0 &0 &-1 &1 &0 &\cellcolor{black}&0 &0 &0 &0 &0 &0 &0 \tabularnewline \cline{2-32}

&( 0 0 0 -1 -1 )&0 &0 &0 &-1 &1 &0 &0 &0 &\cellcolor{black}&0 &0 &-1 &0 &1 &1 &0 &0 &1 &0 &0 &0 &-1 &0 &\cellcolor{black}&-1 &0 &0 &0 &0 &0 \tabularnewline \cline{2-32}

&( -1 -1 -1 0 0 )&-1 &0 &1 &0 &0 &-1 &1 &0 &0 &\cellcolor{black}&0 &0 &1 &0 &-1 &0 &0 &0 &-1 &0 &0 &0 &0 &1 &\cellcolor{black}&0 &0 &0 &0 &0 \tabularnewline \cline{2-32}

&( 0 -1 -1 -1 0 )&0 &-1 &0 &-1 &0 &0 &1 &1 &0 &0 &\cellcolor{black}&0 &1 &1 &0 &-1 &0 &0 &0 &-1 &0 &0 &0 &0 &0 &\cellcolor{black}&0 &0 &0 &0 \tabularnewline \cline{2-32}

&( 0 0 -1 -1 -1 )&0 &0 &1 &0 &-1 &0 &0 &1 &-1 &0 &0 &\cellcolor{black}&0 &1 &-1 &0 &-1 &0 &0 &0 &1 &0 &0 &0 &0 &0 &\cellcolor{black}&0 &0 &0 \tabularnewline \cline{2-32}

&( -1 -1 -1 -1 0 )&-1 &0 &0 &-1 &0 &1 &0 &-1 &0 &1 &1 &0 &\cellcolor{black}&0 &1 &0 &0 &0 &0 &-1 &0 &0 &0 &0 &0 &0 &0 &\cellcolor{black}&0 &0 \tabularnewline \cline{2-32}

&( 0 -1 -1 -1 -1 )&0 &-1 &0 &0 &-1 &0 &-1 &0 &1 &0 &1 &1 &0 &\cellcolor{black}&1 &-1 &0 &0 &0 &0 &0 &0 &0 &0 &0 &0 &0 &0 &\cellcolor{black}&0 \tabularnewline \cline{2-32}

\multirow{-32}{*}{\rotatebox{90}{\textcolor{black}{$\beta$}}}&( -1 -1 -1 -1 -1 )&-1 &0 &0 &0 &-1 &1 &0 &0 &1 &-1 &0 &-1 &1 &1 &\cellcolor{black}&0 &0 &0 &0 &0 &0 &0 &0 &0 &0 &0 &0 &0 &0 &\cellcolor{black}\tabularnewline \cline{2-32}

\hline\end{tabular}}
\end{center}
\caption[Structure constants of $\mathfrak{a}5$]{Structure constants of $\mathfrak{a}5$. The grey-shaded boxes denote \textit{extraspecial pairs} (%
\textit{cfr. e.g.} \protect\cite{Marrani-Riccioni-Romano}, and Refs. therein).}
\label{sl6structconst}
\end{table}\\.
Once the action of each generator on each weight has been defined one can just define the set of stabilizers taking the most general linear combination of generators and acting on a given state. Doing this in the 3-weights case presented in \autoref{section:20ofsl6} we obtain the  semisimple part of stabilizer and define the following identification
\begin{subequations}
\begin{align}
H_{\alpha _{2}}-H_{\alpha _{4}}&\rightarrow H_{\frac{1}{2}\left( \alpha_{2}-\alpha _{4}\right) }, \\
H_{\alpha _{1}}-H_{\alpha _{5}}&\rightarrow H_{\frac{1}{2}\left( \alpha_{1}-\alpha _{5}\right) }, \\
E_{\alpha _{1}+\alpha _{2}+\alpha _{3}}\mp E_{-\left( \alpha _{3}+\alpha_{4}+\alpha _{5}\right) }&\rightarrow E_{\frac{1}{2}\left( \alpha _{1}+\alpha_{2}-\alpha _{4}-\alpha _{5}\right) }, \\
E_{\alpha _{2}+\alpha _{3}+\alpha _{4}}\mp E_{-\left( \alpha _{1}+\alpha_{2}+\alpha _{3}\right) }&\rightarrow E_{\frac{1}{2}\left( \alpha _{4}+\alpha_{5}-\alpha _{1}-\alpha _{2}\right) }, \\
E_{\alpha _{5}}\mp E_{-\alpha _{1}}&\rightarrow E_{\frac{1}{2}\left( \alpha_{5}-\alpha _{1}\right) }, \\
E_{\alpha _{1}}\mp E_{-\alpha _{5}}&\rightarrow E_{\frac{1}{2}\left( \alpha_{1}-\alpha _{5}\right) }, \\
E_{\alpha _{2}+\alpha _{3}}-E_{-\left( \alpha _{3}+\alpha _{4}\right)}&\rightarrow E_{\frac{1}{2}\left( \alpha _{2}-\alpha _{4}\right) }, \\
E_{\alpha _{3}+\alpha _{4}}-E_{-\left( \alpha _{2}+\alpha _{3}\right)}&\rightarrow E_{\frac{1}{2}\left( \alpha _{4}-\alpha _{2}\right) },
\end{align}
\end{subequations}
This identification makes manifest that the semisimple part of the stabilizer corresponds to an  algebra $\mathfrak{sl}(3,\mathbb{R})$
with simple roots $\frac{\alpha _{2}-\alpha _{4}}{2}$ and $\frac{\alpha _{1}-\alpha _{5}}{2}$.\\

For the 4-weight orbit we could proceed by the same way above obtaining for the linear combinations discussed in \autoref{section:20ofsl6} a set of  stabilizers parameterized by the variables $a,b$, The complexification of the stabilizing algebra gives an $\mathfrak{sl}(3,\mathbb{C})$ with generators
\begin{subequations}
\begin{align}
H_{\beta _{1}}&=\frac{1}{2}\left[ H_{\alpha _{1}}-H_{\alpha _{5}}+\sqrt{-ab}\left( F_{\alpha _{2}+\alpha _{3}+\alpha _{4}}^{-ab}-aF_{\alpha _{1}+\alpha_{2}+\alpha _{3}+\alpha _{4}+\alpha _{5}}^{-ab}\right) \right] ; \\
H_{\beta _{2}}&=\frac{1}{2}\left[ H_{\alpha _{2}}-H_{\alpha _{4}}-\sqrt{-ab}\left( F_{\alpha _{3}}^{-ab}+abF_{\alpha _{2}+\alpha _{3}+\alpha_{4}}^{-ab}\right) \right] , \\
H_{\beta _{3}}&=\frac{1}{2}\left[ H_{\alpha _{4}}-H_{\alpha _{2}}-\sqrt{-ab}\left( F_{\alpha _{3}}^{-ab}+abF_{\alpha _{2}+\alpha _{3}+\alpha_{4}}^{-ab}\right) \right] , \\
H_{\beta _{4}}&=\frac{1}{2}\left[ H_{\alpha _{5}}-H_{\alpha _{1}}+\sqrt{-ab}\left( F_{\alpha _{2}+\alpha _{3}+\alpha _{4}}^{-ab}-aF_{\alpha _{1}+\alpha_{2}+\alpha _{3}+\alpha _{4}+\alpha _{5}}^{-ab}\right) \right] , \\
E_{\beta _{1}}&=E_{\alpha _{1}}-aE_{-\alpha _{5}}-\sqrt{-ab}\left( E_{\alpha_{1}+\alpha _{2}+\alpha _{3}+\alpha _{4}}-bE_{-\alpha _{2}-\alpha_{3}-\alpha _{4}-\alpha _{5}}\right) , \\
E_{\beta _{2}}&=E_{\alpha _{2}+\alpha _{3}}-E_{-\alpha _{3}-\alpha _{4}}+\sqrt{-ab}\left( E_{\alpha _{2}}-abE_{-\alpha _{4}}\right) , \\
E_{\beta _{3}}&=E_{\alpha _{3}+\alpha _{4}}-E_{-\alpha _{2}-\alpha _{3}}+\sqrt{-ab}\left( E_{\alpha _{4}}-abE_{-\alpha _{2}}\right) , \\
E_{\beta _{4}}&=E_{\alpha _{5}}-aE_{-\alpha _{1}}-\sqrt{-ab}\left( E_{\alpha_{2}+\alpha _{3}+\alpha _{4}+\alpha _{5}}-bE_{-\alpha _{1}-\alpha_{2}-\alpha _{3}-\alpha _{4}}\right) , \\
E_{-\beta _{1}}&=E_{\alpha _{5}}-aE_{-\alpha _{1}}+\sqrt{-ab}\left( E_{\alpha_{2}+\alpha _{3}+\alpha _{4}+\alpha _{5}}-bE_{-\alpha _{1}-\alpha_{2}-\alpha _{3}-\alpha _{4}}\right) , \\
E_{-\beta _{2}}&=E_{\alpha _{3}+\alpha _{4}}-E_{-\alpha _{2}-\alpha _{3}}-\sqrt{-ab}\left( E_{\alpha _{4}}-abE_{-\alpha _{2}}\right) , \\
E_{-\beta _{3}}&=E_{\alpha _{2}+\alpha _{3}}-E_{-\alpha _{3}-\alpha _{4}}-\sqrt{-ab}\left( E_{\alpha _{2}}-abE_{-\alpha _{4}}\right) , \\
E_{-\beta _{4}}&=E_{\alpha _{1}}-aE_{-\alpha _{5}}+\sqrt{-ab}\left( E_{\alpha_{1}+\alpha _{2}+\alpha _{3}+\alpha _{4}}-bE_{-\alpha _{2}-\alpha_{3}-\alpha _{4}-\alpha _{5}}\right) , \\
E_{\beta _{1}+\beta _{2}}&=E_{\alpha _{1}+\alpha _{2}+\alpha_{3}}-aE_{-\alpha _{3}-\alpha _{4}-\alpha _{5}}-\sqrt{-ab}\left( E_{\alpha_{1}+\alpha _{2}}-bE_{-\alpha _{4}-\alpha _{5}}\right) , \\
E_{\beta _{3}+\beta _{4}}&=E_{\alpha _{3}+\alpha _{4}+\alpha_{5}}-aE_{-\alpha _{1}-\alpha _{2}-\alpha _{3}}-\sqrt{-ab}\left( E_{\alpha_{4}+\alpha _{5}}-bE_{-\alpha _{1}-\alpha _{2}}\right) , \\
E_{-\beta _{1}-\beta _{2}}&=E_{\alpha _{3}+\alpha _{4}+\alpha_{5}}-aE_{-\alpha _{1}-\alpha _{2}-\alpha _{3}}+\sqrt{-ab}\left( E_{\alpha_{4}+\alpha _{5}}-bE_{-\alpha _{1}-\alpha _{2}}\right) , \\
E_{-\beta _{3}-\beta _{4}}&=E_{\alpha _{1}+\alpha _{2}+\alpha_{3}}-aE_{-\alpha _{3}-\alpha _{4}-\alpha _{5}}+\sqrt{-ab}\left( E_{\alpha_{1}+\alpha _{2}}-bE_{-\alpha _{4}-\alpha _{5}}\right)
\end{align}
\end{subequations}
and simple roots $\beta_{1}, \beta_{2}, \beta_{3} $ and $\beta_{4}$. The two resulting real forms $\mathfrak{sl}(3,\mathbb{C})_{\mathbb{R}}$ and $\mathfrak{sl}(3,\mathbb{R})\oplus \mathfrak{sl}(3,\mathbb{R})$ of $\mathfrak{sl}(3,\mathbb{C})$ have the same signature, but they are discriminated by looking at the imaginary units appearing in the Chevalley basis (in particular, for $\mathfrak{sl}(3,\mathbb{C})_{\mathbb{R}}$ there are no imaginary units in the stabilizing algebra).
\FloatBarrier

\subsection{$\mathbf{32}$ of $SO(6,6)$ in $D=4$}\label{appendix:32ofso(6,6)}
\FloatBarrier
In this subsection we dewcribe in detail the structure of the orbits discussed in \autoref{section:32ofso(6,6)}, in particular identifying explicitly the stabilizers. We refer to \autoref{fig:32ofso(6,6)dynkinlabels} and \autoref{fig:32ofso(6,6)orbits} for the weight labels. In \autoref{tab:stabilizer12-32so12} and \autoref{tab:stabilizer34-32so12} we have identified explicitly the stabilizers for the weights $\Lambda_{1}, \Lambda_{2}, \Lambda_{3}$ and $\Lambda_{4}$. From this one can immediately deduce the 1-weight orbit. \\

To construct the stabilizers for multi-weight states we need to identify generators of the algebras annihilating at the same time more than one of weights we have chosen as our single brane. The sets of the common stabilizers appear in \autoref{tab:commonstabilizer-32so12}. It is already interesting to note that for the four weights the common stabilizers form an $[\mathfrak{sl}(2,\mathbb{R})]^3$ algebra. \\

Finally in order to define the full sets of stabilizers for the different bound states we should consider the conjunction stabilizers. These are studied for the general combination $\Lambda_{1}+\Lambda_{2}+a\Lambda_{3}+b\Lambda_{4}$ (where $a,b=\pm 1$) are listed in \autoref{tab:conjunction1-32so12} and \autoref{tab:conjunction2-32so12}.
Collecting all the information we have defined the orbits appearing in \autoref{section:32ofso(6,6)}.
\begin{sidewaystable}[h!]
\renewcommand{\arraystretch}{1.6}
\begin{center}
\resizebox{\textwidth}{!}{
\begin{tabular}{|c|c|}
\hline
\multicolumn{2}{|c|}{\textbf{Stabilizers}}\\
\hline
$\Lambda_{1}(=[-\Sigma_{28}])$&$\Lambda_{2}$\\ \hline

$\alpha_1+2\alpha_2+2\alpha_3+2\alpha_4+\alpha_5+\alpha_6$&$\alpha_1+2\alpha_2+2\alpha_3+2\alpha_4+\alpha_5+\alpha_6$\\
$\alpha_1+\alpha_2+2\alpha_3+2\alpha_4+\alpha_5+\alpha_6$&$\alpha_1+\alpha_2+2\alpha_3+2\alpha_4+\alpha_5+\alpha_6$\\
$\alpha_1+\alpha_2+\alpha_3+2\alpha_4+\alpha_5+\alpha_6\quad \alpha_2+2\alpha_3+2\alpha_4+\alpha_5+\alpha_6$&$\alpha_1+\alpha_2+\alpha_3+2\alpha_4+\alpha_5+\alpha_6\quad \alpha_2+2\alpha_3+2\alpha_4+\alpha_5+\alpha_6$\\

$\alpha_1+\alpha_2+\alpha_3+\alpha_4+\alpha_5+\alpha_6\quad \alpha_2+\alpha_3+2\alpha_4+\alpha_5+\alpha_6$&$\alpha_1+\alpha_2+\alpha_3+\alpha_4+\alpha_5+\alpha_6\quad \alpha_2+\alpha_3+2\alpha_4+\alpha_5+\alpha_6$\\

$\alpha_1+\alpha_2+\alpha_3+\alpha_4+\alpha_5\quad \alpha_2+\alpha_3+\alpha_4+\alpha_5+\alpha_6\quad \alpha_1+\alpha_2+\alpha_3+\alpha_4+\alpha_6\quad \alpha_3+2\alpha_4+\alpha_5+\alpha_6$&$\alpha_1+\alpha_2+\alpha_3+\alpha_4+\alpha_5\quad \alpha_2+\alpha_3+\alpha_4+\alpha_5+\alpha_6\quad \alpha_1+\alpha_2+\alpha_3+\alpha_4+\alpha_6$\\

$\alpha_1+\alpha_2+\alpha_3+\alpha_4\quad \alpha_2+\alpha_3+\alpha_4+\alpha_5\quad \alpha_3+\alpha_4+\alpha_5+\alpha_6\quad \alpha_2+\alpha_3+\alpha_4+\alpha_6$&$\alpha_1+\alpha_2+\alpha_3+\alpha_4\quad \alpha_2+\alpha_3+\alpha_4+\alpha_5\quad \alpha_2+\alpha_3+\alpha_4+\alpha_6$\\

$\alpha_1+\alpha_2+\alpha_3\quad \alpha_2+\alpha_3+\alpha_4\quad \alpha_3+\alpha_4+\alpha_5\quad \alpha_4+\alpha_5+\alpha_6\quad \alpha_3+\alpha_4+\alpha_6$&$\alpha_1+\alpha_2+\alpha_3\quad \alpha_2+\alpha_3+\alpha_4\quad \alpha_3+\alpha_4+\alpha_5$\\

$\alpha_1+\alpha_2\quad\alpha_2+\alpha_3\quad \alpha_3+\alpha_4\quad \alpha_4+\alpha_5\quad \alpha_4+\alpha_6$&$\alpha_1+\alpha_2\quad\alpha_2+\alpha_3\quad \alpha_3+\alpha_4\quad \alpha_4+\alpha_5$\\

$\alpha_1\quad\alpha_2\quad\alpha_3\quad \alpha_4\quad \alpha_5\quad \alpha_6$&$\alpha_1\quad\alpha_2\quad\alpha_3\quad \alpha_4\quad \alpha_5$\\

$H_{\alpha_1}\quad H_{\alpha_2}\quad H_{\alpha_3}\quad H_{\alpha_4}\quad H_{\alpha_5}$&$H_{\alpha_1}\quad H_{\alpha_3}\quad H_{\alpha_4}\quad H_{\alpha_5}\quad  H_{\alpha_2}+H_{\alpha_6}$\\

%%%%%%%%%%%%%%%%%%%%%%%%%%%%%%%%%%%%%%%%%%%%%%%%%5

$-\alpha_1\quad -\alpha_2\quad -\alpha_3\quad - \alpha_4\quad - \alpha_5$&$-\alpha_1\quad -\alpha_3\quad - \alpha_4\quad - \alpha_5\quad - \alpha_6$\\

$-\alpha_1-\alpha_2\quad -\alpha_2-\alpha_3\quad - \alpha_3-\alpha_4\quad - \alpha_4-\alpha_5$&$ - \alpha_3-\alpha_4\quad - \alpha_4-\alpha_5\quad - \alpha_4-\alpha_6$\\

$-\alpha_1-\alpha_2-\alpha_3\quad - \alpha_2-\alpha_3-\alpha_4\quad - \alpha_3-\alpha_4-\alpha_5$&$ - \alpha_3-\alpha_4-\alpha_5\quad - \alpha_4-\alpha_5-\alpha_6\quad - \alpha_3-\alpha_4-\alpha_6$\\

$-\alpha_1-\alpha_2-\alpha_3-\alpha_4\quad - \alpha_2-\alpha_3-\alpha_4-\alpha_5$&$ - \alpha_3-\alpha_4-\alpha_5-\alpha_6\quad - \alpha_2-\alpha_3-\alpha_4-\alpha_6$\\

$-\alpha_1-\alpha_2-\alpha_3-\alpha_4-\alpha_5$&$ - \alpha_2-\alpha_3-\alpha_4-\alpha_5-\alpha_6\quad - \alpha_1-\alpha_2-\alpha_3-\alpha_4-\alpha_6\quad - \alpha_3-2\alpha_4-\alpha_5-\alpha_6$\\

&$-\alpha_1-\alpha_2-\alpha_3-\alpha_4-\alpha_5-\alpha_6\quad - \alpha_2-\alpha_3-2\alpha_4-\alpha_5-\alpha_6$\\

&$-\alpha_1-\alpha_2-\alpha_3-2\alpha_4-\alpha_5-\alpha_6\quad - \alpha_2-2\alpha_3-2\alpha_4-\alpha_5-\alpha_6$\\

&$-\alpha_1-\alpha_2-2\alpha_3-2\alpha_4-\alpha_5-\alpha_6$\\

&\\
\hline
\end{tabular}
}
\caption{Stabilizer for the weights $\Lambda_{1}$ and $\Lambda_{2}$ in the ${\bf 32}$ of $\mathfrak{so}(6,6)$. In the first column we report $\Lambda_{1}(=[-\Sigma_{28}])$ meaning that $\Sigma_{28}$, the lowest weight, has the same stabilizers of the $\Lambda_1$ with different simple root signs. The dyonic orbit is built as combination of these two weights.}\label{tab:stabilizer12-32so12}
\end{center}
\end{sidewaystable}
\begin{sidewaystable}[h!]
\renewcommand{\arraystretch}{1.6}
\begin{center}
\resizebox{\textwidth}{!}{
\begin{tabular}{|c|c|}
\hline
\multicolumn{2}{|c|}{\textbf{Stabilizers}}\\
\hline
$\Lambda_{3}$&$\Lambda_4$\\ \hline

&\\

$\alpha_1+\alpha_2+2\alpha_3+2\alpha_4+\alpha_5+\alpha_6$&\\

$\alpha_1+\alpha_2+\alpha_3+2\alpha_4+\alpha_5+\alpha_6\quad \alpha_2+2\alpha_3+2\alpha_4+\alpha_5+\alpha_6$&\\

$\alpha_2+\alpha_3+2\alpha_4+\alpha_5+\alpha_6$&$\alpha_1+\alpha_2+\alpha_3+\alpha_4+\alpha_5+\alpha_6$\\

$\alpha_1+\alpha_2+\alpha_3+\alpha_4+\alpha_5\quad \alpha_3+2\alpha_4+\alpha_5+\alpha_6$&$ \alpha_2+\alpha_3+\alpha_4+\alpha_5+\alpha_6\quad \alpha_1+\alpha_2+\alpha_3+\alpha_4+\alpha_6$\\

$\alpha_1+\alpha_2+\alpha_3+\alpha_4\quad \alpha_2+\alpha_3+\alpha_4+\alpha_5\quad \alpha_3+\alpha_4+\alpha_5+\alpha_6$&$ \alpha_3+\alpha_4+\alpha_5+\alpha_6\quad \alpha_2+\alpha_3+\alpha_4+\alpha_6$\\

$\alpha_2+\alpha_3+\alpha_4\quad \alpha_3+\alpha_4+\alpha_5\quad \alpha_4+\alpha_5+\alpha_6\quad \alpha_3+\alpha_4+\alpha_6$&$\alpha_1+\alpha_2+\alpha_3\quad \alpha_4+\alpha_5+\alpha_6\quad \alpha_3+\alpha_4+\alpha_6$\\

$\alpha_3+\alpha_4\quad \alpha_4+\alpha_5\quad \alpha_4+\alpha_6$&$\alpha_1+\alpha_2\quad\alpha_2+\alpha_3\quad \alpha_4+\alpha_6$\\

$\alpha_1\quad\alpha_3\quad \alpha_4\quad \alpha_5$&$\alpha_1\quad\alpha_2\quad\alpha_3\quad \alpha_5\quad \alpha_6$\\

$H_{\alpha_1}\quad H_{\alpha_2}+H_{\alpha_4}\quad H_{\alpha_3}\quad H_{\alpha_5}\quad H_{\alpha_4}+H_{\alpha_6}$&$H_{\alpha_1}\quad H_{\alpha_2}\quad H_{\alpha_3}\quad  H_{\alpha_5}\quad H_{\alpha_4}+H_{\alpha_6}$\\

%%%%%%%%%%%%%%%%%%%%%%%%%%%%%%%%%%%%%%%%%%%%%%%%%5

$-\alpha_1\quad -\alpha_2\quad -\alpha_3\quad - \alpha_5\quad - \alpha_6$&$-\alpha_1\quad -\alpha_2\quad -\alpha_3\quad - \alpha_4\quad - \alpha_5$\\

$-\alpha_1-\alpha_2\quad -\alpha_2-\alpha_3\quad - \alpha_4-\alpha_6$&$-\alpha_1-\alpha_2\quad -\alpha_2-\alpha_3\quad - \alpha_3-\alpha_4\quad - \alpha_4-\alpha_5\quad - \alpha_4-\alpha_6$\\

$-\alpha_1-\alpha_2-\alpha_3\quad - \alpha_2-\alpha_3-\alpha_4\quad - \alpha_4-\alpha_5-\alpha_6\quad - \alpha_3-\alpha_4-\alpha_6$&$-\alpha_1-\alpha_2-\alpha_3\quad - \alpha_2-\alpha_3-\alpha_4\quad - \alpha_3-\alpha_4-\alpha_5\quad - \alpha_4-\alpha_5-\alpha_6\quad - \alpha_3-\alpha_4-\alpha_6$\\

$-\alpha_1-\alpha_2-\alpha_3-\alpha_4\quad - \alpha_2-\alpha_3-\alpha_4-\alpha_5\quad - \alpha_3-\alpha_4-\alpha_5-\alpha_6\quad - \alpha_2-\alpha_3-\alpha_4-\alpha_6$&$-\alpha_1-\alpha_2-\alpha_3-\alpha_4\quad - \alpha_2-\alpha_3-\alpha_4-\alpha_5\quad - \alpha_3-\alpha_4-\alpha_5-\alpha_6\quad - \alpha_2-\alpha_3-\alpha_4-\alpha_6$\\

$-\alpha_1-\alpha_2-\alpha_3-\alpha_4-\alpha_5\quad - \alpha_2-\alpha_3-\alpha_4-\alpha_5-\alpha_6\quad - \alpha_1-\alpha_2-\alpha_3-\alpha_4-\alpha_6$&$-\alpha_1-\alpha_2-\alpha_3-\alpha_4-\alpha_5\quad - \alpha_2-\alpha_3-\alpha_4-\alpha_5-\alpha_6\quad - \alpha_1-\alpha_2-\alpha_3-\alpha_4-\alpha_6\quad - \alpha_3-2\alpha_4-\alpha_5-\alpha_6$\\

$-\alpha_1-\alpha_2-\alpha_3-\alpha_4-\alpha_5-\alpha_6\quad - \alpha_2-\alpha_3-2\alpha_4-\alpha_5-\alpha_6$&$-\alpha_1-\alpha_2-\alpha_3-\alpha_4-\alpha_5-\alpha_6\quad - \alpha_2-\alpha_3-2\alpha_4-\alpha_5-\alpha_6$\\

$-\alpha_1-\alpha_2-\alpha_3-2\alpha_4-\alpha_5-\alpha_6\quad - \alpha_2-2\alpha_3-2\alpha_4-\alpha_5-\alpha_6$&$-\alpha_1-\alpha_2-\alpha_3-2\alpha_4-\alpha_5-\alpha_6\quad - \alpha_2-2\alpha_3-2\alpha_4-\alpha_5-\alpha_6$\\
$-\alpha_1-\alpha_2-2\alpha_3-2\alpha_4-\alpha_5-\alpha_6$&$-\alpha_1-\alpha_2-2\alpha_3-2\alpha_4-\alpha_5-\alpha_6$\\
$-\alpha_1-2\alpha_2-2\alpha_3-2\alpha_4-\alpha_5-\alpha_6$&$-\alpha_1-2\alpha_2-2\alpha_3-2\alpha_4-\alpha_5-\alpha_6$\\
\hline
\end{tabular}
}
\caption{Stabilizer for the weights $\Lambda_{3}$ and $\Lambda_{4}$ in the ${\bf 32}$ of $\mathfrak{so}(6,6)$}\label{tab:stabilizer34-32so12}
\end{center}
\end{sidewaystable}

\begin{sidewaystable}[h!]
\renewcommand{\arraystretch}{1.6}
\begin{center}
\resizebox{24cm}{!}{
\begin{tabular}{|c|c|c|}
\hline
\multicolumn{3}{|c|}{\textbf{Common Stabilizers}}\\
\hline
$\Lambda_{1}\Lambda_2$&$\Lambda_{1}\Lambda_2\Lambda_{3}$&$\Lambda_{1}\Lambda_2\Lambda_{3}\Lambda_4$\\ \hline

$\alpha_1+2\alpha_2+2\alpha_3+2\alpha_4+\alpha_5+\alpha_6$&&\\

$\alpha_1+\alpha_2+2\alpha_3+2\alpha_4+\alpha_5+\alpha_6$&$\alpha_1+\alpha_2+2\alpha_3+2\alpha_4+\alpha_5+\alpha_6$&\\

$\alpha_1+\alpha_2+\alpha_3+2\alpha_4+\alpha_5+\alpha_6\quad \alpha_2+2\alpha_3+2\alpha_4+\alpha_5+\alpha_6$&$\alpha_1+\alpha_2+\alpha_3+2\alpha_4+\alpha_5+\alpha_6\quad \alpha_2+2\alpha_3+2\alpha_4+\alpha_5+\alpha_6$&\\

$\alpha_1+\alpha_2+\alpha_3+\alpha_4+\alpha_5+\alpha_6\quad \alpha_2+\alpha_3+2\alpha_4+\alpha_5+\alpha_6$&$ \alpha_2+\alpha_3+2\alpha_4+\alpha_5+\alpha_6$&\\

$\alpha_1+\alpha_2+\alpha_3+\alpha_4+\alpha_5\quad \alpha_2+\alpha_3+\alpha_4+\alpha_5+\alpha_6\quad \alpha_1+\alpha_2+\alpha_3+\alpha_4+\alpha_6$&$\alpha_1+\alpha_2+\alpha_3+\alpha_4+\alpha_5$&\\

$\alpha_1+\alpha_2+\alpha_3+\alpha_4\quad \alpha_2+\alpha_3+\alpha_4+\alpha_5\quad \alpha_2+\alpha_3+\alpha_4+\alpha_6$&$\alpha_1+\alpha_2+\alpha_3+\alpha_4\quad \alpha_2+\alpha_3+\alpha_4+\alpha_5$&\\

$\alpha_1+\alpha_2+\alpha_3\quad \alpha_2+\alpha_3+\alpha_4\quad \alpha_3+\alpha_4+\alpha_5$&$\alpha_2+\alpha_3+\alpha_4\quad \alpha_3+\alpha_4+\alpha_5$&\\

$\alpha_1+\alpha_2\quad\alpha_2+\alpha_3\quad \alpha_3+\alpha_4\quad \alpha_4+\alpha_5$&$ \alpha_3+\alpha_4\quad \alpha_4+\alpha_5$&\\

$\alpha_1\quad\alpha_2\quad\alpha_3\quad \alpha_4\quad \alpha_5$&$\alpha_1\quad\alpha_3\quad \alpha_4\quad \alpha_5$&$\alpha_1\quad\alpha_3\quad \alpha_5$\\

$H_{\alpha_1}\quad H_{\alpha_3}\quad H_{\alpha_4}\quad H_{\alpha_5}$&$H_{\alpha_1}\quad H_{\alpha_3}\quad H_{\alpha_5}$&$H_{\alpha_1}\quad H_{\alpha_3}\quad H_{\alpha_5}$\\

%%%%%%%%%%%%%%%%%%%%%%%%%%%%%%%%%%%%%%%%%%%%%%%%%5

$-\alpha_1\quad -\alpha_3\quad - \alpha_4\quad - \alpha_5$&$-\alpha_1\quad -\alpha_3\quad - \alpha_5$&$-\alpha_1\quad -\alpha_3\quad - \alpha_5$\\

$ - \alpha_3-\alpha_4\quad - \alpha_4-\alpha_5$&&\\

$ - \alpha_3-\alpha_4-\alpha_5$&&\\

\hline
\end{tabular}
}
\end{center}
\caption{Common stabilizer in the ${\bf 32}$ of $\mathfrak{so}(6,6)$}\label{tab:commonstabilizer-32so12}
\end{sidewaystable}

\begin{table}[h!]
\renewcommand{\arraystretch}{1.6}
\begin{center}
%\resizebox{\textwidth}{!}{
\begin{tabular}{|c|c|c|}
\hline
\multicolumn{3}{|c|}{{\bf Conjunction Stabilizers}}\\
\hline\hline
$\Lambda_{1}+\Lambda_2$&$a\Lambda_{3}$&$b\Lambda_{4}$\\ \hline

$E_{\alpha_3+2\alpha_4+\alpha_5+\alpha_6}-E_{-\alpha_6}$&-&$b(\Sigma_{20}-\Sigma_{28})$\\
$E_{\alpha_3+\alpha_4+\alpha_5+\alpha_6}-E_{-\alpha_4-\alpha_6}$&-&-\\
$E_{\alpha_3+\alpha_4+\alpha_6}-E_{-\alpha_4-\alpha_5-\alpha_6}$&-&-\\
$E_{\alpha_4+\alpha_5+\alpha_6}-E_{-\alpha_3-\alpha_4-\alpha_6}$&-&-\\
$E_{\alpha_4+\alpha_6}-E_{-\alpha_3-\alpha_4-\alpha_5-\alpha_6}$&-&-\\
$E_{\alpha_6}-E_{-\alpha_3-2\alpha_4-\alpha_5-\alpha_6}$&$a(\Sigma_{20}-\Sigma_{28})$&-\\ \hline

$\Lambda_{1}+a\Lambda_3$&$\Lambda_{2}$&$b\Lambda_{4}$\\ \hline
$E_{\alpha_1+2\alpha_2+2\alpha_3+2\alpha_4+\alpha_5+\alpha_6}-aE_{-\alpha_6}$&-&$b\Sigma_{7}-ab\Sigma_{28}$\\
$E_{\alpha_1+\alpha_2+\alpha_3+\alpha_4+\alpha_5+\alpha_6}-aE_{-\alpha_2-\alpha_3-\alpha_4-\alpha_6}$&-&-\\
$E_{\alpha_2+\alpha_3+\alpha_4+\alpha_5+\alpha_6}-aE_{-\alpha_1-\alpha_2-\alpha_3-\alpha_4-\alpha_6}$&-&-\\
$E_{\alpha_2+\alpha_3+\alpha_4+\alpha_6}-aE_{-\alpha_1-\alpha_2-\alpha_3-\alpha_4-\alpha_5-\alpha_6}$&-&-\\
$E_{\alpha_6}-aE_{-\alpha_1-2\alpha_2-2\alpha_3-2\alpha_4-\alpha_5-\alpha_6}$&$\Sigma_{7}-a\Sigma_{28}$&-\\
$E_{\alpha_1+\alpha_2+\alpha_3+\alpha_4+\alpha_6}-aE_{-\alpha_2-\alpha_3-\alpha_4-\alpha_5-\alpha_6}$&-&-\\ \hline

$\Lambda_{1}+b\Lambda_4$&$\Lambda_{2}$&$a\Lambda_{3}$\\ \hline
$E_{\alpha_3+2\alpha_4+\alpha_5+\alpha_6}-bE_{-\alpha_1-2\alpha_2-2\alpha_3-2\alpha_4-\alpha_5-\alpha_6}$&$\Sigma_{1}-b\Sigma_{28}$&-\\
$E_{\alpha_2+\alpha_3+2\alpha_4+\alpha_5+\alpha_6}-bE_{-\alpha_1-\alpha_2-2\alpha_3-2\alpha_4-\alpha_5-\alpha_6}$&-&-\\
$E_{\alpha_1+\alpha_2+\alpha_3+2\alpha_4+\alpha_5+\alpha_6}-bE_{-\alpha_2-2\alpha_3-2\alpha_4-\alpha_5-\alpha_6}$&-&-\\
$E_{\alpha_2+2\alpha_3+2\alpha_4+\alpha_5+\alpha_6}-bE_{-\alpha_1-\alpha_2-\alpha_3-2\alpha_4-\alpha_5-\alpha_6}$&-&-\\
$E_{\alpha_1+\alpha_2+2\alpha_3+2\alpha_4+\alpha_5+\alpha_6}-bE_{-\alpha_2-\alpha_3-2\alpha_4-\alpha_5-\alpha_6}$&-&-\\
$E_{\alpha_1+2\alpha_2+2\alpha_3+2\alpha_4+\alpha_5+\alpha_6}-bE_{-\alpha_3-2\alpha_4-\alpha_5-\alpha_6}$&-&$a\Sigma_{1}-ab\Sigma_{28}$\\

\hline
\end{tabular}
%}
\end{center}
\caption{Conjunction stabilizers for the weights in the ${\bf 32}$ of $\mathfrak{so}(6,6)$. In the first column we list the combination of generators annihilating the corresponding pair of state, in the last two columns their effect on the other weights we have chosen to study the orbit.  }\label{tab:conjunction1-32so12}
\end{table}
\begin{table}
\renewcommand{\arraystretch}{1.6}
\begin{center}
%\resizebox{\textwidth}{!}{
\begin{tabular}{|c|c|c|}
\hline
\multicolumn{3}{|c|}{{\bf Conjunction Stabilizers}}\\
\hline\hline
$\Lambda_{2}+a\Lambda_{3}$&$\Lambda_{1}$&$b\Lambda_{4}$\\ \hline

$E_{\alpha_2}-aE_{-\alpha_1-\alpha_2-\alpha_3}$&-&-\\
$E_{\alpha_2+\alpha_3}-aE_{-\alpha_1-\alpha_2}$&-&-\\
$E_{\alpha_1+\alpha_2}-aE_{-\alpha_2-\alpha_3}$&-&-\\
$E_{\alpha_1+\alpha_2+\alpha_3}-aE_{-\alpha_2}$&-&-\\
$E_{\alpha_3+2\alpha_4+\alpha_5+\alpha_6}-aE_{\alpha_1+2\alpha_2+2\alpha_3+2\alpha_4+\alpha_5+\alpha_6}$&-&$b\Sigma_{20}-ab\Sigma_{7}$\\
$E_{-\alpha_1-2\alpha_2-2\alpha_3-2\alpha_4-\alpha_5-\alpha_6}-aE_{-\alpha_3-2\alpha_4-\alpha_5-\alpha_6}$&$\Sigma_{20}-a\Sigma_{7}$&-\\ \hline

$\Lambda_{2}+b\Lambda_{4}$&$\Lambda_{1}$&$a\Lambda_{3}$\\ \hline

$E_{\alpha_2+\alpha_3+\alpha_4}-bE_{-\alpha_1-\alpha_2-\alpha_3-\alpha_4-\alpha_5}$&-&-\\
$E_{\alpha_1+\alpha_2+\alpha_3+\alpha_4}-bE_{-\alpha_2-\alpha_3-\alpha_4-\alpha_5}$&-&-\\
$E_{\alpha_1+\alpha_2+\alpha_3+\alpha_4+\alpha_5}-E_{-\alpha_2-\alpha_3-\alpha_4}$&-&-\\
$E_{\alpha_2+\alpha_3+\alpha_4+\alpha_5}-bE_{-\alpha_1-\alpha_2-\alpha_3-\alpha_4}$&-&-\\
$E_{-\alpha_6}-bE_{-\alpha_1-2\alpha_2-2\alpha_3-2\alpha_4-\alpha_5-\alpha_6}$&$\Sigma_{1}-b\Sigma_{20}$&-\\
$E_{\alpha_6}-bE_{\alpha_1+2\alpha_2+2\alpha_3+2\alpha_4+\alpha_5+\alpha_6}$&-&$a\Sigma_{20}-ba\Sigma_{1}$\\ \hline

$a\Lambda_{3}+b\Lambda_{4}$&$\Lambda_{1}$&$\Lambda_{2}$\\ \hline

$aE_{-\alpha_6}-bE_{-\alpha_3-2\alpha_4-\alpha_5-\alpha_6}$&$a\Sigma_{1}-b\Sigma_{7}$&-\\
$aE_{\alpha_4}-bE_{-\alpha_3-\alpha_4-\alpha_5}$&-&-\\
$aE_{\alpha_3+\alpha_4}-bE_{-\alpha_4-\alpha_5}$&-&-\\
$aE_{\alpha_4+\alpha_5}-bE_{-\alpha_3-\alpha_4}$&-&-\\
$bE_{\alpha_3+\alpha_4+\alpha_5}-aE_{-\alpha_4}$&-&-\\
$aE_{\alpha_6}-bE_{\alpha_3+2\alpha_4+\alpha_5+\alpha_6}$&-&$a\Sigma_{7}-b\Sigma_{1}$\\

\hline
\end{tabular}
%}
\end{center}
\caption{Conjunction stabilizers for the weights in the ${\bf 32}$ of $\mathfrak{so}(6,6)$. In the first column we list the combination of generators annihilating the corresponding pair of state, in the last two columns their effect on the other weights we have chosen to study the orbit.}\label{tab:conjunction2-32so12}
\end{table}
\FloatBarrier


\begin{thebibliography}{99}
\bibitem{Gunaydin:1983rk} M.~G\"{u}naydin, G.~Sierra, P.~K.~Townsend,
\textit{Exceptional Supergravity Theories and the Magic Square}, Phys.\
Lett.\ \textbf{B133} (1983) 72.

\bibitem{Gunaydin:1983bi} M.~G\"{u}naydin, G.~Sierra, P.~K.~Townsend, T%
\textit{he Geometry of }$\mathcal{N}\mathit{=2}$\textit{\ Maxwell-Einstein
Supergravity and Jordan Algebras}, Nucl.\ Phys.\ \textbf{B242} (1984) 244.

\bibitem{Gunaydin:1984ak} M.~G\"{u}naydin,, G.~Sierra, P.~K.~Townsend,
\textit{Gauging the }$\mathit{d=5}$\textit{\ Maxwell-Einstein Supergravity
Theories: More on Jordan Algebras}, Nucl.\ Phys.\ \textbf{B253} (1985) 573.

\bibitem{Squaring-Magic} S.~L.~Cacciatori, B.~L.~Cerchiai, A.~Marrani,
\textit{Squaring the Magic},\ Adv.\ Theor.\ Math.\ Phys.\ \textbf{19} (2015)
923, \texttt{arXiv:1208.6153 [math-ph]}.

\bibitem{GZ} M. G\"{u}naydin, M. Zagermann, \textit{Unified Maxwell-Einstein
and Yang-Mills-Einstein supergravity theories in five-dimensions}, JHEP
\textbf{0307} (2003) 023, \texttt{hep-th/0304109}. M. G\"{u}naydin, M.
Zagermann, \textit{Unified }$\mathcal{N}\mathit{=2}$\textit{\
Maxwell-Einstein and Yang-Mills-Einstein supergravity theories in four
dimensions}, JHEP \textbf{0509} (2005) 026, \texttt{hep-th/0507227}.

\bibitem{Cremmer:1978ds} E.~Cremmer, B.~Julia, \textit{The }$\mathcal{N}%
\mathit{=8}$\textit{\ Supergravity Theory. 1. The Lagrangian}, Phys.\ Lett.\
\textbf{B80} (1978) 48.

\bibitem{Cremmer:1979up} E.~Cremmer, B.~Julia, \textit{The }$\mathit{SO(8)}$%
\textit{\ Supergravity}, Nucl.\ Phys.\ \textbf{B159} (1979) 141.

\bibitem{Hull:1994ys} C.~M.~Hull, P.~K.~Townsend, \textit{Unity of
superstring dualities}, Nucl.\ Phys.\ \textbf{B438} (1995) 109, \texttt{%
hep-th/9410167}.

\bibitem{FRT} H. Freudenthal, \textit{Lie groups in the foundations of
geometry},\ Adv. Math. \textbf{1} (1963) 145. J. Tits, \textit{Alg\`{e}bres
alternatives, alg\`{e}bres de Jordan et alg\`{e}bres de Lie exceptionnelles.
I. Construction},\ (in French), Nederl. Akad. Wetensch. Proc. Ser. \textbf{A
69} (1966) 223. B. A. Rozenfeld, \textit{Geometrical interpretation of the
compact simple Lie groups of the class }$\mathit{E}$,\ (in Russian), Dokl.
Akad. Nauk. SSSR \textbf{106} (1956) 600.

\bibitem{Ferrara:1997uz} S.~Ferrara, M.~G\"{u}naydin, \textit{Orbits of
exceptional groups, duality and BPS states in string theory}, Int.\ J.\
Mod.\ Phys.\ \textbf{A13} (1998) 2075, \texttt{hep-th/9708025}.

\bibitem{Borsten:2011ai} L.~Borsten, M.~J.~Duff, S.~Ferrara, A.~Marrani,
W.~Rubens, Small Orbits,\ Phys.\ Rev.\ \textbf{D85} (2012) 086002, \texttt{%
arXiv:1108.0424 [hep-th]}.

\bibitem{Marrani-Riccioni-Romano} A.~Marrani, F.~Riccioni, L.~Romano,
\textit{Real weights, bound states and duality orbits}, Int.\ J.\ Mod.\
Phys.\ \textbf{A31} (2016) no.01, 1550218, \texttt{arXiv:1501.06895 [hep-th]}%
.

\bibitem{Riccioni:2008jz} F.~Riccioni, A.~Van Proeyen, P.~C.~West, \textit{%
Real forms of very extended Kac-Moody algebras and theories with eight
supersymmetries}, JHEP \textbf{0805} (2008) 079, \texttt{arXiv:0801.2763
[hep-th]}.

\bibitem{Barton:2000ki} C.~H.~Barton and A.~Sudbery, \textit{Magic squares
of Lie algebras}, \texttt{math/0001083 [math-ra]}.

\bibitem{GNK} M. G\"{u}naydin, K. Koepsell, H. Nicolai, \textit{Conformal
and Quasi-Conformal Realizations of Exceptional Lie Groups}, Commun. Math.
Phys. \textbf{221}, 57 (2001), \texttt{hep-th/0008063}.

\bibitem{West:2001as} P.~C.~West, $\mathit{E}_{11}$\textit{\ and M theory},
Class.\ Quant.\ Grav.\ \textbf{18} (2001) 4443, \texttt{hep-th/0104081}.

\bibitem{Schnakenburg:2001he} I.~Schnakenburg, P.~C.~West, \textit{Kac-Moody
symmetries of IIB supergravity}, Phys.\ Lett.\ \textbf{B517} (2001) 421,
\texttt{hep-th/0107181}.

\bibitem{Riccioni:2007au} F.~Riccioni, P.~C.~West, \textit{The }$\mathit{E}%
_{11}$\textit{\ origin of all maximal supergravities}, JHEP \textbf{0707}
(2007) 063, \texttt{arXiv:0705.0752 [hep-th]}.

\bibitem{Bergshoeff:2007qi} E.~A.~Bergshoeff, I.~De Baetselier, T.~A.~Nutma,
$\mathit{E}_{11}$\textit{\ and the embedding tensor}, JHEP \textbf{0709}
(2007) 047, \texttt{arXiv:0705.1304 [hep-th]}.

\bibitem{Kleinschmidt:2003mf} A.~Kleinschmidt, I.~Schnakenburg, P.~C.~West,
\textit{Very extended Kac-Moody algebras and their interpretation at low
levels}, Class.\ Quant.\ Grav.\ \textbf{21} (2004) 2493, \texttt{%
hep-th/0309198}.

\bibitem{FG-1} S. Ferrara, M. G\"{u}naydin, \textit{Orbits of Exceptional
Groups, Duality and BPS States in String Theory}, Int. J. Mod. Phys. \textbf{%
A13} (1988) 1075, \texttt{hep-th/9708025}.

\bibitem{LPS-1} H. Lu, C. N. Pope, K. S. Stelle, \textit{Multiplet
structures of BPS solitons}, Class. Quant. Grav. \textbf{15} (1998) 537,
\texttt{hep-th/9708109}.

\bibitem{F-Maldacena-1} S. Ferrara, J. M. Maldacena, \textit{Branes, central
charges and U duality invariant BPS conditions}, Class. Quant. Grav. \textbf{%
15} (1998) 749, \texttt{hep-th/9706097}.

\bibitem{ICL-1} L. Borsten, D. Dahanayake, M.J. Duff, S. Ferrara, A.
Marrani, W. Rubens, \textit{Observations on Integral and Continuous
U-duality Orbits in }$\mathcal{N}\mathit{=8}$\textit{\ Supergravity}, Class.
Quant. Grav. \textbf{27}, 185003 (2010), \texttt{arXiv:1002.4223 [hep-th]}.

\bibitem{FG-2} S. Ferrara, M. G\"{u}naydin, \textit{Orbits and Attractors
for }$\mathcal{N}\mathit{=2}$\textit{\ Maxwell-Einstein Supergravity
Theories in Five Dimensions}, Nucl. Phys. \textbf{B759}, 1 (2006), \texttt{%
arXiv:hep-th/0606108}.

\bibitem{Marrani-Pradisi-Riccioni-Romano} A. Marrani, G. Pradisi, F.
Riccioni, L. Romano, \textit{Nonsupersymmetric magic theories and Ehlers
truncations}, Int. J. Mod. Phys. \textbf{A32} (2017) no.19n20, 1750120,
\texttt{arXiv:1701.03031 [hep-th]}.

\bibitem{Englert:2003zs} F.~Englert, L.~Houart, A.~Taormina, P.~C.~West,
\textit{The Symmetry of M theories}, JHEP \textbf{0309} (2003) 020, \texttt{%
hep-th/0304206}.

\bibitem{Ferrara:2012zc} S.~Ferrara, A.~Marrani, M.~Trigiante, \textit{%
Super-Ehlers in Any Dimensio}n,\ JHEP \textbf{1211} (2012) 068, \texttt{%
arXiv:1206.1255 [hep-th]}.

\bibitem{triality} T. Springer, F. D. Veldkamp : \textit{\textquotedblleft
Octonions, Jordan Algebras and Exceptional Groups"}, Springer, 2013.

\bibitem{BGM-1} P. Breitenlohner, G. W. Gibbons, D. Maison, \textit{%
4-Dimensional Black Holes from Kaluza-Klein Theories}, Commun. Math. Phys.
\textbf{120}, 295 (1988).

\bibitem{G-Pavlyk-1} M.~G\"{u}naydin, O.~Pavlyk, \textit{Quasiconformal
Realizations of }$\mathit{E}_{6(6)}$\textit{, }$\mathit{E}_{7(7)}$\textit{, }%
$\mathit{E}_{8(8)}$\textit{\ and }$\mathit{SO(n+3,m+3)}$\textit{, }$\mathcal{%
N}\geq \mathit{4}$\textit{\ Supergravity and Spherical Vectors}, Adv.\
Theor.\ Math.\ Phys.\ \textbf{13} (2009) no.6, 1895, \texttt{arXiv:0904.0784
[hep-th]}.

\bibitem{F-Dual} L.~Borsten, M.~J.~Duff, S.~Ferrara, A.~Marrani, \textit{%
Freudenthal Dual Lagrangians}, Class.\ Quant.\ Grav.\ \textbf{30} (2013)
235003, \texttt{arXiv:1212.3254 [hep-th]}.

\bibitem{Bossard-N=8} G.~Bossard, \textit{Octonionic black holes}, JHEP
\textbf{1205} (2012) 113, \texttt{arXiv:1203.0530 [hep-th]}.

\bibitem{Cosmo} C.~Gao, X.~Chen, Y.~G.~Shen, \textit{Quintessence and
phantom emerging from the split-complex field and the split-quaternion field}%
,\ Gen.\ Rel.\ Grav.\ \textbf{48} (2016) no.1, 11, \texttt{arXiv:1501.03220
[gr-qc]}.

\bibitem{FTS} H. Freudenthal, \textit{Beziehungen der E7 und E8 zur
oktavenebene I-II}, Nederl. Akad. Wetensch. Proc. Ser. \textbf{57} (1954)
218--230. H. Freudenthal, \textit{Beziehungen der E7 und E8 zur oktavenebene
IX}, Nederl. Akad. Wetensch. Proc. Ser. \textbf{A62} (1959) 466--474. K.
Meyberg, \textit{Eine theorie der freudenthalschen tripelsysteme. i, ii},
Nederl. Akad. Wetensch. Proc. Ser. \textbf{A71} (1968) .

\bibitem{BS} C. H. Barton, A. Sudbery, \textit{Magic squares and matrix
models of Lie algebras}, Adv. in Math. \textbf{180}, 596 (2003), \texttt{%
math/0203010 [math.RA]}.

\bibitem{F-D=6} L. Andrianopoli, R. D'Auria , S. Ferrara, M.A. Lled\'{o},
\textit{Horizon geometry, duality and fixed scalars in six dimensions},
Nucl. Phys. \textbf{B528}, 218 (1998), \texttt{hep-th/9802147}.

\bibitem{Groups-type-E7} R. B. Brown, \textit{Groups of type E}$_{7}$, J.
Reine Angew. Math. \textbf{236} (1969) 79--102. S. Ferrara, R. Kallosh, A.
Marrani, \textit{Degeneration of Groups of Type E}$_{7}$\textit{\ and
Minimal Coupling in Supergravity}, JHEP \textbf{1206}, 074 (2012), \texttt{%
arXiv:1202.1290 [hep-th]}.

\bibitem{Sato-Kimura} M. Sato, T. Kimura, \textit{A classification of
irreducible prehomogeneous vector spaces and their relative invariants},
Nagoya Math. J. \textbf{65}, 1 (1977).

\bibitem{Igusa} J. Igusa, \textit{A classification of spinors up to
dimension twelve}, Am. J. of Math. \textbf{92} (1970), 997--1028.

\bibitem{Kac-80} V. G. Kac, \textit{Some Remarks on Nilpotent Orbits}, J. of
Algebra \textbf{64}, 190 (1980).

\bibitem{Garibaldi-1} H. Bermudez, S. Garibaldi, V. Larsen, \textit{Linear
preservers and representations with a 1-dimensional ring of invariants},
Trans. Amer. Math. Soc., vol. \textbf{366}, \#9 (2014), 4755-4780, \texttt{%
arXiv:1204.2840 [math.RT]}.

\bibitem{Garibaldi-2} S. Garibaldi, R. Guralnick, \textit{Simple groups
stabilizing polynomials}, Forum of Mathematics, Pi (2015), vol. \textbf{3},
e3, \texttt{arXiv:1309.6611 [math.GR]}.

\bibitem{rank-J} J. C. Ferrar, \textit{Strictly Regular Elements in
Freudenthal Triple Systems}, Trans. Am. Math. Soc. \textbf{174}, 313 (1972).

\bibitem{rank-FTS} S. Krutelevich, \textit{Jordan algebras, exceptional
groups, and Bhargava composition}, J. Algebra \textbf{314} (2007) no. 2,
924--977, \texttt{arXiv:math/0411104}.

\bibitem{Pure-Spinors} S. Giler, P. Kosinski, J. Rembielinski, \textit{On }$%
\mathit{SO(p,q)}$\textit{\ Pure Spinors}, Acta Phys. Pol. Vol. \textbf{B18},
no. 8, 713 (1987).

\bibitem{Gunaydin-rev} M. G\"{u}naydin, \textit{Lectures on Spectrum
Generating Symmetries and U-duality in Supergravity, Extremal Black Holes,
Quantum Attractors and Harmonic Superspace}, Springer Proc. Phys. \textbf{134%
}, 31 (2010), \texttt{arXiv:0908.0374 [hep-th]}.

\bibitem{Pasquier} B. Pasquier, \textit{Introduction to spherical varieties
and description of special classes of spherical varieties}, Lectures given
at KIAS (Seoul, South Korea), available at : \texttt{www.math.univ-montp2.fr/%
\symbol{126}pasquier/KIAS.pdf}.

\bibitem{Rios} M. Rios, \textit{Extremal Black Holes as Qudits}, \texttt{%
arXiv:1102.1193 [hep-th]}.

\bibitem{AM-Refs} S. Ferrara, R. Kallosh, A. Strominger, $\mathcal{N}\mathit{%
=2}$ \textit{extremal black holes}, Phys. Rev. \textbf{D52}, 5412 (1995),
\texttt{hep-th/9508072}. A. Strominger, \textit{Macroscopic entropy of }$%
\mathcal{N}\mathit{=2}$ \textit{extremal black holes}, Phys. Lett. \textbf{%
B383}, 39 (1996), \texttt{hep-th/9602111}. S. Ferrara, R. Kallosh, \textit{%
Supersymmetry and attractors}, Phys. Rev. \textbf{D54}, 1514 (1996), \texttt{%
hep-th/9602136}. S. Ferrara, R. Kallosh, \textit{Universality of
supersymmetric attractors}, Phys. Rev. \textbf{D54}, 1525 (1996), \texttt{%
hep-th/9603090}.S. Ferrara, G.W. Gibbons, R. Kallosh, \textit{Black Holes
and Critical Points in Moduli Space}, Nucl. Phys. \textbf{B500}, 75 (1997),
\texttt{hep-th/9702103}.

\bibitem{Sudbery} A. Sudbery, \textit{Division Algebras, (Pseudo)Orthogonal
Groups and Spinors}, J. Phys. \textbf{A17}, 939 (1984).

\bibitem{Kugo-Townsend} T. Kugo, P. K. Townsend, \textit{Supersymmetry and
the Division Algebras}, Nucl. Phys. \textbf{B221}, 357 (1983).

\bibitem{FLM-1} R. Fioresi, E. Latini and A. Marrani, \textit{Klein and
Conformal Superspaces, Split Algebras and Spinor Orbits}, Rev. Math. Phys.
\textbf{29} (2017) 0011, \texttt{arXiv:1603.09063 [hep-th]}.
\end{thebibliography}
\end{document}